\documentclass[twocolumn,aps,superscriptaddress,showpacs,nofootinbib,floatfix]{revtex4}
\usepackage{epsfig,bm,feynmf}
\usepackage{graphics}
\usepackage{amsmath}
\usepackage[normalem]{ulem}
\usepackage[dvipsnames]{color}
\usepackage{morefloats}
\usepackage{array}
\usepackage{subfigure}

\begin{document}

\title{Probing the hot and dense nuclear matter with $K^*, \bar K^*$ vector mesons}


\author{Andrej Ilner}\email{ilner@fias.uni-frankfurt.de}
\affiliation{Institut f\"ur Theoretische Physik, Johann Wolfgang Goethe-Universit\"at Frankfurt am Main, 60438 Frankfurt am Main, Germany}
\affiliation{Frankfurt Institute for Advanced Studies (FIAS), 60438 Frankfurt am Main, Germany}

\author{Justin Blair}
\affiliation{The University of Texas at Austin, Physics Department, Austin, Texas, USA}

\author{Daniel Cabrera}
\affiliation{Instituto de F\'{\i}sica Corpuscular (IFIC), Centro Mixto Universidad de Valencia - CSIC, Institutos de Investigaci\'on de Paterna, Ap. Correos 22085, E-46071 Valencia, Spain.}

\author{Christina Markert}
\affiliation{The University of Texas at Austin, Physics Department, Austin, Texas, USA}

\author{Elena Bratkovskaya}
\affiliation{GSI Helmholtzzentrum f\"{u}r Schwerionenforschung GmbH Planckstrasse 1, 64291 Darmstadt, Germany}
\affiliation{Institut f\"ur Theoretische Physik, Johann Wolfgang Goethe-Universit\"at Frankfurt am Main, 60438 Frankfurt am Main, Germany}

\begin{abstract}
We investigate the possibility of probing the hot and dense nuclear matter - 
created in relativistic heavy-ion collisions (HIC) - with strange vector mesons ($K^{*}, \bar K^*$).
Our analysis is based on the non-equilibrium  Parton-Hadron-String Dynamics (PHSD) 
transport approach which incorporates  partonic and hadronic degrees-of-freedom
and describes the full dynamics of HIC on a microscopic level --
starting from the primary nucleon-nucleon collisions to the formation
of the strongly interacting Quark-Gluon-Plasma (QGP), followed by dynamical hadronization
of (anti-)quarks as well as final hadronic elastic and inelastic interactions.
This allows to study the ${K}^{*}$ and $\bar{K}^{*}$ meson
formation from the QGP as well as the in-medium effects related to
the modification of their spectral properties during the propagation
through the dense and hot hadronic environment in the expansion phase.
We employ relativistic Breit-Wigner spectral functions for the $K^{*},\bar{K}^{*}$ mesons
with self-energies obtained from a self-consistent
coupled-channel G-matrix approach to study the role of in-medium effects
on the ${K}^{*}$ and $\bar{K}^{*}$ meson dynamics in heavy-ion collisions
from  FAIR/NICA to  LHC energies.
According to our analysis most of the final ${K}^{*}/\bar K^*$'s, that can be 
observed experimentally by reconstruction of the invariant mass of $\pi + K(\bar K)$ pairs,  
are produced during the late hadronic phase and originate dominantly from the 
$K (\bar K) + \pi \to K^*(\bar K^*)$ formation channel. The amount of ${K}^{*}/\bar K^*$'s 
originating from the QGP channel is comparatively small even at LHC energies and those 
${K}^{*}/\bar K^*$'s can hardly be reconstructed experimentally due to the rescattering 
of final pions and (anti-)kaons. This mirrors the results from our previous study on the strange vector-meson production in heavy-ion collisions at RHIC energies. 
We demonstrate that ${K}^{*}/\bar K^*$ in-medium effects should be visible at 
FAIR/NICA and BES RHIC  energies, where the production of ${K}^{*}/\bar K^*$'s occurs at larger net-baryon densities.
Finally, we present the experimental procedures to extract the information on 
the resonance masses and widths by fitting the final mass spectra at LHC energies.
\end{abstract}


\maketitle

\section{Introduction}\label{sec:intro}

The properties of hot and dense matter under extreme conditions, the origin of the phase
transition from hadronic to partonic matter and the formation of the Quark-gluon Plasma (QGP)  are
the subjects of extensive theoretical and experimental studies in the last decades.
Such conditions - realized in nature during the Big Bang at
the beginning of our universe - can be achieved nowadays in the laboratory
in the collisions of heavy ions. There are presently several experimental facilities like
the Schwerionensynchrotron (SIS) at the Gesellschaft f\"ur Schwerionenforschung (GSI),
the Relativistic Heavy Ion Collider (RHIC) at the Brookhaven National Laboratory (BNL),
the Super-Proton Synchrotron (SPS) and the Large Hadron Collider (LHC)
at the Conseil Europ\'een pour la Recherche Nucl\'eaire (CERN)
which cover the range in invariant energy $\sqrt{s_{NN}}$ from a few GeV at SIS to $\sim$ 5 TeV at LHC.
Moreover, two further accelerators are under construction --
the Facility for Antiproton and Ion Research (FAIR)
as well as the Nuclotron-based Ion Collider fAcility (NICA) --
which will become operational in the next years.

Due to  confinement the QGP can not be observed directly in experiments,
which measure the final hadrons and leptons, and thus one needs reliable observables
which carry information about the initial stages when the QGP has been created.
Electromagnetic probes have the advantage that they practically do not suffer from final
state interactions with the matter, however, they are very rare and it is hard to detect
them experimentally with high statistics. On the other hand hadrons are very abundant and rather easy to detect, however, they participate in strong hadronic interactions after their creation which distort
the information about their origin to some extent.

In view of many hadronic probes the strange hadrons are of special interest since strangeness is not
initially present in the colliding nuclei, but created during their collisions. Thus one
hopes that it is easier to keep track of their production mechanism. In particular
the strange vector-meson resonances ${K}^{*}$ and $\bar{K}^{*}$ have been proposed as
possibly sensitive probes   \cite{Kstar_probe,Markert:2002rw}.
The ${K}^{*}/\bar K^*$'s are expected to be produced at the partonic freeze-out \cite{Markert:2008jc}
at relativistic energies and thus carry information about final QGP as well as
the hadronic phase due to the final-state interactions.

Apart from detailed studies that have been performed by
the STAR collaboration at RHIC \cite{Adams:2004ep,Aggarwal:2010mt,Kumar:2015uxe},
the ${K}^{*}/\bar K^*$ production has also been studied by the ALICE experiment at the
LHC \cite{Abelev:2012hy,Abelev:2014uua,Pulvirenti:2011xs,KarasuUysal:2012zz,Badala:2013dma,Nicassio:2013lla,Knospe:2013ir,Knospe:2013tda,Singha:2013spa,Knospe:2013lla,Abelev:2014uua,Fragiacomo:2014naa,Badala:2014bca,Ortiz:2015jto,Bellini:2015tra,Knospe:2015rja,Badala:2015vaa,Badala:2015qma,Knospe:2016hae,Adam:2016bpr}.
The measurement of the strange vector mesons ${K}^{*}/\bar K^*$ is quite challenging;
the reconstruction goes via the decay channel $K^*(\bar K^*) \to \pi + K(\bar K)$ by measuring 
the invariant mass of final pions and (anti-)kaons pairs. However, $K^*/\bar K^*$'s are rather 
short-lived resonances and, even if they are produced at the hadronization of the QGP, they
decay in the hadronic medium during the expansion of the system. The decay products - pions and (anti-)kaons -
suffer from hadronic final-state interactions - rescattering and absorption - which  leads to a distortion
of the $K^*/\bar K^*$'s spectra; furthermore,  a sizeable fraction of the decayed $K^*/\bar K^*$'s can not be reconstructed at all.
This is especially visible at LHC energies due to the large multiplicity of the final hadronic states.
Additionally, due to the high meson density the $K^*/\bar K^*$'s can be often produced in the hadronic medium
by formation of (anti-)kaons and pions $\pi + K(\bar K) \to {K}^{*}(\bar K^*)$. 
Moreover, this mechanism turns out to be dominant
as compared to the other production mechanisms of the ${K}^{*}/\bar K^*$ and the hadronisation from the QGP in particular.

Thus, in order to understand $K^*/\bar K^*$ production and to provide a robust interpretation of
the experimental results, one needs to address theoretical models.
The most suitable models for that are  transport approaches since only they can cover
the whole complexity of the $K^*/\bar K^*$ dynamics from a microscopic point of view.
We recall that such studies have been performed early with UrQMD \cite{Bleicher:2002dm,Vogel:2010pb}
and EPOS3 \cite{Knospe:2015nva} in Pb+Pb collisions at center-of-mass energies
of $\sqrt{{s}_{NN}} = 17.3$~GeV and $\sqrt{{s}_{NN}} = 2.76$~TeV, respectively.

Recently, we have studied the ${K}^{*}/\bar K^*$ production within the PHSD transport approach for
relativistic energies of $\sqrt{{s}_{NN}}=200$~GeV in Au+Au collisions at RHIC conditions \cite{Ilner:2016xqr}.
There we have investigated the different mechanisms for the $K^*/\bar K^*$ production and have shown that most of the
$K^*/\bar K^*$ measured experimentally at RHIC energies originate from $\pi + K(\bar K)$ annihilation and overshine the direct
production from the QGP which makes it quite difficult to use the $K^*/\bar K^*$ as a probe of the late partonic phase.
Additionally, rescattering and absorption of final pions and (anti-)kaons from $K^*/\bar K^*$ decay significantly distort
the final spectra. Moreover, in Ref. \cite{Ilner:2016xqr} we have studied for the first time the influence
of in-medium effects on the $K^*/\bar K^*$ dynamics and final observables in the hadronic phase  which are related to the modification
of the $K^*/\bar K^*$ spectral properties during the propagation through the dense and hot hadronic medium.
Such in-medium effects have been predicted by  chiral models and G-matrix approaches and
successfully been used for the description of the kaon $K$ and anti-kaon $\bar{K}$ production
\cite{Lutz:1997wt,Ramos:1999ku,Tolos:2000fj,Tolos:2006ny,Lutz:2007bh,Tolos:2008di}.
In Ref. \cite{Ilner:2016xqr} we have implemented in the PHSD the in-medium effects of $K^*/\bar K^*$ resonances based on
a G-matrix approach \cite{PHSDrev} as a function of the nuclear density \cite{Ilner:2013ksa}.
We found that the influence of in-medium effects on the final $K^*/\bar K^*$ spectra is rather modest since
most of $K^*/\bar K^*$'s are produced in the hadronic phase by $\pi + K(\bar K)$ annihilation when the net-baryon density and the temperature have become low due to the rapid expansion of the system at $\sqrt{{s}_{NN}} = 200$~GeV.

The aim of this study is to investigate the dynamics of the ${K}^{*}/\bar K^*$ vector mesons
in a wide energy range - from FAIR to LHC energies.
We address the following questions in our study: \\
i) Is the fraction of the $K^*/\bar K^*$'s produced from the QGP hadronization visible in the final observables?
Indeed, due to the larger volume of the QGP at the LHC (compared to the RHIC energies),
the total production of $K^*/\bar K^*$'s from the QGP is enhanced. However, the meson multiplicity at LHC
is also much larger than at RHIC, thus the $\pi + K(\bar K)$ annihilation mechanism would be also enhanced.
Accordingly, the task is to give a quantitative answer to this issue.\\
ii)  What is the role of the final-state interaction on the $K^*/\bar K^*$ decay products, i.e. pions and (anti-)kaons at the LHC? \\
iii) What is the quantitative effect of in-medium modifications of the $K^*/\bar K^*$ properties
on the final observables in HIC? \\
iv) Which conditions, i.e. colliding energies, are best suited to study the  $K^*/\bar K^*$ in-medium effects?
In order to answer the last question we will go down in energy and study the $K^*/\bar K^*$ production
also at the energies of the future FAIR/NICA and BES program at RHIC.\\
V) How one can subtract the information on the medium effects from the observables
using  experimental methods?
For that we will consider PHSD events in the same fashion as experimental data and apply
the experimental extraction and fitting procedures.

Throughout this paper the following convention will be used: strange vector mesons consisting of an anti-strange quark, i.e. ${K}^{*+} = (u\bar{s})$ and ${K}^{*0} = (d \bar{s})$, will be referred to as ${K}^{*} = ({K}^{*+},{K}^{*0})$, while for mesons with a strange quark, i.e. ${K}^{*-} = (\bar{u}s)$ and $\bar{K}^{*0} = (\bar{d} s)$, the convention $\bar{K}^{*} = ({K}^{*-},\bar{K}^{*0})$ will be used.
Often we will use a notation $K^*/\bar K^*$ for $K^*$ and $\bar K^*$.
Also in the text we often express masses in the units of GeV or MeV assuming GeV$/c^2$ or MeV$/c^2$.

The structure of this study is as follows:
In Section \ref{sec:phsd} we will give a short recall of the PHSD transport approach. In Section \ref{sec:ksmed} the ${K}^{*}/\bar K^*$ vector-meson resonance in-medium properties will be shortly discussed, i.e. spectral functions calculated from a state-of-the-art G-matrix model as well as the implementation of these spectral functions into PHSD. In Section \ref{sec:ksphsd} the properties and the dynamics of the ${K}^{*}/\bar K^*$ vector-meson resonance are investigated within PHSD for Pb+Pb collisions at LHC energies. Furthermore, the different production channels, the actual baryon densities and the in-medium effects of the ${K}^{*}/\bar K^*$ spectral functions  are analysed in detail. In Section \ref{sec:results} our results are presented and compared to the experimental data; in the first part we compare our results with data from p+p collisions at LHC; the second part contains a detailed comparison of PHSD results with data from Pb+Pb collisions at LHC energies from the ALICE collaboration. In Section \ref{sec:fair} we use PHSD to obtain results for lower bombarding energies 
4.5, 6, 8, 10.7 and 15 AGeV to give  predictions for the ${K}^{*}/\bar K^*$ in-medium dynamics 
at the future FAIR and NICA. 
In Section \ref{sec:experim} we discuss the experimental procedure to extract in-medium mass 
and width of a resonance.
Finally, we summarise our findings in Sec. \ref{sec:summary}.

\section{The PHSD transport approach}\label{sec:phsd}

Our study is based on the Parton-Hadron-String Dynamics (PHSD) approach, which  is a microscopic covariant dynamical approach for strongly interacting systems in and out-of equilibrium \cite{Cassing:2009vt,Bratkovskaya:2011wp}. The PHSD incorporates both partonic and hadronic degrees-of-freedom as well as  the transition
from the hadronic to the partonic phase, the QGP phase in terms of strongly interacting quasiparticles with further dynamical hadronization and final hadronic interactions in
the late stage;  thus, PHSD covers the full time evolution of a relativistic
heavy-ion collision on a microscopic level.
The dynamical description of the strongly interacting system is realized by solving
the generalised off-shell Cassing's transport equations which are obtained
from the Kadanoff-Baym equations \cite{Kadanoff1962,Juchem:2004cs,Cassing:2007nb} in first-order gradient expansion and go beyond the mean-field and on-shell Boltzmann approximation for the collision terms.

The theoretical description of the partonic degrees-of-freedom (quarks and gluons)
is realized in line with the Dynamical-Quasi-Particle Model (DQPM)
\cite{Cassing:2007yg,Cassing:2007nb} and describes the properties of QCD in terms
of resummed single-particle Green's functions. The three parameters of the DQPM are fitted to reproduce lQCD results in thermodynamical equilibrium \cite{Aoki:2009sc,Cheng:2007jq}
such as energy density, pressure and entropy density;
the real and imaginary parts of the parton self-energies are used to define the widths and pole positions of the spectral functions of quarks and gluons taken in relativistic Breit-Wigner form.
The DQPM provides the properties of the partons, i.e. masses and widths in their spectral functions as well as the mean fields for gluons/quarks and their effective 2-body interactions that are implemented in the PHSD. For details about the DQPM model and the off-shell transport approach we refer the reader to the reviews in Refs. \cite{Cassing:2008nn,PHSDrev}.
We mention, that in equilibrium the PHSD reproduces the partonic transport coefficients such as shear and bulk viscosities or the electric conductivity from lattice QCD (lQCD) calculations as well \cite{PHSDrev,Cassing2013}.

The hadronic part is governed by the Hadron-String-Dynamics (HSD) part of the transport approach \cite{Ehehalt:1996uq,Cassing:1999es}; the hadronic degrees-of-freedom include the baryon octet and decouplet, the ${0}^{-}$ and ${1}^{-}$ meson nonets as well as higher resonances.
In the beginning of relativistic heavy-ion collisions color-neutral strings (described by the LUND model~\cite{Andersson:1992iq}) are produced in highly energetic scatterings of nucleons from the impinging nuclei, i.e. two strings can form through primary NN collisions.  These strings are dissolved into 'pre-hadrons', i.e. unformed hadrons
with a formation time of $\tau_F \sim$ 0.8 fm/c in the rest frame of the corresponding string, except for the 'leading hadrons'. Those are the fastest residues of the string ends, which can re-interact (practically instantly) with hadrons with a reduced cross sections in line with quark counting rules.
If the energy density is below the critical value for the phase transition, 
which is taken to be $\mathcal{E}_C = 0.5$~$GeV/{fm}^{-3}$ (e.g. in p+p reactions or in the hadronic corona), 'pre-hadrons' become real hadrons after the formation time $t_F=\tau_F \gamma$ ($\gamma$ is the Lorentz gamma factor of the pre-hadron) in the calculational frame
(center-of-mass system of A+A) and interact with hadronic cross sections.
If the local energy density is larger than the critical value for the phase transition
$\mathcal{E}_C$, the pre-hadrons melt into (colored) effective quarks and antiquarks in their self-generated repulsive mean-field as defined by the DQPM~\cite{Cassing:2008nn}. In the DQPM the quarks, antiquarks and gluons are dressed quasi-particles and have temperature-dependent effective masses and widths which have been fitted to lattice thermal quantities such as energy density, pressure and entropy density.

For the time evolution of the QGP phase off-shell transport equations with self-energies and cross-sections from the DQPM are used. With the expansion of the fireball the probability that the partons hadronize increases strongly close to the phase boundary. The hadronisation is carried out on the basis of covariant transition rates. The resulting hadronic system is then governed by the off-shell HSD dynamics with optionally incorporated self-energies for the hadronic degrees-of-freedom \cite{HSDK}.

To summarise: the full evolution of a relativistic heavy-ion collision, from the initial hard NN collisions out-of equilibrium up to the hadronisation and final interactions of the resulting hadronic particles is fully realised in the PHSD approach. We recall that PHSD has been successfully employed for p+p, p+A and A+A reactions ranging from SIS to LHC energies (cf. Ref. \cite{PHSDrev} and references therein).
Furthermore, in Ref. \cite{Ilner:2016xqr} we have extended the PHSD approach to the explicit
$K^*/\bar K^*$ resonance dynamics  by implementing the in-medium effects in terms of density and temperature dependent spectral functions at the hadronization, production and propagation of $K^*, \bar K^*$'s.

\section{Reminder of ${K}^{*}$, $\bar{K}^{*}$ in-medium effects and implementation in PHSD}\label{sec:ksmed}

In this Section we briefly recall our approach used for the implementation of in-medium effects 
in the PHSD for the off-shell dynamics of the strange vector-meson resonances ${K}^{*}$ 
and $\bar{K}^{*}$ \cite{Ilner:2016xqr,Ilner:2013ksa}.

The in-medium properties strange mesons in dense and hot nuclear matter are determined by the meson 
self-energies  calculated based on chirally motivated effective field models 
implemented using the G-matrix approach. The G-matrix approach is a unitary 
self-consistent coupled-channel approach which, in this case, involves also 
vector mesons \cite{Bando:1985rf,Oset:2009vf,Oset:2012ap}
within the Hidden Local Symmetry approach 
\cite{Bando:1984ej,Bando:1987br,Harada:2003jx,Meissner:1987ge}. 
We use G-matrix approach from Ref. \cite{Tolos:2010fq} to calculate the self-energy of 
the  $\bar{K}^{*}$. 
The in-medium spectral functions for ${K}^{*}$ and $\bar{K}^{*}$
then are approximated by relativistic Breit-Wigner spectral functions
\cite{Ilner:2013ksa} with density or temperature dependent effective masses and widths
related to the real and imaginary parts of the G-matrix self-energies.
These Breit-Wigner spectral functions are used in the PHSD for the off-shell production 
and propagation of the strange vector-meson resonances ${K}^{*}$ and $\bar{K}^{*}$ in heavy-ion collisions \cite{Ilner:2016xqr}.

We remind here that in Ref. \cite{Ilner:2013ksa} the ${K}^{*}$ meson self-energy at 
threshold energy has been obtained within an effective Lagrangian \cite{Oset:2009vf}.
The collisional part of the self-energy of the ${K}^{*}$ stems from
summing the forward $K^*N$ scattering amplitude over the allowed nucleon states in the medium, schematically $\Pi^{\rm coll}_{K^*}=\sum_{\vec{p}}n(\vec{p}\,) T_{K^*N}$. Due to the absence of resonant states nearby, a $t\rho$ approximation is well justified at energies sufficiently 
close to threshold
\begin{align}
  \Pi^{\rm coll}_{K^{*}} 
 \simeq \alpha \frac{M_K}{M_{K^*}} M_{K^*}^2 \left(\frac{\rho}{\rho_{0}}\right),
\label{eq:Kstar-trho}
\end{align}
with $\alpha\simeq 0.13$.

The $\bar{K}^{*}$ collisional self-energy part was implemented by using a parametrisation of the $\bar{K}^{*}$ self-energy and spectral function from Ref.  \cite{Tolos:2010fq}, 
where a detailed analysis of the medium corrections and self-consistent evaluation 
of the in-medium $\bar{K}^{*} N$ scattering has been carried out.

For a realistic description of the production rates the decay width of the ${K}^{*}$ and $\bar{K}^{*}$ 
needs to be accounted including the in-medium modifications of the decay products, too.
Such effects are readily incorporated for the $\bar K^*$ within the $G$-matrix approach which we parametrize 
from Ref.~\cite{Tolos:2010fq}. For the $K^*$, instead, we evaluate explicitly its medium-modified 
$K^*\to K \pi$ decay width as follows \cite{Ilner:2013ksa},
\begin{align}
&  \Gamma_{V,\, \textrm{dec}} (M, \rho) =   \label{eq:vmdw}\\
&= \Gamma_{V}^{0} \left( \frac{M_{V}}{M} \right)^{2} 
  \frac{\int_{0}^{M - m_{\pi}} q^3 (M, M_j, M_\pi) A_{j}(M_j,\rho) \, dM_j}{\int_{M_{\rm min}}^{M_{V} - m_{\pi}} 
  q^3 (M_{V}, M_j, M_\pi) A_{j} (M_j,0) \, dM_j }. \nonumber
 \end{align}  
The indices are $j = K$ and $V = {K}^{*}$. Furthermore  
$q(M, M_j, M_\pi) = \sqrt{\lambda(M, M_j, M_{\pi})}/ 2 M$ with 
$\lambda(x,y,z) = \left[ x^{2} - (y + z)^{2} \right] \left[ x^{2} - (y - z)^{2} \right]$. 
${\Gamma}_{V}^{0}$ is the partial vector meson decay width and 
${M}_{V}$ is the pole mass of the resonance in vacuum. 
We use $\Gamma_{K^{*}}^{0} = 42$~MeV and $M_V=892$~MeV \cite{Beringer:1900zz} 
(and the same for $\bar{K}^{*}$), $M_\pi$ is the pion mass.
Eq.~(\ref{eq:vmdw}) accounts for the in-medium modification of the resonance decay  width by its decay products. In particular, we consider the fact that kaons (and anti-kaons) may acquire a broad spectral function in the medium, $A_{j}(M,\rho)$. As discussed in Ref. \cite{Ilner:2013ksa}, the kaon spectral function $A_K$ in Eq.~(\ref{eq:vmdw}) is a delta function in vacuum since the kaon is stable in vacuum with respect to the strong interaction, and to a good approximation the same can be kept at finite nuclear density by using an effective kaon mass $M_K^{*\, 2}(\rho) = M_K^2 + \Pi_K(\rho)$ with $\Pi_K(\rho) \simeq 0.13 M_K^2 (\rho/\rho_0)$ \cite{Kaiser:1995eg,Oset:1997it,Tolos:2008di}. 

Pions are assumed to stay as narrow quasiparticles with vacuum properties 
in the evaluation of $\Gamma_{V,dec}$ throughout this work.
However, we mention that the in-medium modification of the pion spectral function 
in hot and dense hadronic matter is an unsettled question (cf. \cite{Ko:1996yy}):
while from the experimental side there is no clear evidence for the pion broadening 
in heavy-ion collisions, there are many different models with quite contradictory 
results - such in Ref. \cite{Haglin94} a very large width of pions at zero momentum
(up to 200 MeV at $T=150$~MeV) was predicted
due to there interaction in a hot meson gas while 
according to Refs.\cite{Goity:1989gs,Schenk:1993ru} the pions are expected to 
experience small changes up to temperatures $T \sim 100$ MeV in hot matter with 
low baryonic content, which supports our approximation of on-shell pions.
This topic requires further theoretical (as well as experimental) 
investigation which is beyond the scope of our present study.

The spectral function is proportional to the imaginary part of the vector-meson in-medium propagator and has the following form:
\begin{align}
&S_{V} (\omega, \vec{q}; \rho) = - \frac{1}{\pi} {\rm Im} D_V(\omega, \vec{q}; \rho) \label{eq:gmsf} \\
  &=-\frac{1}{\pi} \frac{\textrm{Im}\,\Pi_{V} (\omega,\vec{q};\rho)}{\left[ \omega^{2} - \vec{q}\,^{2} -  M_{V}^{2} - \textrm{Re}\,  \Pi_{V} (\omega, \vec{q}; \rho)  \right]^{2} + \left[ \textrm{Im}\,  \Pi_{V} (\omega, \vec{q}; \rho) \right]^{2}} ,  \nonumber 
\end{align}
where $V={K}^{*}, \bar{K}^{*}$ and ${\Pi}_{V}$ is the sum of the decay and collisional self-energy.

In spite that the G-matrix spectral function (\ref{eq:gmsf}) contains the 
full information on the in-medium properties of strange vector mesons, 
for the practical purpose of implementation of in-medium effects in the microscopic transport approach PHSD (${K}^{*}$ and $\bar{K}^{*}$ production and off-shell propagation)
we approximate the G-matrix spectral function (\ref{eq:gmsf}) by the 
relativistic Breit-Wigner spectral function  
($A_{V} (M,\rho), \ V=K^*,\bar K^*$) within the assumption of small momentum 
$\vec q \to 0$ -- see \cite{Bratkovskaya:2007jk,Ilner:2013ksa}: 
\begin{align}
  A_{V} (M,\rho) = C_{1} \, \frac{2}{\pi} \frac{M^{2} \Gamma_{V}^{*} (M,\rho)}{\left(M^{2} - {M_V^{*}}^{2} (\rho) \right)^{2} + \left( M \Gamma_{V}^{*} (M,\rho) \right)^{2}} \ ,
\label{eq:sfrbw}
\end{align}
where $M$ is the invariant mass and ${C}_{1}$ is a constant fixed by normalization:
\begin{align}
  \int_{0}^{\infty} A_V \left( M, \rho \right) ~ dM = 1.
\label{eq:sumrule}
\end{align}
Note, that the dimension of $A(M)$ is GeV$^{-1}$ since $M$ is included
in (\ref{eq:sumrule}).

As mentioned above the Breit-Wigner spectral function (\ref{eq:sfrbw}) follows from
the spectral function from the G-Matrix approach (\ref{eq:gmsf}) when setting the three-momentum of the vector meson to zero,
\begin{align}
  A_V (M, \rho) = 2 \cdot {C}_{1} \cdot M \cdot S_V (M, \vec{0}, \rho),
\label{eq:sfs}
\end{align}
i.e. neglecting the explicit momentum-dependence. However, the self-energy is evaluated in such a way that it is at rest in the nuclear medium which is consistent with the aforementioned approximation. The in-medium mass ${M}_{V}^{*}$ and the width ${\Gamma}_{V}^{*}$ of the spectral function are related to the real and imaginary part of the self-energy in the following way,
\begin{align}
  (M_V^*)^2 &= M_V^2+\textrm{Re}\,\Pi_V(M_V^*,\rho) \ , \nonumber \\
  \Gamma_V^*(M,\rho) &= - \frac{1}{2 M} \cdot \textrm{Im} \, \Pi_V(M,\rho) \ ,
\label{eq:mass-width}
\end{align}
where ${M}_{V}$ is the pole mass of the resonance in vacuum.

\begin{figure}[h]
  \centerline{\includegraphics[width=9.5cm]{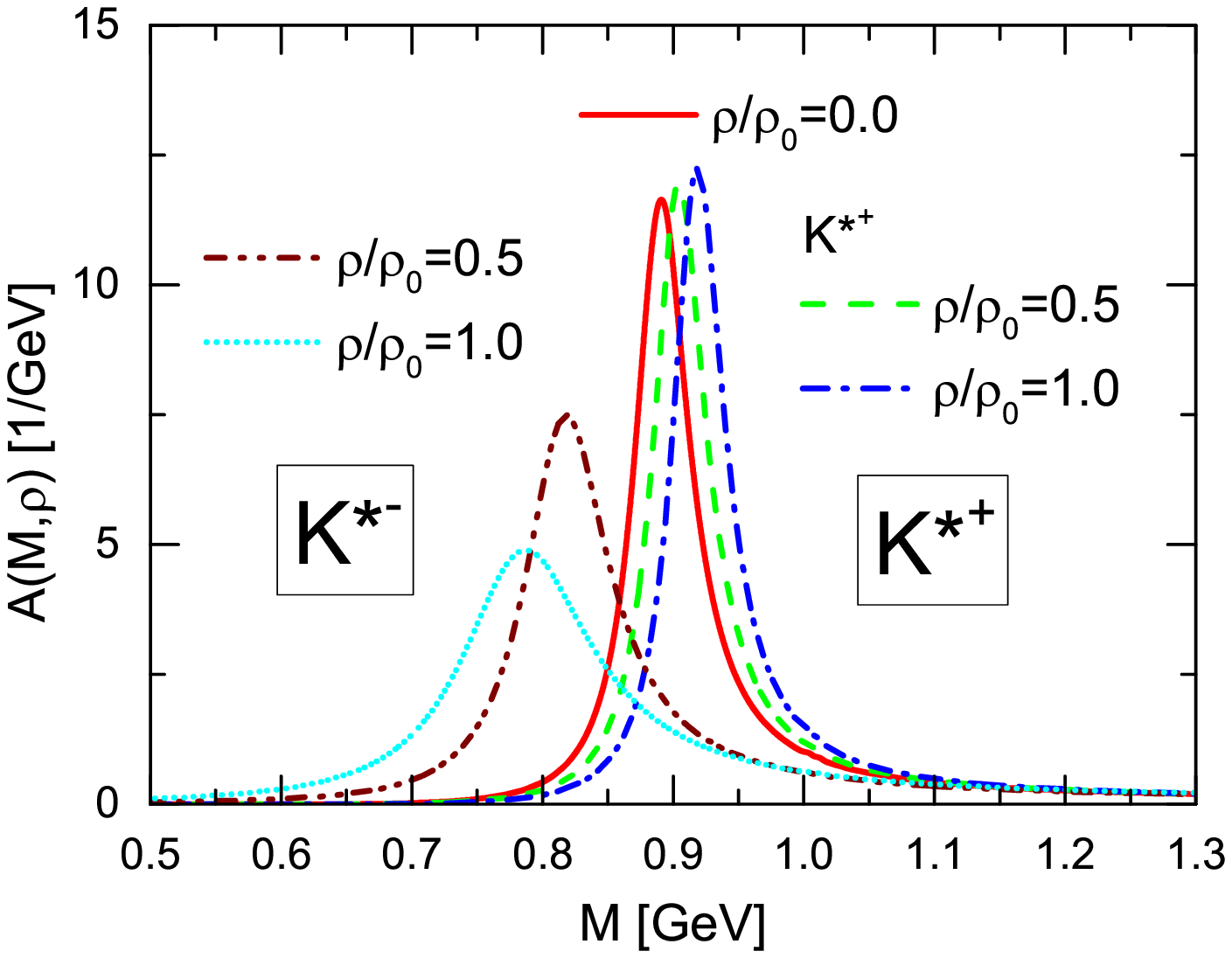}}
  \caption{The relativistic Breit-Wigner spectral functions $A(M,\rho)$ of the ${K}^{*}$ and the $\bar{K}^{*}$ versus the invariant mass $M$ for different nuclear baryon densities and in vacuum. The red solid line shows the vacuum ${K}^{*}/\bar{K}^{*}$ spectral functions, the green dashed line stands for the spectral function at baryon density of $\rho/{\rho}_{0}=0.5$,
 the blue dash-dotted line -- at $\rho/{\rho}_{0}=1.0$, the wine-coloured dash-dot-dotted line --
at $\rho/{\rho}_{0}=0.5$, the light blue short-dotted line -- at $\rho/{\rho}_{0}=1.0$.}
\label{fig:specfun}
\end{figure}
\begin{figure}[h]
  \centerline{\includegraphics[width=9.5cm]{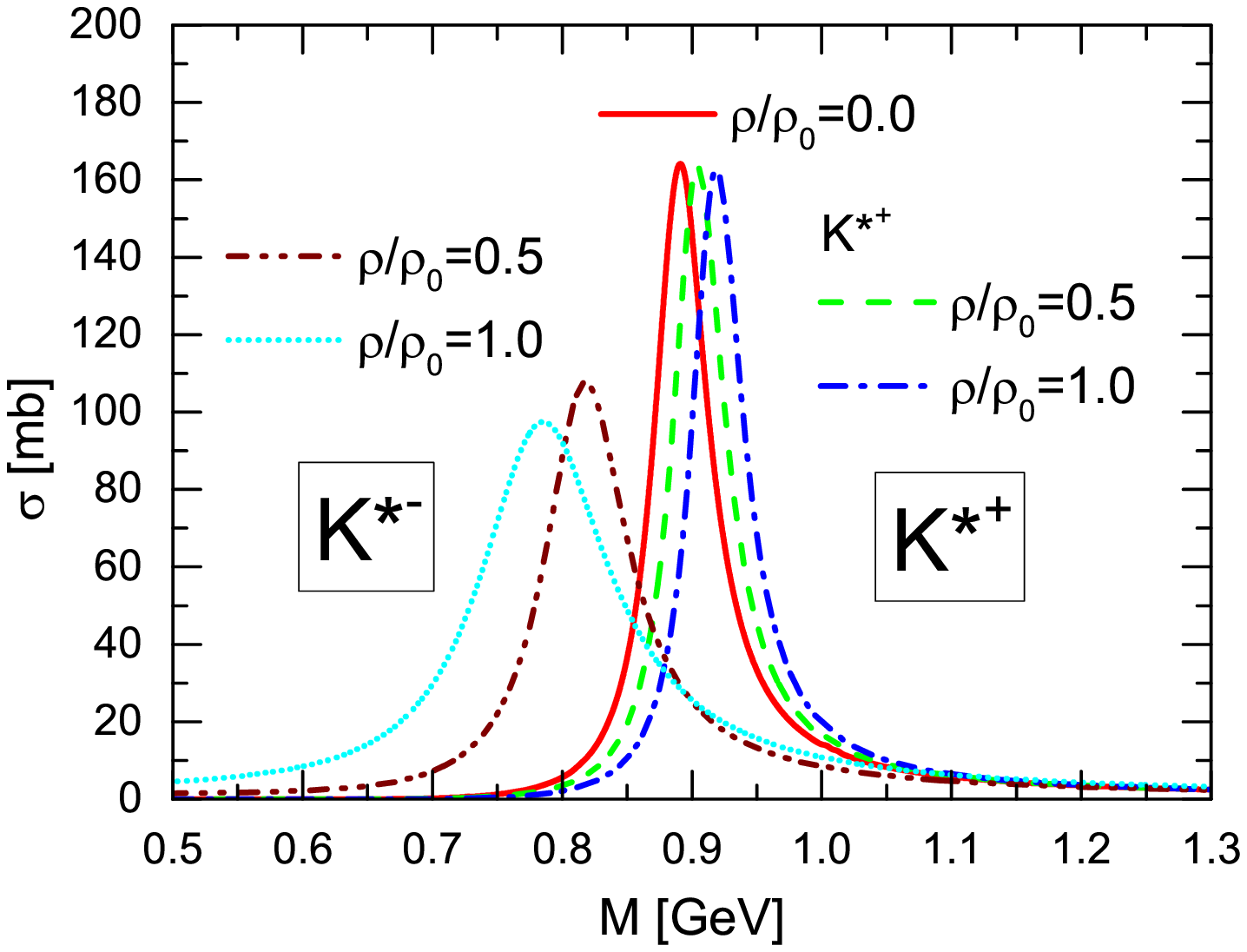}}
  \caption{The cross-section $\sigma$ for $K^{*}$ production/annihilation is shown as
a function of the invariant mass $M$ for different baryon densities and in vacuum.
The red solid line shows the vacuum ${K}^{*}/\bar{K}^{*}$ cross sections. The green dashed line shows the cross sections of ${K}^{*}/\bar{K}^{*}$ for a baryon density of $\rho/{\rho}_{0}=0.5$, the blue dash-dotted line -- at $\rho/{\rho}_{0}=1.0$, the wine-coloured dash-dot-dotted line -- at $\rho/{\rho}_{0}=0.5$, the light blue short-dotted line -- at $\rho/{\rho}_{0}=1.0$.}
  \label{fig:crosssec}
\end{figure}

Fig. \ref{fig:specfun} shows the spectral function for both the ${K}^{*}$s and $\bar{K}^{*}$s in vacuum and at finite nuclear density. While the vacuum spectral function has a width of $42$~MeV and is centred around its respective pole mass of $892$~MeV, a shift to higher and lower invariant masses can be seen for ${K}^{*}$ and $\bar K^*$ at non-zero nuclear density. The ${K}^{*}$ experiences a slight shift to higher invariant masses, the width is anti-proportional to the nuclear density since the kaon also gets heavier. The threshold energy for ${K}^{*}$ creation follows ${M}_{th} = {M}_{K} + {M}_{\pi} + \Delta M (\rho) \approx 0.633$ GeV + $\Delta M (\rho)$, with $\Delta M (\rho) \simeq \Pi_K(\rho)/2M_K$, which is approximately $0.06\,M_K$ at normal matter density.

The $\bar{K}^{*}$ on the other hand experiences a strong attraction and the spectral function is therefore shifted to lower invariant masses. The width gets considerably broader with increasing nuclear density. The threshold energy for the creation of a $\bar{K}^{*}$ decreases to  ${M}_{th} \sim 2 {M}_{\pi}$, i.e. an off-shell $\bar{K}^{*}$ can be created also at  low invariant masses.

These in-medium effects have also an effect on the production rates of the ${K}^{*}$ 
and the $\bar{K}^{*}$ in the hadronic phase of a heavy-ion collision. 
For the production cross section of ${K}^{*}$ and $\bar{K}^{*}$ via
$K+\pi$ or $\bar K +\pi$ annihilation we use the Breit-Wigner
cross section \cite{BW,PDG,Haglin94,Ko96} in relativistic form which can be written
with the help of the in-medium spectral function (\ref{eq:sfrbw}) as
\begin{align}
  \sigma_{K^*(\bar K^*)}(M,\rho) &= \frac{6 {\pi}^{2} ~ A_{K^*(\bar K^*)}(M,\rho) ~ {\Gamma}_{K^*(\bar K^*)}^{*} (M,\rho)}{{q (M,{M}_{K},{M}_{\pi})}^{2}} \ .
\label{eq:cross-sec-med}
\end{align}

Fig. \ref{fig:crosssec} shows the production cross-section of ${K}^{*}/\bar{K}^{*}$ vector mesons at different baryon densities. The structure is similar to  Fig. \ref{fig:specfun} depicting
the spectral functions, i.e. the vacuum cross section is at the center, while the cross-sections shifted to higher invariant masses corresponds to the ${K}^{*}$ and the cross-section shifted to lower invariant masses corresponds to the $\bar{K}^{*}$. The  $\bar{K}^{*}$ 
cross-section follows the same trend as the spectral function, i.e. the cross-section becomes smaller with increasing nuclear baryon density. However, for the ${K}^{*}$ a reversed effect emerges as compared to the spectral function, i.e. the cross-section of the ${K}^{*}$ slightly decreases with
increasing nuclear density due to a reduction of the phase space when the
mass increases.

We note, that in the present study we have omitted 'thermal' in-medium effects
related to a coupling to the hot mesonic medium since according to our
previous study \cite{Ilner:2013ksa} such effects are small. 
This topic requires further detailed investigation since the meson density at the LHC energies is very high.

\section{$K^{*}/\bar K^*$ dynamics in PHSD}\label{sec:ksphsd}

In this section we present our results on the on- and off-shell dynamics of the ${K}^{*}/\bar K^*$ within the PHSD transport approach at center-of-mass energies of $\sqrt{{s}_{NN}}=2.76$~TeV in central Pb+Pb collisions. We investigate the evolution of the ${K}^{*}/\bar K^*$ abundance in time as well as the main contributing production channels. Furthermore, we compare the baryon density during the creation of the ${K}^{*}/\bar K^*$'s at different energies. For a better understanding of the effects of experimental cuts we refer to our last study in Ref. \cite{Ilner:2016xqr}.

\begin{figure}[h]
  \centerline{\includegraphics[width=8.0cm]{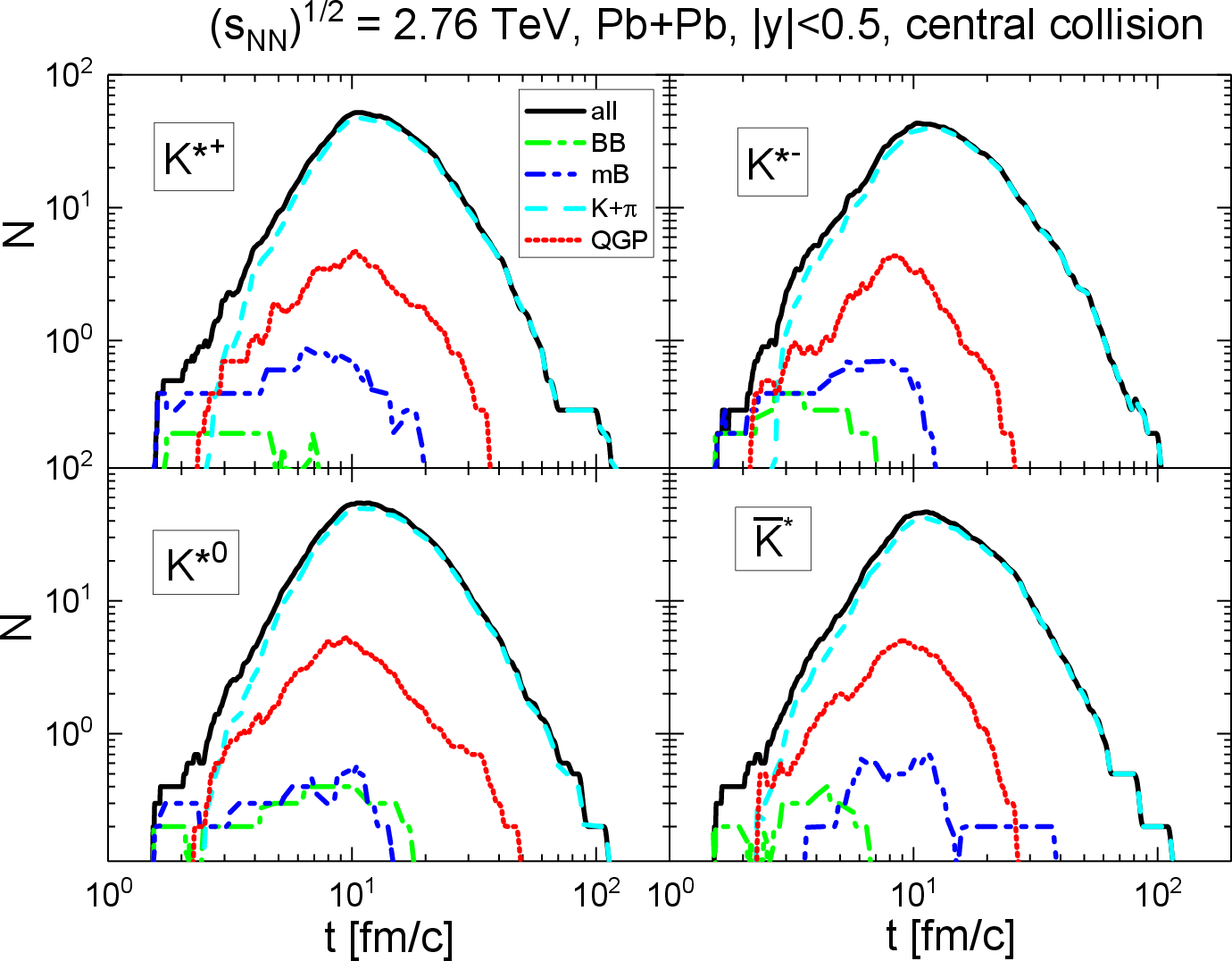}}
  \caption{The number $N(t)$ of formed strange vector mesons  versus the time $t$
for different production channels at mid-rapidity $y$ in a central Pb+Pb collision at
a center-of-mass energy of $\sqrt{{s}_{NN}}=2.76$~TeV from  PHSD calculations.
The upper left panel (a) shows the channel decomposition for the ${K}^{*+}$,
the upper right panel (b) -- for the ${K}^{*-}$, the lower left panel (c) --
for the ${K}^{*0}$ and the lower right panel (d) -- for the $\bar{K}^{*0}$.
The color coding is the same for all of the four panels:
the sum over all the production channels at given time $t$ is shown as a black solid line.
The dash-dotted green line shows the ${K}^{*}/\bar K^*$'s produced from baryon-baryon strings,
the dash-dot-dotted blue line -- from meson-baryon strings,
the dashed line  -- from $K(\bar K) + \pi$ annihilations and
the short-dotted red line -- from the hadronization of the QGP.}
\label{fig:nvst}
\end{figure}

We start with the time evolution of the ${K}^{*}/\bar{K}^{*}$ abundances.
As seen from Fig. \ref{fig:nvst} the total number of formed ${K}^{*}/\bar K^*$'s from all isospin channel is roughly the same, i.e. there is an approximately equal number of ${K}^{*}$'s and $\bar{K}^{*}$'s present at every time during the collision and expansion. 
The production by strings dominates at the very early stages of the collisions and very slowly decreases, although its relative contribution to the total number of ${K}^{*}$'s becomes negligibly small after about $10$~fm/c. We note, that the mesons coming from string decay 
are mainly dissolved to the QGP partons, they also can be 'leading' mesons if they
come from the ends of breaking string or stay as   under the formation time
in corona of the collisions where the energy density is not high enough to form the QGP. 
Since in Fig. \ref{fig:nvst} only the formed mesons are shown, 
pre-hadrons and headings are not accounted here.
The contribution of the QGP is not very large either, especially when compared to the main production channel, i.e. the $K(\bar K) + \pi$ annihilation. The contribution of the QGP starts a few fm/c later in the collision, however, its overall effect does not exceed the contribution from strings. Almost at the same time, when the QGP starts to contribute to the total number of particles, ${K}^{*}/\bar K^*$'s also start to emerge from $K(\bar K) + \pi$ annihilations. However, this channel rises quickly and stays the dominant channel throughout the collision.
Thus, we found that even at the LHC energy, in spite of the large volume of the created QGP,
 most of the $K^*$'s are produced by the resonant final-state interactions of $\pi + K (\bar K) $, similar
to the RHIC energies \cite{Ilner:2016xqr}.

\begin{figure}[h]
  \centerline{\includegraphics[width=8.5cm]{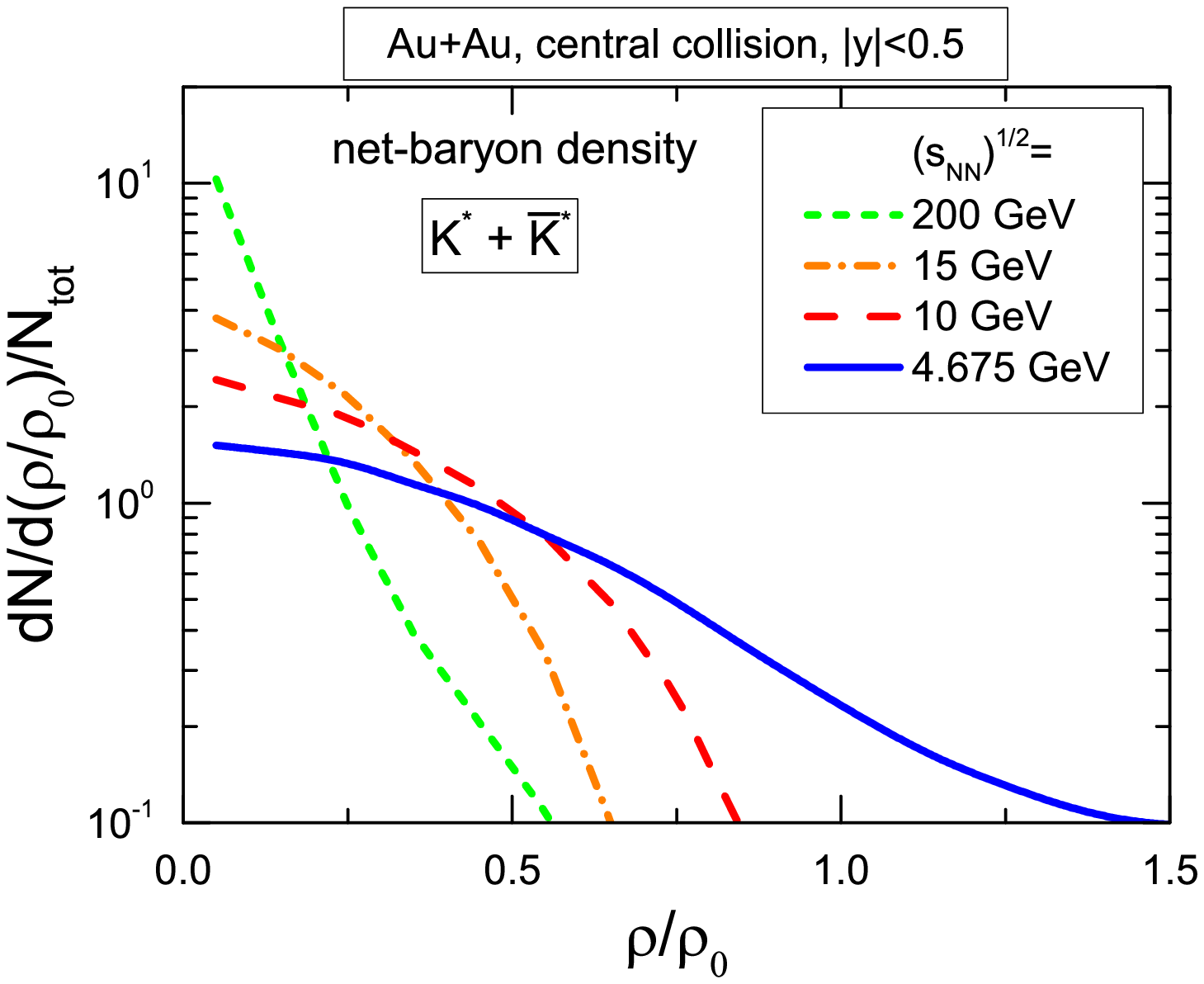}}
  \caption{The normalized net-baryon density distribution $dN/d\left(\rho/{\rho}_{0}\right)/{N}_{tot}$
at the production point of $K^*+\bar K^*$'s  for different collision  energies  
at midrapidity ($|y|<0.5$) as obtained from the PHSD calculations.
The short-dashed green line shows the result for Au+Au at 200~GeV, 
short-dot-dashed orange line -- at 15 GeV,
dashed red line -- at 10 GeV,
solid blue line -- at  4.675~GeV.}
\label{fig:dndrho}
\end{figure}

In view of the fact that $K^*$'s are produced dominantly in the final hadronic phase,
the questions arise: i) Which baryon density is probed with $K^*$'s?
ii) Can one observe an in-medium modification of the $K^*$ properties in the hadronic
environment and iii) which energies are more suited for a robust observation? In order to answer these questions
we show in Fig. \ref{fig:dndrho} the normalized net-baryon density distribution
$dN/d\left(\rho/{\rho}_{0}\right)/{N}_{tot}$  at the $K^*+\bar K^*$'s production point
for central Au+Au collisions at midrapidity ($|y|<0.5$)
for different collision energies of $\sqrt{{s}_{NN}}$=4.675~GeV, 10 GeV, 15 GeV
and 200 GeV .
As follows from Fig. \ref{fig:dndrho}, at high collision energies
the $K^*/\bar K^*$'s are produced at rather low net-baryon density since the dominant
production proceeds via $\pi + K$ annihilation (cf. Fig. \ref{fig:nvst}) when the system
is dominated by mesons with only a low amount of baryons and antibaryons. However, when decreasing the energy the fraction of $K^*/\bar K^*$'s created at larger density increases, such that for $\sqrt{{s}_{NN}}=4.675$~GeV one can probe
baryon densities even above normal nuclear matter density. Correspondingly, the in-medium
effects are expected to be more pronounced at low energies (e.g. at FAIR and NICA or low BES RHIC).

\section{Results from PHSD at LHC energies}\label{sec:results}

In this section we will present the results for $K^*/\bar K^*$ production in heavy-ion collisions 
and $pp$ reactions from the PHSD transport approach.
We compare these results to experimental data measured by the ALICE collaboration at the LHC
\cite{Abelev:2012hy,Abelev:2014uua,Pulvirenti:2011xs,KarasuUysal:2012zz,Badala:2013dma,Nicassio:2013lla,Knospe:2013ir,Knospe:2013tda,Singha:2013spa,Knospe:2013lla,Fragiacomo:2014naa,Badala:2014bca,Ortiz:2015jto,Bellini:2015tra,Knospe:2015rja,Badala:2015vaa,Badala:2015qma,Knospe:2016hae,Adam:2016bpr}.
Since the net-baryon density at mid-rapidity is very small at the LHC energies, 
we discard here a consideration of the in-medium effects and present the results for the 'free' case here.
We recall that in our in-medium scenario the modification of the $K^*/\bar K^*$ properties occurs 
due to the coupling to the baryonic medium while the coupling to anti-baryons is discarded.
However, at the LHC energies the anti-baryon density at midrapidity is close to the
baryon density, thus the expected in-medium modification of $K^*$ and $\bar K^*$
properties would be small since the coupling to the anti-baryons
lead to the opposite effect compared to the coupling to the baryons for the vector interaction.
A consistent consideration of the coupling to anti-baryons requires a further extension of the in-medium model which is a subject 
of future studies.

\subsection{'Decay' spectra versus 'reconstructed' spectra}

Some remarks on the $K^*/\bar K^*$  reconstruction procedure -- experimental as well as theoretical --
have to be made first:\\
i) The  ALICE Collaboration measures the ${K}^{*0}/\bar K^*$ vector mesons through the  hadronic decay channel:
${K}^{*0} \rightarrow {\pi}^{\pm} +  {K}^{\mp}$. The daughter particles of the ${K}^{*}$
can be tracked in the time projection chamber (TPC) which has a finite acceptance,
i.e. the resolution and accuracy of the detector also needs to be taken into account.
The main problem here is related to the fact that the decay products -- pions and kaons --
suffer from  final-state interactions (FSI) during the expansion phase:
they can rescatter or may be absorbed. This leads to a substantial distortion of the
reconstructed spectra which makes the physical interpretation of experimental
results rather difficult.
In our previous study \cite{Ilner:2016xqr} we have analysed the influence
of the FSI effects at RHIC energies, which were found to be significant.
It is expected also to play an important role at the LHC energy. 
\\
ii) The other problem with the experimental reconstruction of $K^*/\bar K^*$'s is related to the
background subtraction.
The ${K}^{*}/\bar K^*$ signal is obtained by taking all the viable decay channels into account and
combining all pions and (anti-)kaons from a single event. This will lead to the
'real' ${K}^{*}/\bar K^*$ signal on a top of a very large background (uncorrelated spectrum)
which is subtracted by a combinatorical method. The  signal and the shape
of the final $K^*/\bar K^*$ distribution is sensitive to the selected region around the 
$K^*/\bar K^*$ peak position
in the mass distribution. However, as discussed above, due to the FSI the signal is distorted,
thus a large fraction of $K^*/\bar K^*$'s cannot be reconstructed. 
Also experimental procedure to fit the signal and extract the information on
$K^*$'s is model dependent as will be discussed in Section VII.

\begin{figure}[t]
  \centerline{\includegraphics[width=8.0cm]{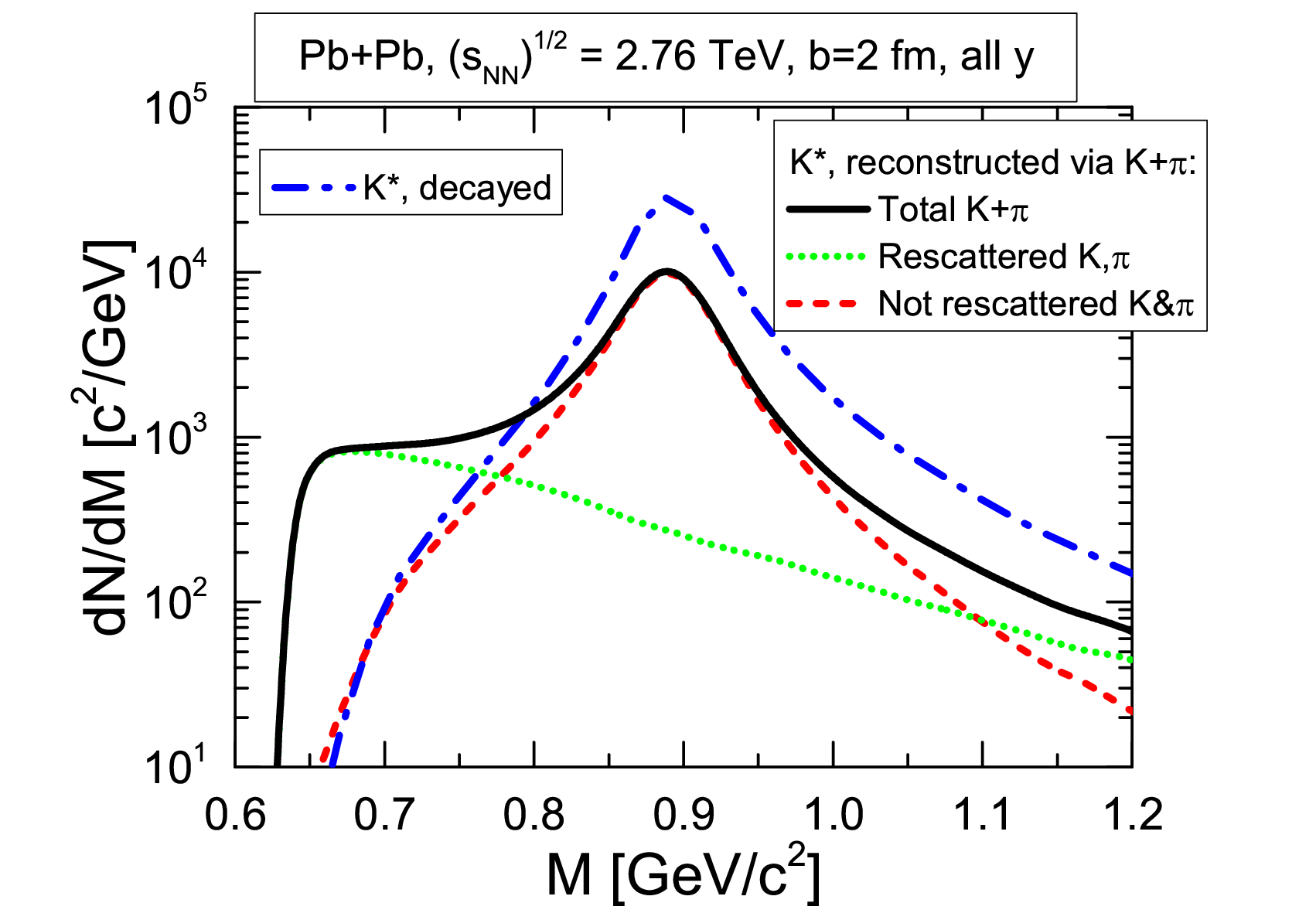}}
  \caption{The 'reconstructed'  $K^*={K}^{*+}+ K^{*0}$ mass distribution $dN/dM$ via $K+\pi$ 
  for Pb+Pb collision at impact parameter $b=2$~fm without cut in rapidity $y$
   at a center-of-mass energy  $\sqrt{{s}_{NN}} = 2.76$~TeV from the PHSD calculation: 
  the green dotted line shows the reconstructed $K+\pi$ mass distribution for the case 
  without rescattering of pions and kaons while the red dashed line stands for the
 case when pions and/or kaons rescatter elastically in the medium;
 the black solid line shows the sum of both contributions. 
 The blue dot-dashed line stands for the 'decayed' $K^*$ spectra.}
\label{fig:rescat}
\end{figure}

Contrary to  experiment, in theoretical calculations we can follow (within the microscopic
transport models) all $K^*/\bar K^*$'s in  their history
of  production and decay. In Ref. \cite{Ilner:2016xqr} we have compared the PHSD
spectra of $K^*/\bar K^*$'s at the decay point, i.e. 'true' $K^*/\bar K^*$'s, with the reconstructed $K^*/\bar K^*$'s
from the $\pi + K(\bar K)$ decay channel as in experiment. 
Following the same strategy, in our present theoretical study we will call 
the $K^*/\bar K^*$'s spectra obtained directly at the decay point as a 'decayed'
spectra. When $K^*/\bar K^*$ decay to pions and (anti-)kaons, we trace the 
collision history of the decay products: if pions and/or (anti-)kaons scatter elastically,
and finally both escape the fireball, we account them in the 'reconstructed' 
$K^*/\bar K^*$ spectra by summing their 4-momenta.
On the contrary, the pions and/or (anti-)kaons which approach inelastic (or quasielastic charge exchange) reactions are lost for the reconstruction
-- for our theoretical model and for the real experimental reconstruction -- since they did not 
escape the fireball and thus, did not reach the detector.
We note, however, that in our 'reconstructed' case - contrary to experiment - 
we do not have a 'loss' of the signal due to misidentification,
background subtraction or experimental acceptances. Thus, we use here our analysis
to illustrate the distortion of the spectra due to the final-state interaction in the hadronic phase.

\begin{figure}[h!]
  \centerline{\includegraphics[width=8cm]{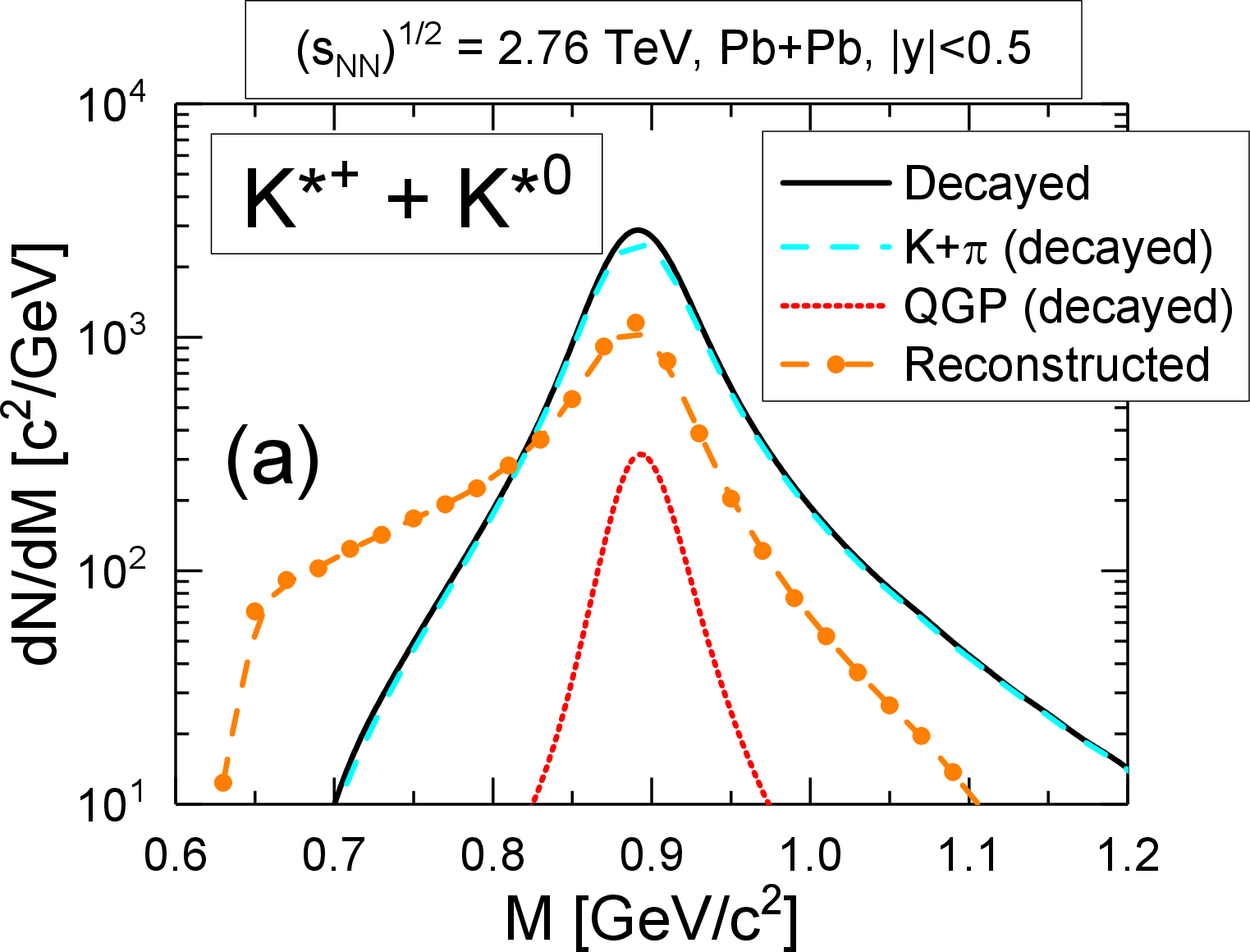}}
    \centerline{\includegraphics[width=8cm]{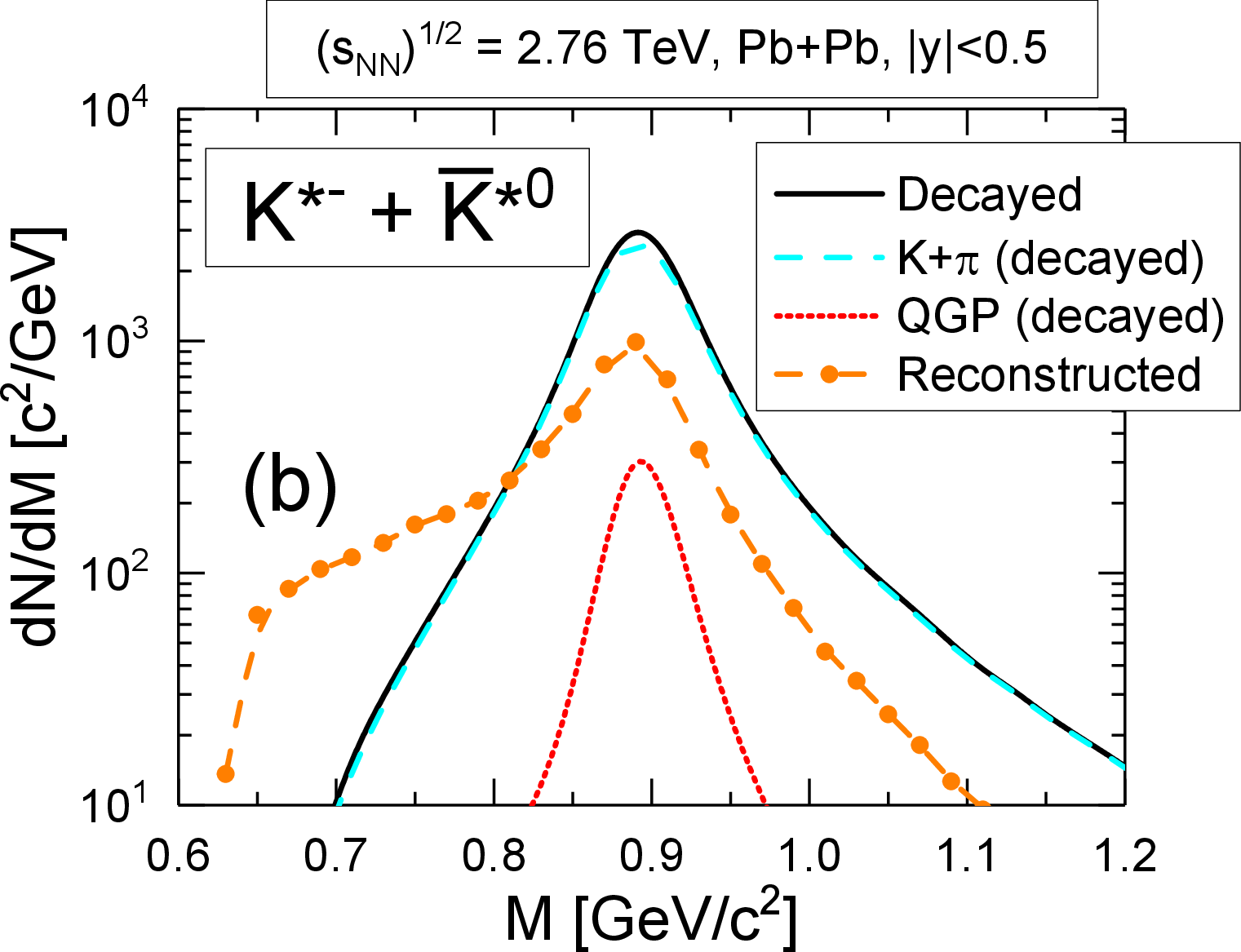}}
  \caption{The differential mass distribution $\frac{dN}{dM}$ for the vector kaons
${K}^{*+}+K^{*0}$ (a, upper part) and for vector anti-kaons $\bar{K}^{*-}+\bar K^{*0}$ (b, lower part)
for central Pb+Pb collisions at a center-of-mass energy of $\sqrt{{s}_{NN}} = 2.76$~TeV at midrapidity ($|y|<0.5$) from the PHSD calculations.
The solid orange lines with circles show ${K}^{*}$'s and $\bar{K}^{*}$'s reconstructed
from final kaon and pion pairs while all of the other lines represent the different production
channels at the decay point of the ${K}^{*}$ and $\bar{K}^{*}$, i.e. the black lines
show the total number of the ${K}^{*}$'s and $\bar{K}^{*}$'s at their decay points,
while the light blue dashed lines show the decayed ${K}^{*}$'s and $\bar{K}^{*}$'s
that stem from the $\pi + K$ annihilation and the short-dotted red lines indicate
the decayed ${K}^{*}$'s and $\bar{K}^{*}$'s which have been produced during
the hadronisation of the QGP.}
  \label{fig:dndm_chdec}
\end{figure}

In Fig. \ref{fig:rescat} we show the 'reconstruction' procedure
for the $K^*$ mass distribution $dN/dM$ for Pb+Pb collision at impact 
parameter $b=2 fm$  at a center-of-mass energy $\sqrt{{s}_{NN}} = 2.76$~TeV. 
For this illustrative plot we show the spectra without cut in rapidity (all $y$)
in order to accommodate better statistics, however, we note that the cut in rapidity 
reduces the spectra by about a factor of 9.
The green dotted line in Fig. \ref{fig:rescat} shows the reconstructed 
$K+\pi$ mass distribution for the case without rescattering of pions and kaons 
while the red dashed line stands for the case when pions and/or kaons rescatter 
elastically in the medium. One can see that the shape of 'not rescattered' $K + \pi$ 
mass spectrum are close to the shape of 'true', i.e. decayed $K^*$ spectra 
(blue dot-dashed line). On the other hand the shape of the decayed $K^*$ spectra
follows the shape of the $K^*$ spectral function weighted with the occupation probability,
(a Boltzmann factor) which tilts the spectra a bit
to lower masses and suppresses the high tail of the spectral function.

The 'rescattered' $K + \pi$ contribution is rather flat and shifted to the lower
mass region. This part of the spectrum can not be identified experimentally
in the $\bar K^*$ observables since it contributes to the subtracted combinatorial 
background which is fitted to the low and high mass region of the measured 
$\bar K + \pi$ mass distribution.
Thus, experimentally the 'signal' is selected under the peak of the $K + \pi$ 
mass distribution while the rest is not included in measured observables.
In order to be close to the experimental situation we selected the mass region 
$M=[0.8,1.0]$~GeV when comparing to the experimental data. In this mass range the
contribution of the 'rescattered' $K + \pi$ is less then 7\% to the total 
reconstructed spectra. On the other hand only about 40\% of the decayed $K^*$'s 
with the masses 0.8-1.0~GeV  end up in the 'reconstructed' spectra 
at the same mass region.
Later we will show examples
 of 'reconstructed' mass distributions keeping all 
'true' $K + \pi$ pairs from the decay of $K^*/\bar K^*$ resonances.

\begin{figure*}[t!]
  \centering
  \raggedleft
  \subfigure[Reconstructed spectrum for the ${K}^{*}$]
  {\includegraphics[scale=0.32]{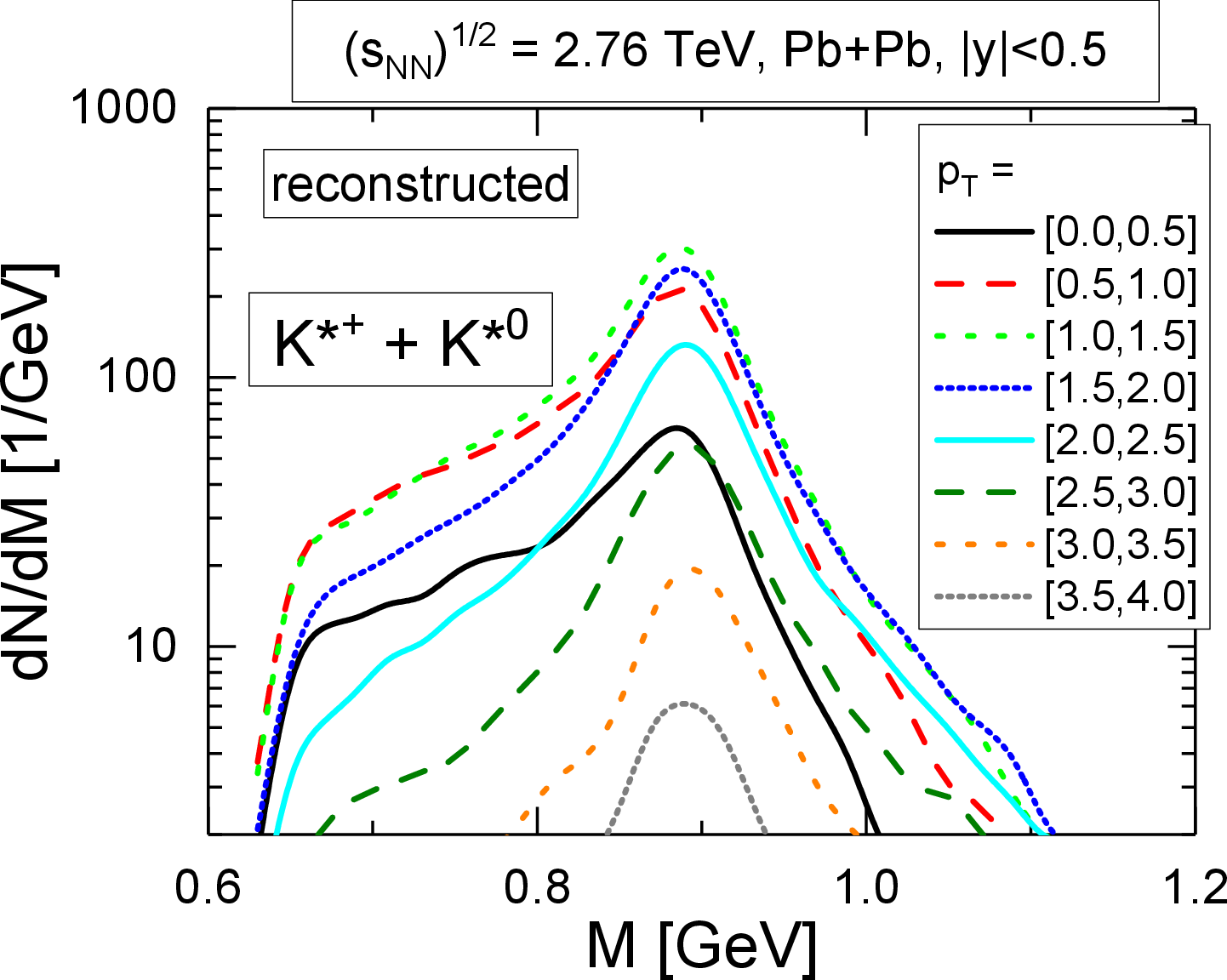}}\hspace{35pt}
  \raggedright
  \subfigure[Reconstructed spectrum for the $\bar{K}^{*}$]{\includegraphics[scale=0.32]{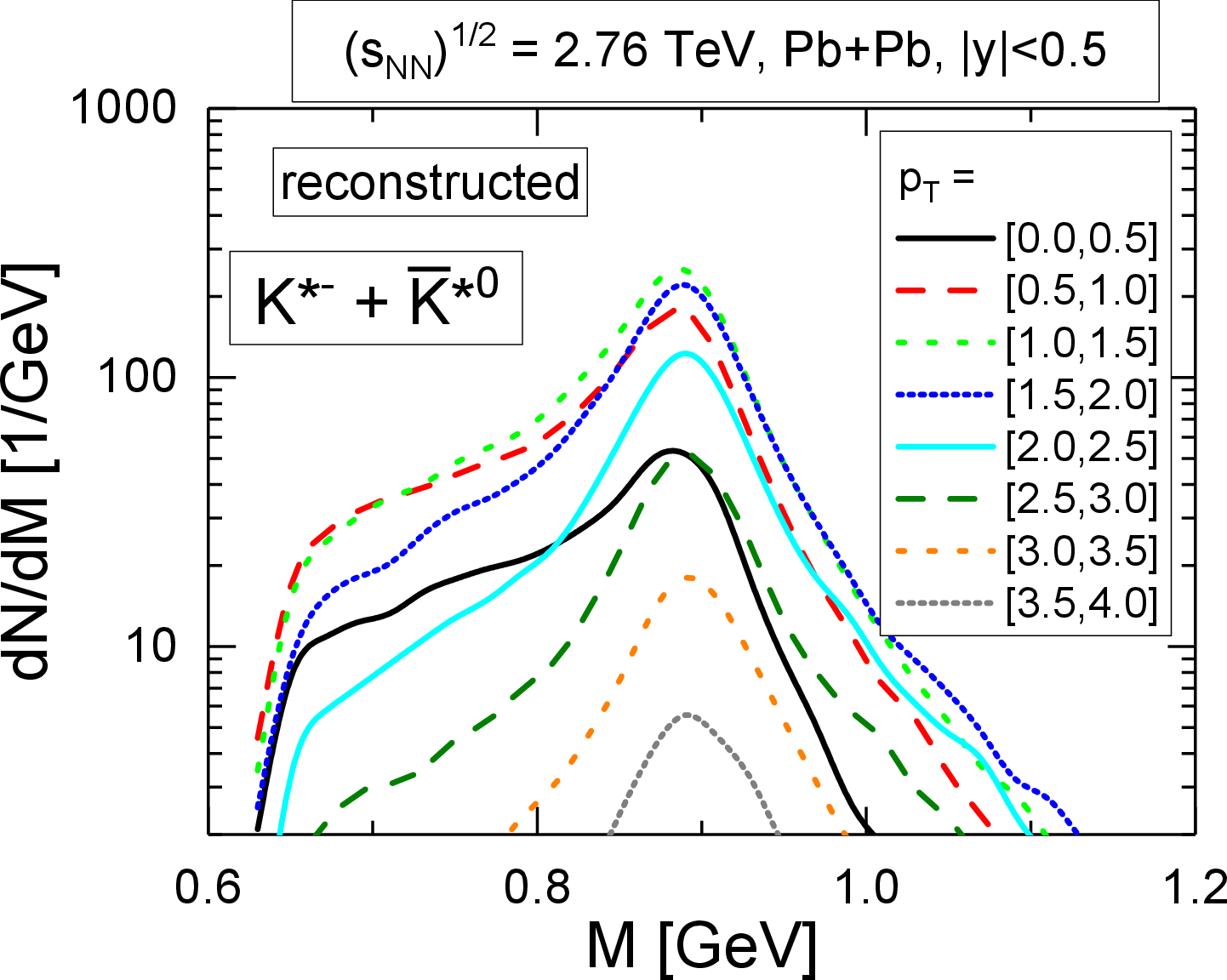}}
  \subfigure[Decayed spectrum for the ${K}^{*}$]{\includegraphics[scale=0.32]{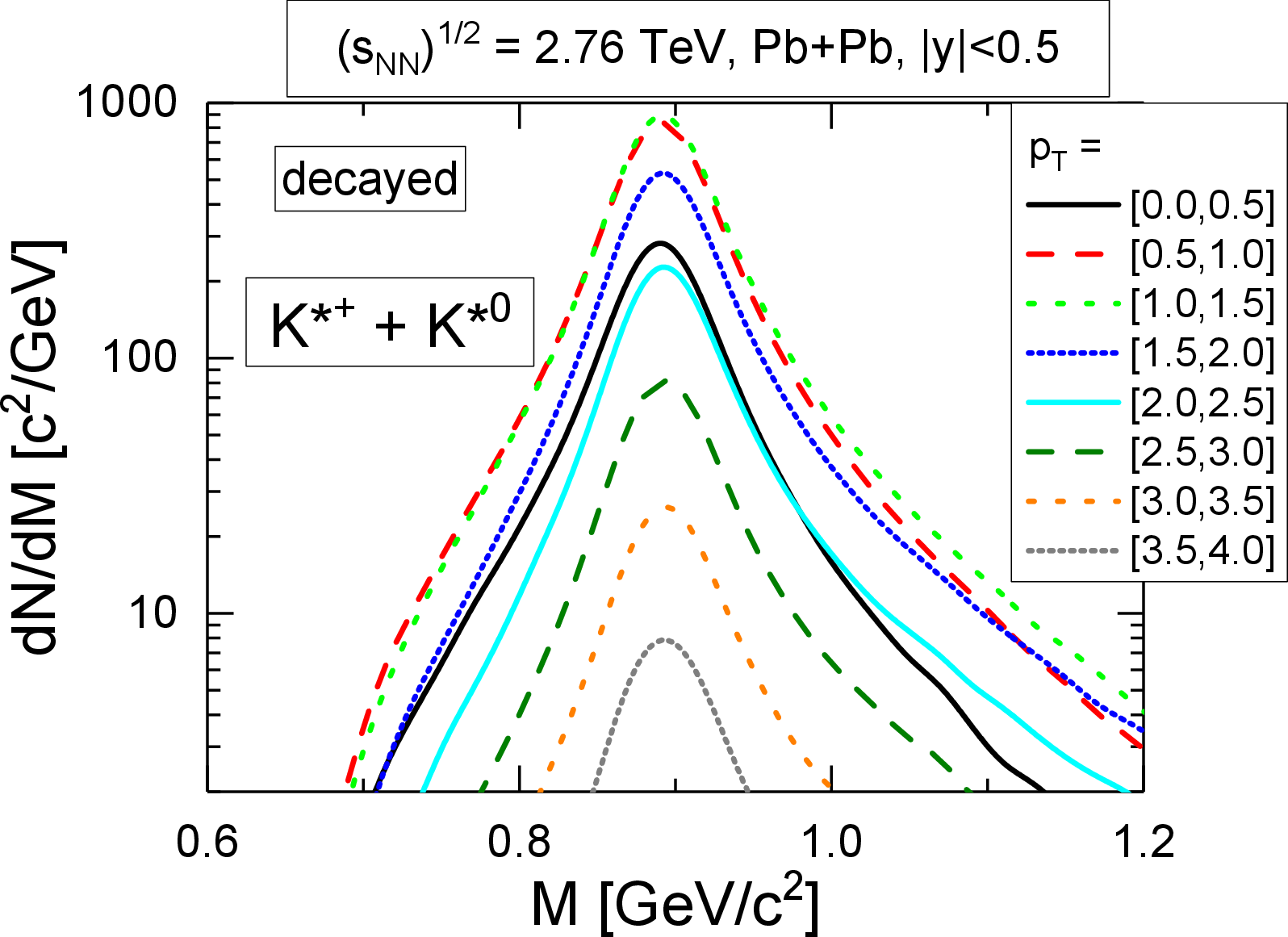}}\hspace{10pt}
  \subfigure[Decayed spectrum for the $\bar{K}^{*}$]{\includegraphics[scale=0.32]{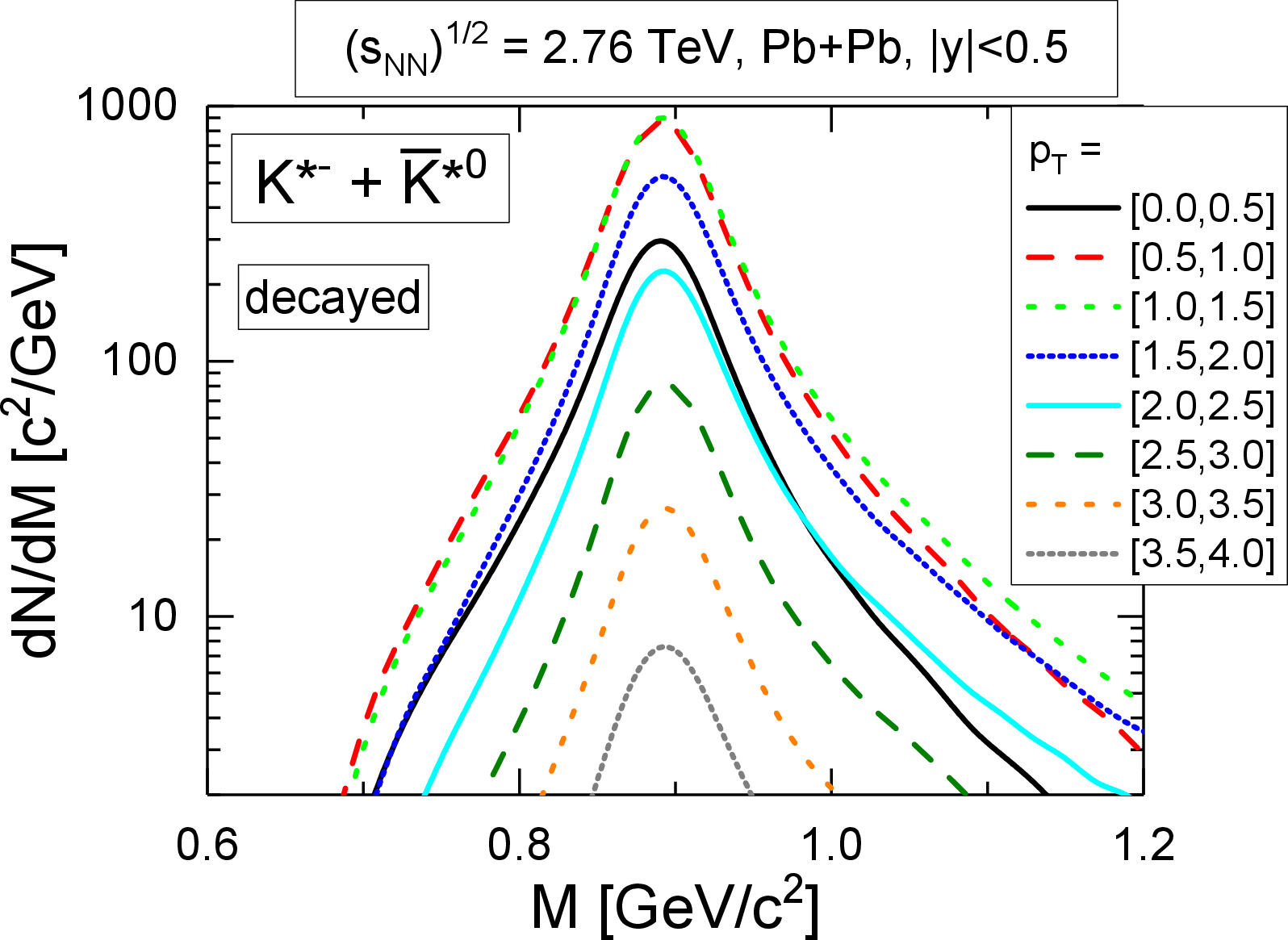}}
  \caption{The differential $K^*/\bar K^*$ mass spectrum $dN/dM$ for central Pb+Pb collision at midrapidity
($|y|<0.5$) at a center-of-mass energy $\sqrt{{s}_{NN}} = 2.76$~TeV from the PHSD calculation
for different $p_T$ bins.
The upper panels (a) and (b) show the spectrum for the reconstructed ${K}^{*}/\bar{K}^{*}$s while
the lower panels (c) and (d) show the spectrum for the decayed ${K}^{*}/\bar{K}^{*}$s.
The panels (a) and (c) show results for the ${K}^{*}$ mesons while the panels (b) and (d) show results for
the $\bar{K}^{*}$ mesons. All panels show the differential mass spectrum taken from the ${K}^{*}$'s and $\bar{K}^{*}$'s in specific transverse momentum ranges of ${p}_{T} = [0.0,0.5]$~GeV (solid black line), ${p}_{T} = [0.5,1.0]$~GeV (dashed red line), ${p}_{T} = [1.0,1.5]$~GeV (short-dashed green line), ${p}_{T} = [1.5,2.0]$~GeV (short-dotted blue line), ${p}_{T} = [2.0,2.5]$~GeV (solid light blue line), ${p}_{T} = [2.5,3.0]$~GeV (dashed olive line), ${p}_{T} = [3.0,3.5]$~GeV (short-dashed orange line), ${p}_{T} = [3.5,4.0]$~GeV (short-dotted grey line).}
  \label{fig:ptbins}
\end{figure*}

In Fig. \ref{fig:dndm_chdec} we present  the differential mass distribution ${dN}/{dM}$
for the vector kaons ${K}^{*+}+K^{*0}$ (a, upper part) and for vector anti-kaons $\bar{K}^{*-}+\bar K^{*0}$ (b, lower part) for central Pb+Pb collisions at $\sqrt{{s}_{NN}} = 2.76$~TeV at midrapidity
($|y|<0.5$) from the PHSD calculations. Here we show the 'true' $K^*/\bar{K}^{*}$ spectra,
i.e. obtained directly at their decay point (solid black lines) and the reconstructed spectra
from the final pions and kaons (solid orange lines with circles). Similar to Fig. \ref{fig:rescat} 
one can see a strong modification of the mass spectra: a shift to lower invariant masses and a reduction
of the yield at the vacuum peak position.  As said above, this modification arises from the rescattering and absorption
of the final pions and kaons. Moreover, in Fig. \ref{fig:dndm_chdec} we show
that at the LHC energy the main source of $K^*/\bar{K}^{*}$ mesons is the $\pi +K$
annihilation which is substantially larger than the fraction of $K^*/\bar{K}^{*}$
produced in the hadronization of the QGP. We note that the relative fraction
of the $K^*/\bar{K}^{*}$ from the annihilation at LHC is even larger than at RHIC since
the total abundance of mesons is much larger at the LHC energy.

Furthermore, in Fig. \ref{fig:ptbins} we show the differential $K^*/\bar K^*$ mass spectrum $dN/dM$ for central
Pb+Pb collisions at midrapidity ($|y|<0.5$) and  $\sqrt{{s}_{NN}} = 2.76$~TeV from the PHSD calculation for different $p_T$ bins. The upper panels (a) and (b) show the spectrum for the
reconstructed ${K}^{*}/\bar{K}^{*}$s while the lower panels (c) and (d) show the spectrum for
the decayed ${K}^{*}/\bar{K}^{*}$s. One can see that the shape of the 'decayed' $K^*/\bar K^*$'s
mass spectra are rather similar for the different $p_T$ bins while the 'reconstructed'
spectra changes differently in different $p_T$ bins: the distortion of the spectra is strongest 
for low $p_T$. That is due to the fact that the low $p_T$ pions and kaons suffer from stronger 
and more frequent rescattering and absorption.

\begin{figure}[h!]
  \centerline{\includegraphics[width=8.0cm]{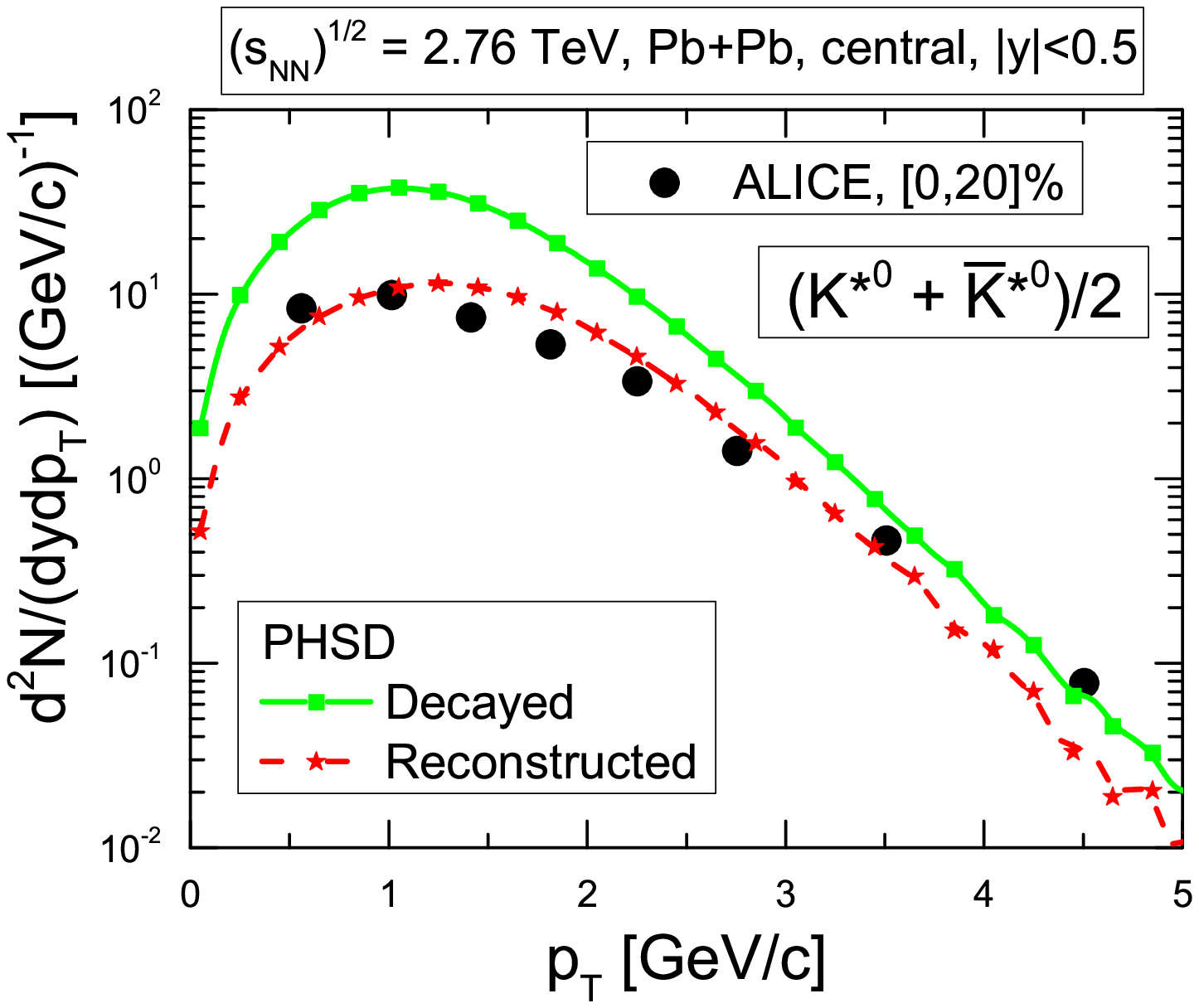}}
  \caption{The transverse momentum spectrum ${d}^{2}N/\left( dyd{p}_{T}\right)$
of $(K^{*0}+\bar K^{*0})/2$ for a central Pb+Pb collision at midrapidity $|y|<0.5$
and $\sqrt{{s}_{NN}} = 2.76$~GeV. The solid black circles show data from
the ALICE collaboration \cite{Abelev:2014uua}, while the lines with symbols show the results
from the PHSD: the solid green line with squares shows the spectrum calculated
from ${K}^{*}$'s at the decay point while the dashed red line with stars shows
the reconstructed spectrum from final (matching) (anti-)kaons and pions. }
  \label{fig:recvsdec}
\end{figure}

The influence of the experimental reconstruction procedure and experimental cuts - 
due to the detector acceptance - on the $p_T$ spectra is shown in Fig. \ref{fig:recvsdec}.
Here the PHSD results for the transverse momentum spectrum ${d}^{2}N/\left( dyd{p}_{T}\right)$
of $(K^{*0}+\bar K^{*0})/2$ for a central Pb+Pb collision at midrapidity $|y|<0.5$
at  $\sqrt{{s}_{NN}} = 2.76$~GeV
are shown in comparison to the experimental data from the
ALICE Collaboration \cite{Abelev:2014uua}: the solid green line with squares
shows the spectrum calculated from ${K}^{*}/\bar K^*$'s at the decay point while
the dashed red line with stars shows the reconstructed spectrum from final
(matching) kaons and pions. Similar to our finding (at RHIC energies) in Ref. \cite{Ilner:2016xqr}
we see a strong reduction of the low $p_T$ spectra which is due to the 'loss' of signal
stemming from the rescattering/absorbtion mechanisms and the experimental cuts applied; in particular
the restriction on the invariant mass range for the $K^*/\bar K^*$ reconstruction contributes to this distortion.
Finally the reconstructed spectrum is much harder than the 'true' ('decayed) one.
Furthermore, we find  that at the LHC energy - similar to the RHIC energies - the influence
of the FSI on the final $K^*/\bar K^*$ spectra is getting smaller with decreasing 
system size and becomes negligible for $p+p$ collisions where the reconstructed
and decayed spectra are almost identical.
\begin{figure}[h!]
  \centerline{\includegraphics[width=8.0cm]{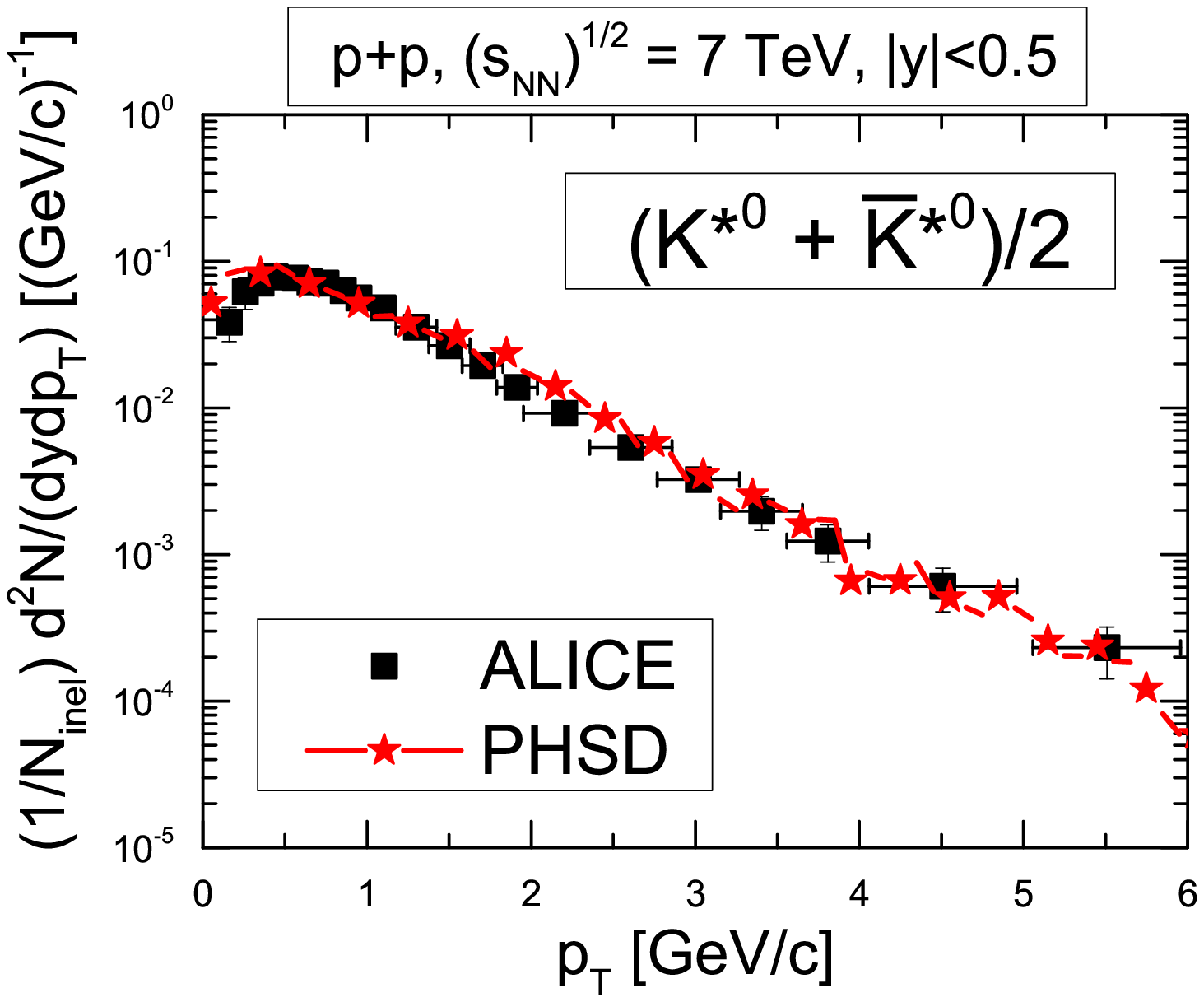}}
  \caption{The transverse momentum spectrum $\frac{1}{{N}_{inel}} \frac{d^{2}N}{dyd{p}_{T}}$
of  $({K}^{*0}+\bar K^{*0})/2$ mesons for p+p collisions at mid-rapidity $|y|<0.5$
at the LHC energy of $\sqrt{{s}_{NN}}=7$~TeV. The line stands for the PHSD results while
the black symbols show the experimental data from the ALICE Collaboration \cite{Abelev:2012hy}.}
  \label{fig:pt_pp}
\end{figure}

In Fig. \ref{fig:pt_pp} we show the comparison of the PHSD results (red solid line with stars)
for the transverse momentum spectrum $\frac{1}{{N}_{inel}} \frac{d^{2}N}{dyd{p}_{T}}$
of the neutral $({K}^{*0}+\bar K^{*0})/2$
mesons for p+p collisions at mid-rapidity $|y|<0.5$ at LHC energies of $\sqrt{{s}_{NN}}=7$~TeV
versus the ALICE data (black symbols) from Ref. \cite{Abelev:2012hy}.
The $K^*$ momenta  have been obtained by reconstruction from the final $\pi + K$ mesons.
As mentioned above, the final hadronic interaction in p+p collisions is very small,
thus the reconstructed and decay spectra for p+p collisions are practically identical.
As seen from Fig. \ref{fig:pt_pp}, the elementary spectra are rather well reproduced
by PHSD which provides a solid basis for the interpretation of the heavy-ion results, too.
We recall that p+p collisions in the PHSD are  based on the PYTHIA
even generator \cite{LUND}.

\subsection{Comparison of the PHSD results with experimental data}

We now step on to a comparison of  PHSD results for strange vector mesons from
 heavy-ion collisions with  experimental observables at the LHC energy.
To compare with the experimental data, we have used the experimental reconstruction method
for the theoretical spectra by matching
the 4-momentum of the final pions and (anti-)kaons stemming from the same $K^*/\bar K^*$ decay vertex.
As has been shown above, this implies that the final 'reconstructed' spectra
differ from the 'true' or 'decay' $K^*/\bar K^*$ spectra due to the final-state interaction
in the hadronic phase.
\begin{figure}[ht!]
  \centerline{\includegraphics[width=8.0cm]{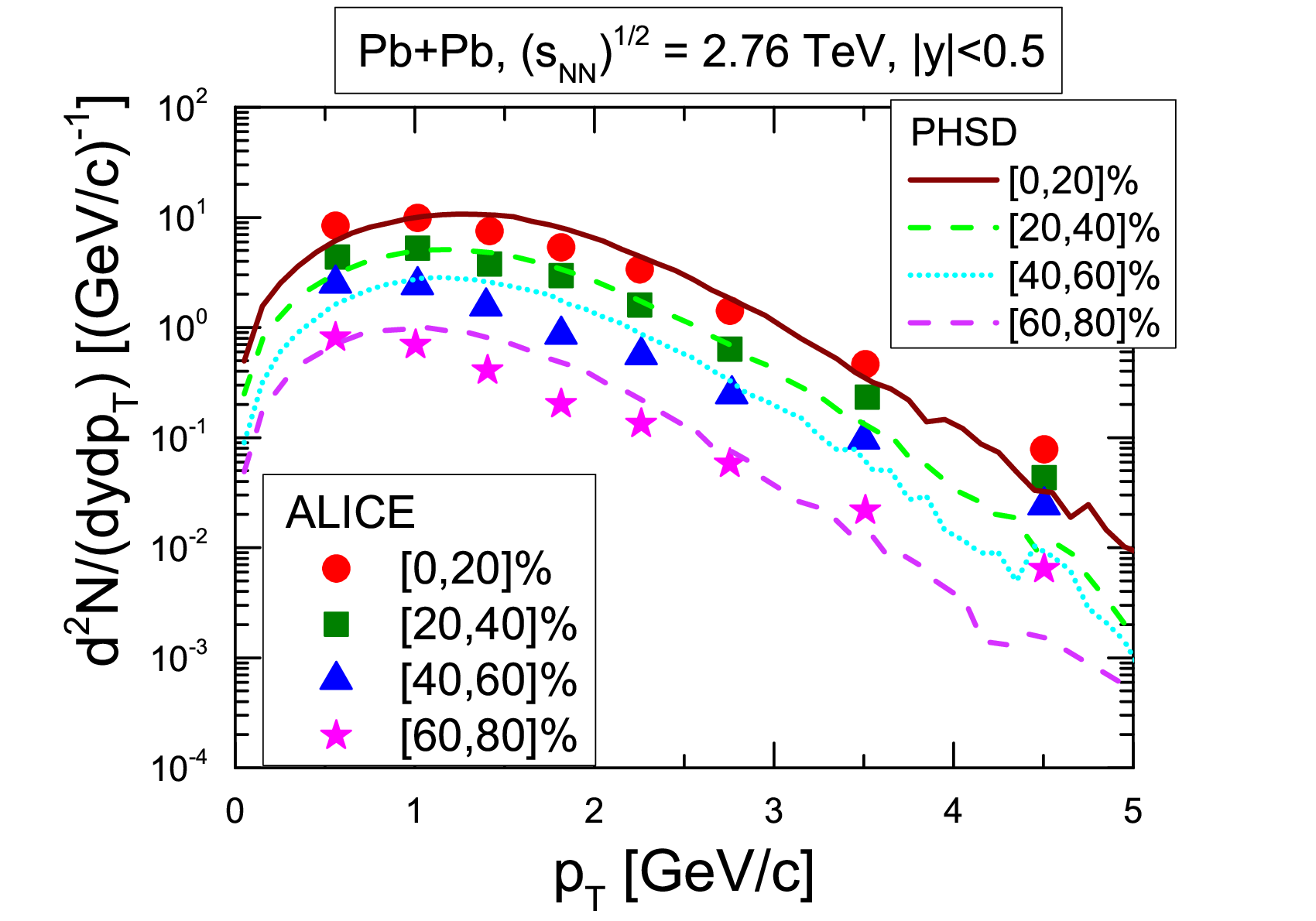}}
  \caption{The transverse momentum spectra $\frac{d^{2}N}{dyd{p}_{T}}$
of  $({K}^{*0}+\bar K^{*0})/2$ as a function of the transverse momentum ${p}_{T}$
for Pb+Pb collisions  at midrapidity ($|y|<0.5$)
 and $\sqrt{{s}_{NN}}=2.76$~TeV for different centralities, ranging from very central to peripheral collisions. The lines correspond to results obtained from PHSD while the symbols represent experimental data from the ALICE collaboration \cite{Abelev:2014uua}. The solid red circles correspond to a centrality of $[0,20]\%$, the red solid green squares correspond to a centrality of $[20,40]\%$, the solid blue triangles correspond to a centrality of $[40,60]\%$ and the solid magenta stars correspond to a centrality of $[60,80]\%$. The solid dark red line corresponds to PHSD results for a centrality of $[0,20]\%$, the dashed green line corresponds to a centrality of $[20,40]\%$, the short-dotted light blue line corresponds to a centrality of $[40,60]\%$ and the dashed violet line corresponds to a centrality of $[60,80]\%$. }
  \label{fig:pt_PbPb}
\end{figure}
In Fig. \ref{fig:pt_PbPb} we compare PHSD results for the transverse momentum
spectra $\frac{d^{2}N}{dyd{p}_{T}}$ of $({K}^{*0}+\bar K^{*0})/2$
for  Pb+Pb collisions at midrapidity ($|y|<0.5$)
with the ALICE data at a center-of-mass energy of $\sqrt{{s}_{NN}}=2.76$~TeV for different centralities:
[0-20]\%, [20,40]\%, [40,60]\% and [60,80]\%.
As seen from Fig. \ref{fig:pt_PbPb} the PHSD calculations reasonably reproduce 
the ALICE data for all centralities at lower transverse momenta up to about 
${p}_{T}\approx3$~GeV/c, however,  underestimate the high $p_T$ part of the 
experimental spectra.
This discrepancy for peripheral Pb+Pb collisions at large $p_T \ge 4$~GeV --
in spite of a good agreement for pp collisions -- can be understood as follows:
in pp collisions $K^*$'s are produced directly from the string decays
while in (semi-)peripheral collisions ([60-80\%]) most $K^*$'s at midrapidity
come from  $K+\pi$ annihilation; thus the final $K^*$ spectra are
sensitive to the 'bulk dynamics' at larger $p_T$ and less to the initial 
momentum distribution from strings.
We note, that in midrapidity (semi-)peripheral collisions  at LHC energies 
the energy density is very large and far above the critical energy density 
for the phase transition.
Thus, in the overlapping region (i.e mid-rapidity) of colliding nuclei,
a QGP is formed. In the PHSD approach the properties of quarks
and gluons are defined by the DQPM model, which is fitted to the
lQCD data in equilibrium, and depend only on temperature $T$. The same holds
for the partonic interaction cross sections.
The quark momentum distribution is transferred via hadronization 
to the mesons (pions, kaons, $K^*$'s etc.) and finally is reflected in $K^*$'s 
formed in the hadronic phase by $K+\pi$ annihilation.
Thus, an underestimation of the slope of the final kaons and pions
at large $p_T$  in the PHSD is showing up in the underestimation of the
$K^*$'s slope, too. 
We note that an extension of the DQPM with respect to energy dependent
interaction cross sections of partons (which is under development) 
might improve the agreement with heavy-ion data at larger $p_T$.

\begin{figure}[t]
  \centerline{\includegraphics[width=9cm]{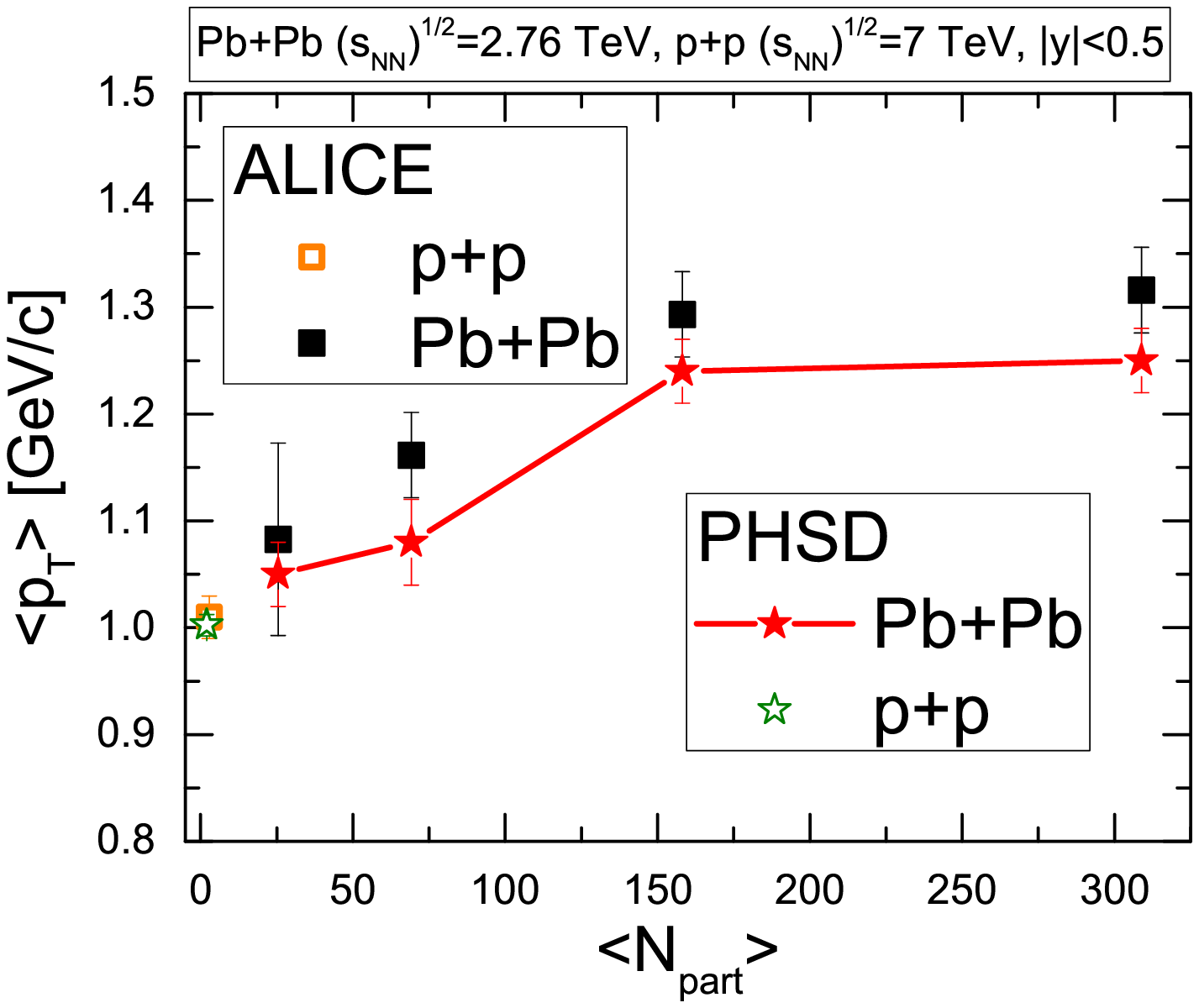}}
  \caption{The average transverse momentum $<{p}_{T}>$ of $({K}^{*0}+\bar K^{*0})/2$
  as a function of the average number
 of participants $<{N}_{part}>$ for p+p and Pb+Pb collisions at $\sqrt{{s}_{NN}}=7$~TeV and
 $\sqrt{{s}_{NN}}=2.76$~TeV, respectively. The solid red line with full stars stands
 for the PHSD results for Pb+Pb, the open star shows the PHSD result for p+p collisions.
 Experimental data from the ALICE collaboration are shown as solid black
 squares for Pb+Pb and open squares for p+p and taken from Ref. \cite{Abelev:2014uua}.}
  \label{fig:avpt_npart}
\end{figure}
The information on the centrality dependence of the transverse spectra of
$({K}^{*0}+\bar K^{*0})/2$ can be viewed also in terms of the averaged $<p_T>$ at each centrality bins.
Figure \ref{fig:avpt_npart} shows the average transverse momentum of
the ${K}^{*0}$ and $\bar{K}^{*0}$ mesons from the PHSD calculations for  Pb+Pb at midrapidity
as a function of the average number of participants $<N_{]part}>$ in comparison to
the ALICE data \cite{Abelev:2014uua}. Also the results for p+p collisions is shown.
One can see that the average $<p_T>$ grows from p+p to peripheral and to central collisions
and then saturates. The PHSD results are in a good agreement with the ALICE measurements.

\begin{figure}[h]
  \centerline{\includegraphics[width=8.0cm]{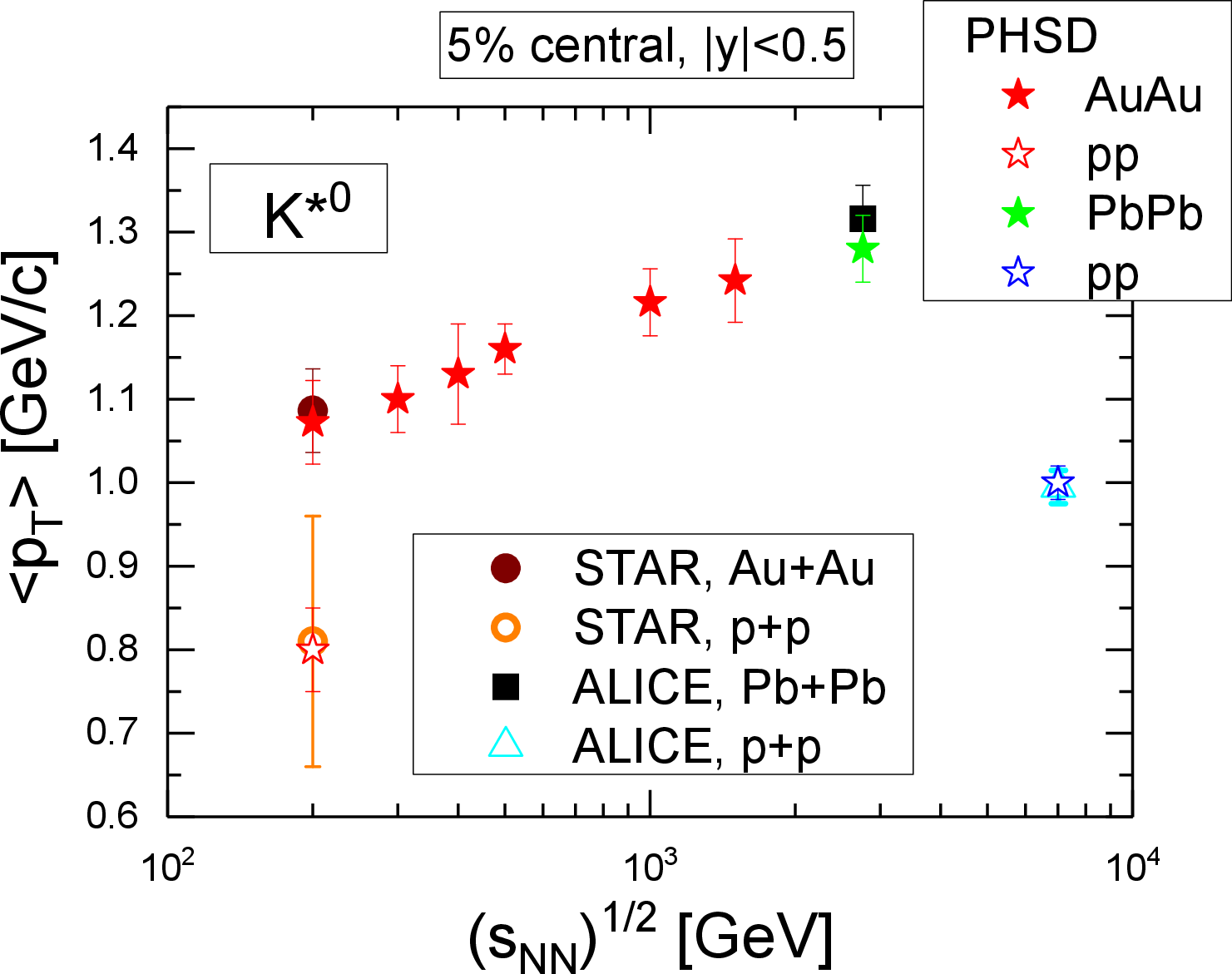}}
  \caption{The average transverse momentum $<{p}_{T}>$ of $K^{*0}$ as a function of the invariant center-of-mass energy $\sqrt{{s}_{NN}}$ ranging from central Au+Au collisions at RHIC energies up to central Pb+Pb
and p+p collisions at LHC energies. The PHSD results are shown as  stars
(open - for p+p, solid for Au+Au/Pb+Pb), the full dot shows the STAR data, the full
square -- ALICE data for Pb+Pb and open square - for p+p.
  The experimental data are taken from Ref. \cite{Abelev:2014uua}.}
  \label{fig:avpt_snn}
\end{figure}
Figure \ref{fig:avpt_snn} shows the average transverse momentum $ <p_T>$ of the ${K}^{*0}$
and $\bar{K}^{*0}$ from the PHSD calculations as a function of the center-of-mass energy $\sqrt{{s}_{NN}}$ --
from RHIC to LHC -- for p+p and Au+Au/Pb+Pb collisions in comparison to the experimental data
from STAR and ALICE. One can see that the experimental data show
that $<{p}_{T}>$ from central Au+Au/Pb+Pb is larger than those from p+p for both
(RHIC and LHC) energies; on the other hand  the $<{p}_{T}>$ grows with the energy 
for p+p as well as heavy-ion collisions. These tendencies are reproduced by the PHSD calculations
which predict a monotonic increase of $<{p}_{T}>$ with energy.

\begin{figure}[t]
  \centerline{\includegraphics[width=8.0cm]{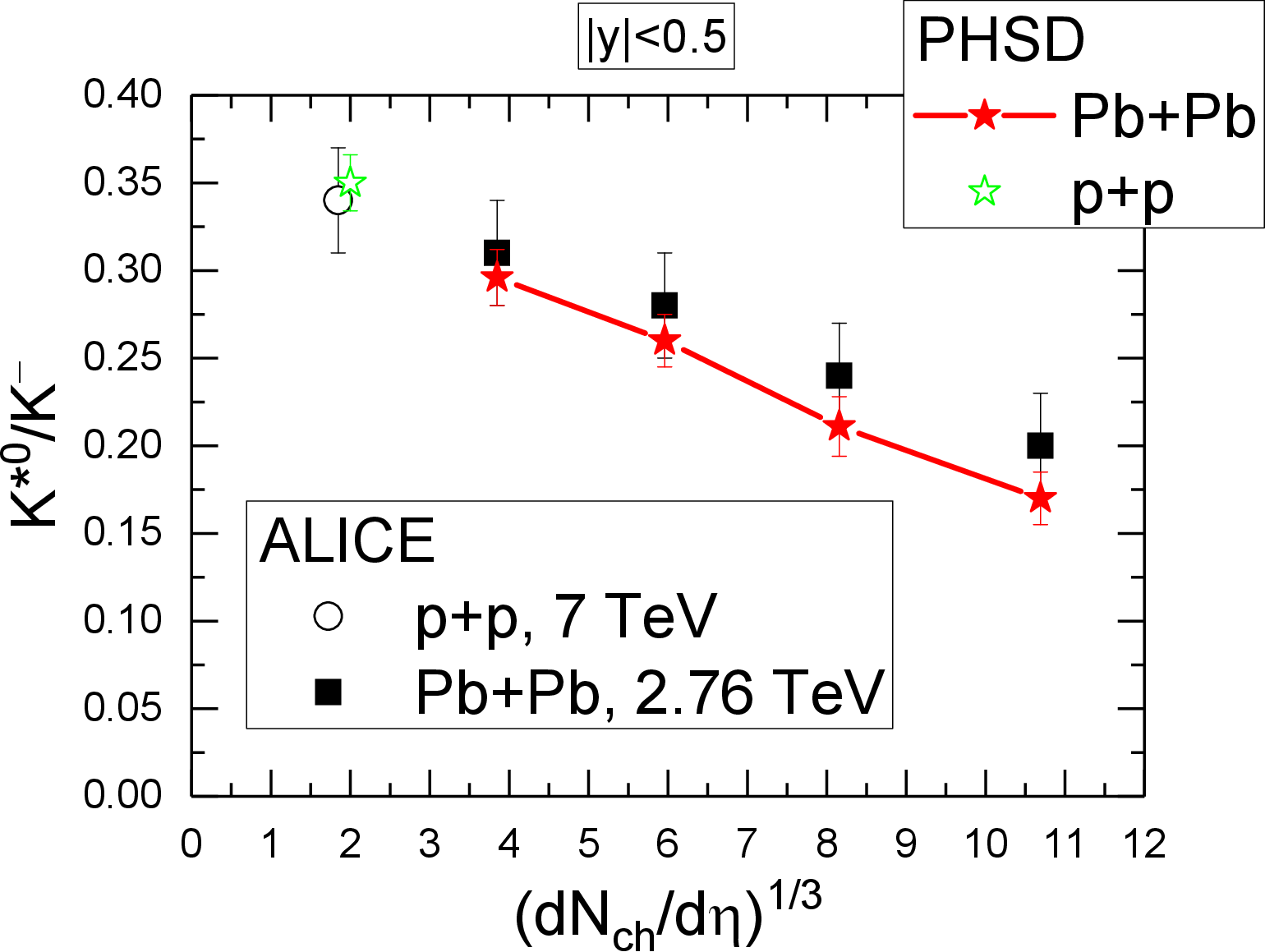}}
  \caption{The particle ratio ${K}^{*0}/{K}^{-}$ versus ${\left( d{N}_{ch}/d\eta\right)}^{1/3}$ for p+p and Pb+Pb collisions at the LHC energies of $\sqrt{{s}_{NN}}=7$~TeV and $\sqrt{{s}_{NN}}=2.76$~TeV, respectively.  The PHSD results are shown by the red line with starts for central Pb+Pb and the open star for p+p
 collisions. The experimental data from the ALICE Collaboration are displayed as solid squares for Pb+Pb and
 the open dot for p+p as taken from Ref. \cite{Abelev:2014uua}.}
  \label{fig:pr1}
\end{figure}
\begin{figure}[h!]
  \centerline{\includegraphics[width=8.0cm]{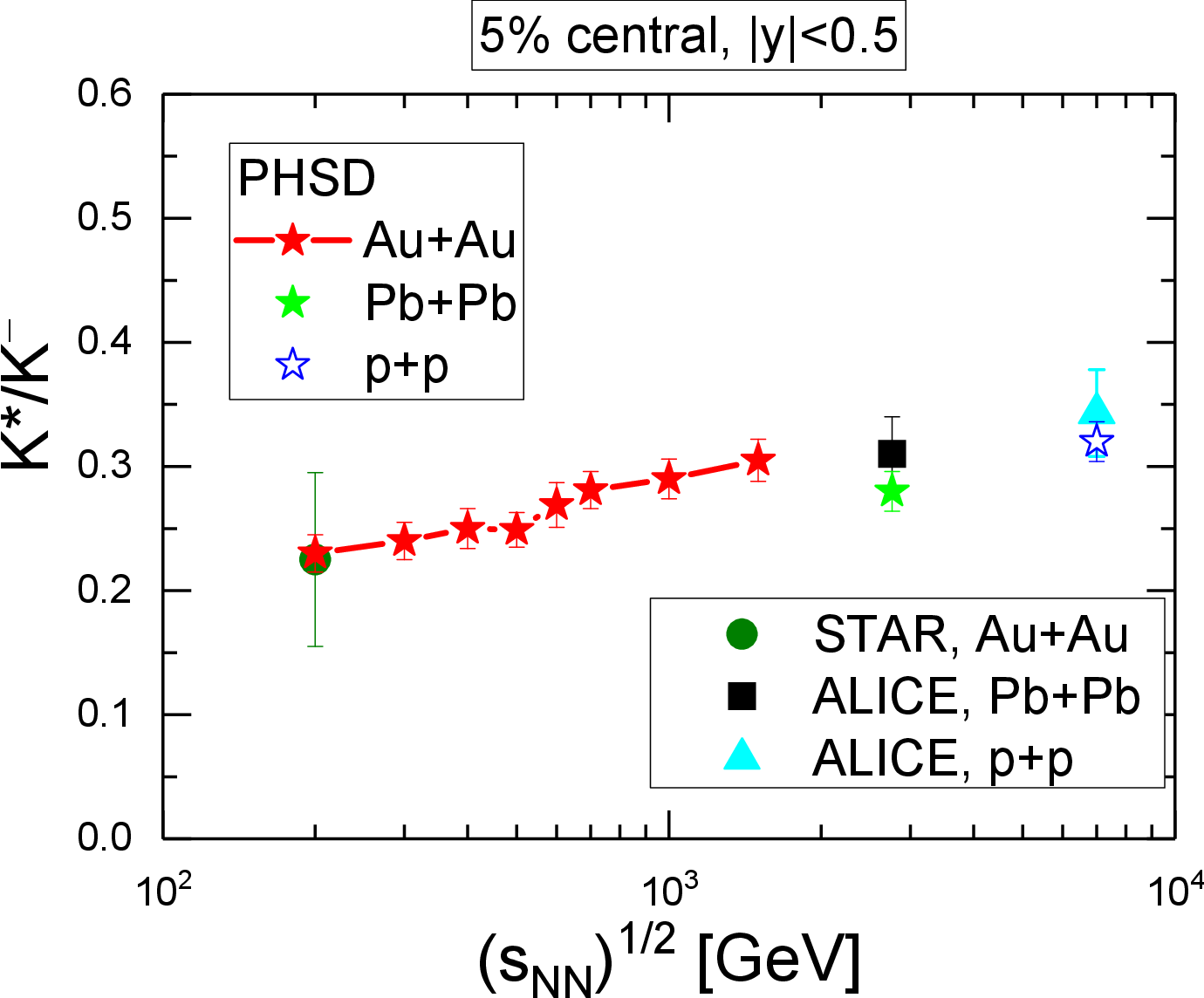}}
  \caption{The particle ratio ${K}^{*-}/{K}^{-}$ as a function of the invariant center-of-mass
 energy energy $\sqrt{{s}_{NN}}$ ranging from central Au+Au collisions at RHIC energies
 up to central Pb+Pb collisions at the LHC energy of $\sqrt{{s}_{NN}}=2.76$~TeV.
 The PHSD results are shown by the stars (open - for p+p, solid for Au+Au/Pb+Pb,
 connected by the red line);
 the full dot shows the STAR data, the full square corresponds to ALICE data for Pb+Pb and
 full triangles  for p+p.
 The experimental data are taken from Ref. \cite{Abelev:2014uua}.}
  \label{fig:pr2}
\end{figure}

Now we step to the particle ratios: we show in Fig. \ref{fig:pr1} the ${K}^{*0}/{K}^{-}$ ratios
as a function of $(dN\-(ch)/d\eta)^{1/3}$ for p+p and Pb+Pb collisions at LHC energies of 
$\sqrt{{s}_{NN}}=7$~TeV and $\sqrt{{s}_{NN}}=2.76$~TeV, respectively.
 The results from PHSD (using a 'reconstruction' method for $K^{*0}$ and applying
 the experimental cuts on the mass of $K^*$) 
 are compared to experimental data from the ALICE collaboration
 \cite{Abelev:2014uua}.  The experimental as well as theoretical ratios decrease
 with increasing centrality due to the stronger final-state interaction
 effect in central collisions compared to peripheral reactions. Since
 the hadron density is large in the central region at the LHC, the $K^-$'s 
 as well as the pions and kaons  from ${K}^{*0}$ decays rescatter very often.
  We note that the meson rescattering and absorption effects are stronger at LHC
 than at RHIC which leads to the decrease of the ${K}^{*0}/{K}^{-}$ ratios at LHC
 for central collisions compared to a rather flat ratio at RHIC
 (cf. Fig. 15 in \cite{Ilner:2013ksa}).

Furthermore, Fig. \ref{fig:pr2} shows the PHSD calculations for the
${K}^{*-}/{K}^{-}$ ratio as a function of the invariant center-of-mass
 energy energy $\sqrt{{s}_{NN}}$ ranging from central Au+Au collisions at RHIC energies
 up to central Pb+Pb collisions at LHC energies 
in comparison to the experimental data from STAR (full dot) and ALICE (full square for Pb+Pb
 and full triangle for p+p). One can see that the ratio very smoothly increases with energy
 and is larger for p+p than for Pb+Pb at LHC. That is due to a practically negligible
 final-state interaction in p+p compared to  Au+Au/Pb+Pb collisions.

\section{Predictions for FAIR/NICA}\label{sec:fair}

\begin{figure}[t!]
  \centerline{\includegraphics[width=7.5cm]{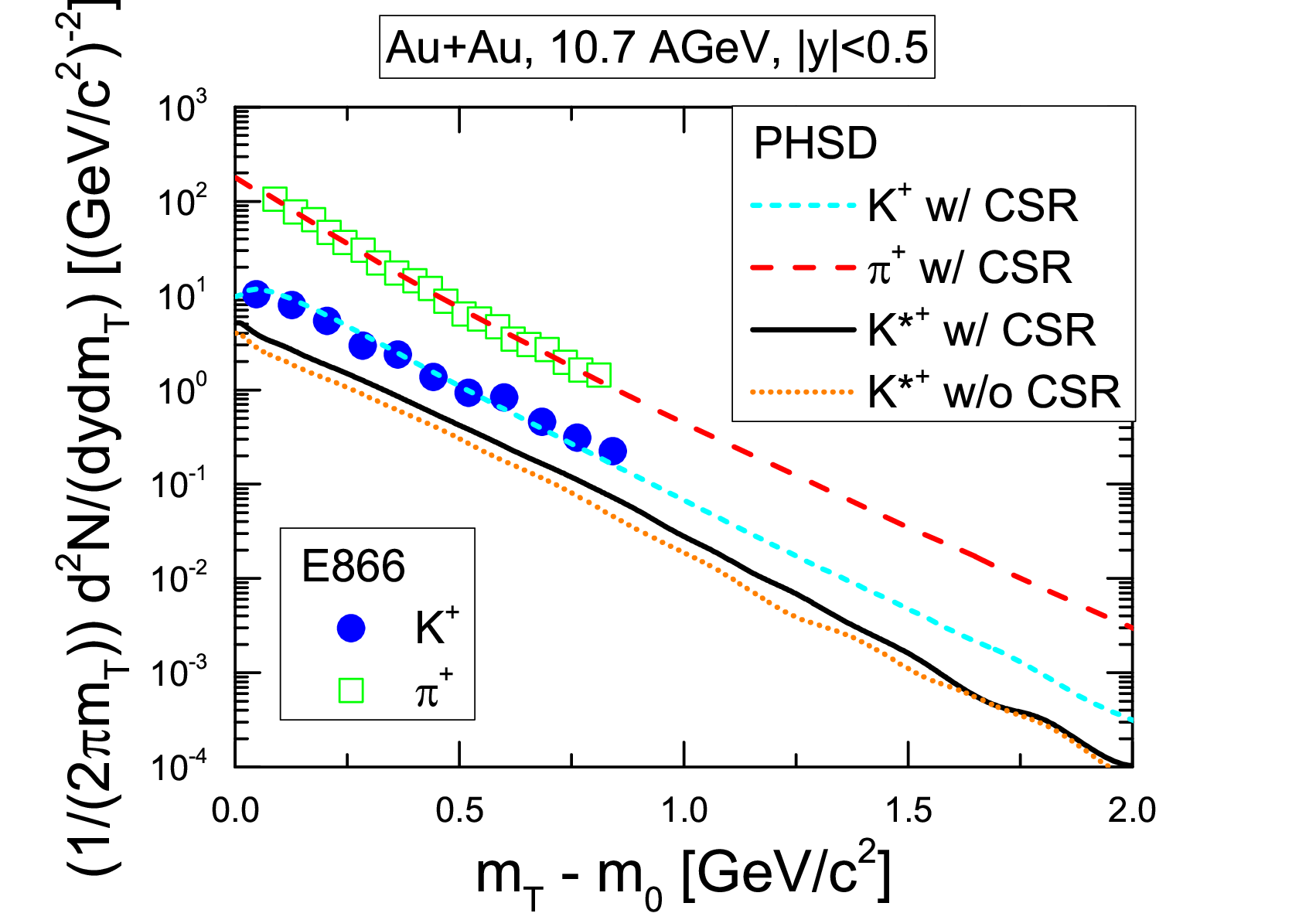}}
  \caption{Transverse mass spectra $1/(2 \pi {m}_{T}) dN/d{m}_{T}$ as a function
  of the reduced transverse mass ${m}_{T} - {m}_{0}$ for central Au+Au collisions
  at bombarding energy of $E = 10.7$~AGeV at midrapidity ($|y|<0.5$).
The lines show results from PHSD while the symbols show experimental data
from the E866 collaboration \cite{Ahle:1999uy}.
The solid blue circles display experimental data for  ${K}^{+}$
while the open green squares show the experimental data for  ${\pi}^{+}$ mesons.
The corresponding theoretical results from PHSD including Chiral Symmtery Restoration (CSR)
are the light blue short-dashed line for the ${K}^{+}$ and the red dashed line for the ${\pi}^{+}$.
The solid black line shows the results for the ${K}^{*+}$ with CSR turned on in PHSD
while the short-dotted orange line shows results for the ${K}^{*+}$ without including CSR. }
\label{fig:lowmt}
 \phantom{a}\vspace*{1mm}
  \centerline{\includegraphics[width=8.0cm]{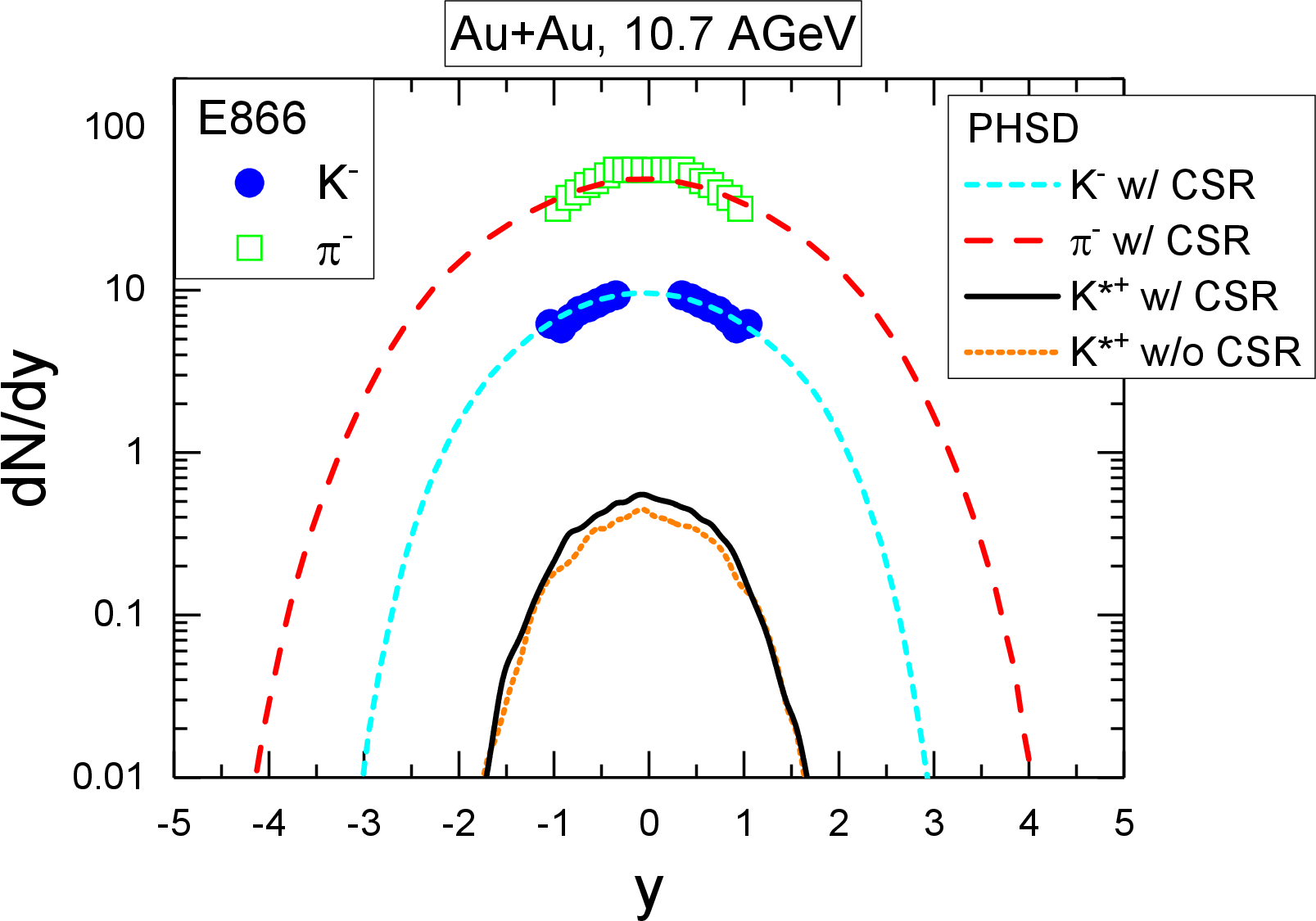}}
  \caption{Rapidity spectra $dN/dy$  versus the rapidity $y$ for a Au+Au collision at
 bombarding energy of $E = 10.7$~AGeV.  The lines show results from PHSD while the symbols display experimental data from
the E866 collaboration \cite{Ahle:1999uy}. The solid blue circles show data for
the ${K}^{-}$ mesons while the open green squares show data for  ${\pi}^{-}$.
The corresponding theoretical results from PHSD are the light blue short-dashed line
for the ${K}^{-}$ and the red dashed line for the ${\pi}^{-}$ mesons.
The solid black line shows the results for the ${K}^{*-}$ with CSR included
while the short-dotted orange line shows results for the ${K}^{*-}$ without CSR. }
\label{fig:lowrap}
\end{figure}

In the this section we show the results from the PHSD for strange vector meson
production at much lower energies - from a few AGeV to few tens of AGeV,
which will be achievable by the future FAIR and NICA or the BES program at RHIC.
This energy range is very interesting since \\
i) an interplay between the deconfined and chiral transitions is expected to happen in this energy range.
Recently, the consequences of the chiral symmetry restoration (CSR) on observables
in HICs has been studied within the PHSD approach \cite{Cassing:2015owa,Palmese:2016rtq}.
The CSR has been incorporated in the PHSD via the Schwinger mechanism for the quark-antiquark
production in the string decay and related to the dressing of the quark masses in the medium due
to a linear coupling to the quark condensate $\langle {\bar q} q \rangle$. It has been shown 
that the inclusion of CSR effects provides
a microscopic explanation for the 'horn' structure in the excitation
function of the $K^+/\pi^+$ ratio: the CSR in the hadronic phase produces the steep increase
of this particle ratio up to $\sqrt{s}\sim 7 GeV$, while the drop at higher energies
is associated to the appearance of a deconfined partonic medium.
In this section we additionally investigate the effect of CSR on the production of the
strange vector mesons $K^*$ and $\bar K^*$.

\begin{figure}[t!]
  \centerline{\includegraphics[width=7.0cm]{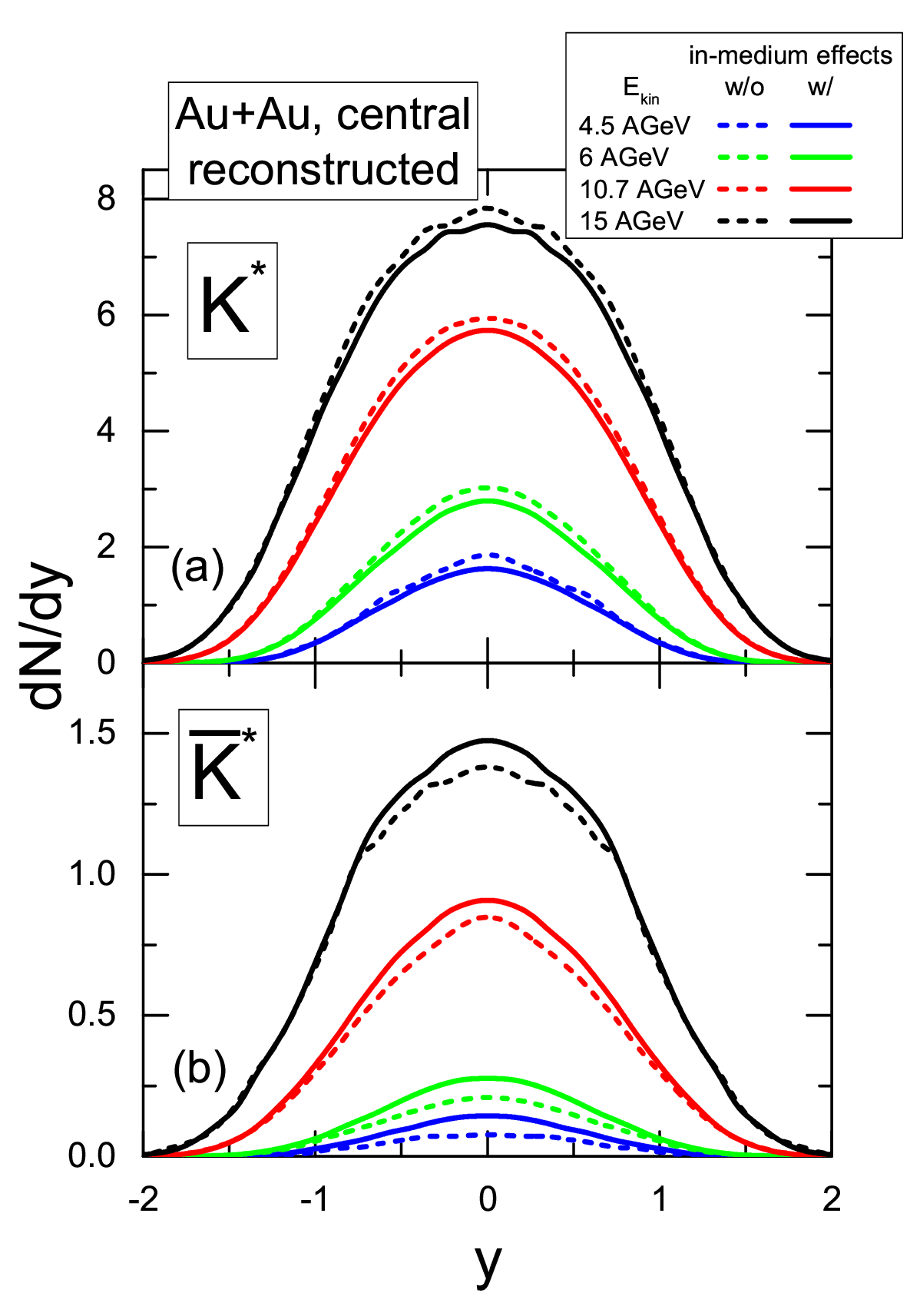}}
  \caption{ PHSD predictions for the 'reconstructed' rapidity spectra 
of $K^*=K^{*+}+K^{*0}$ (upper panel (a)) 
and $\bar K^*=K^{*-}+\bar K^{*0}$ (lower panel (b))  
for  central Au+Au collisions  at bombarding energies
of 4.5, 6, 8, 10.7 and 15 A GeV. The dashed lines show the PHSD results without
including the in-medium effects for $K^{*}$ and $\bar K^{*}$
while the solid lines correspond to the case with in medium effects. }
\label{fig:yFAIR}
\end{figure}

ii) at FAIR/NICA energies the medium effects - related to the modification of hadron properties
at high baryon densities - are expected to be more visible
than at RHIC or LHC energies due to the slower fireball expansion and larger net-baryon densities achieved - cf. Fig. \ref{fig:dndrho}.
Thus, we investigate here the energy range which would be most appropriate for a study of
the in-medium effects with $K^*, \bar K^*$ mesons.

We start by showing the $m_T$- and rapidity  spectra of ${K}^{*0}$ and $\bar{K}^{*0}$ mesons
for central Au+Au collisions calculated with and without CSR effects at a center-of-mass energy
of $\sqrt{{s}_{NN}}=4.765$~GeV which is equivalent to the laboratory energy of $E = 10.7$~AGeV
-- cf. Figs. \ref{fig:lowmt} and \ref{fig:lowrap}.
Here the experimental data (from the E866 Collaboration) are also available
for pions and kaons \cite{Ahle:1999uy}.

As seen from Figs. \ref{fig:lowmt} and \ref{fig:lowrap} the PHSD calculations provide a good description
of pion and kaon spectra when the CSR effect is included; we refer the reader to the detailed study on this issue
to  Refs. \cite{Cassing:2015owa,Palmese:2016rtq}. The inclusion of the CSR increases
the yield of ${K}^{*0}$ and $\bar{K}^{*0}$ by about 15-20\%. That is mainly
due to the increase of the kaon and antikaon yield when including the CSR.

\begin{figure*}[h!]
  \centering
 \raggedleft
  {\includegraphics[width=7.0cm]{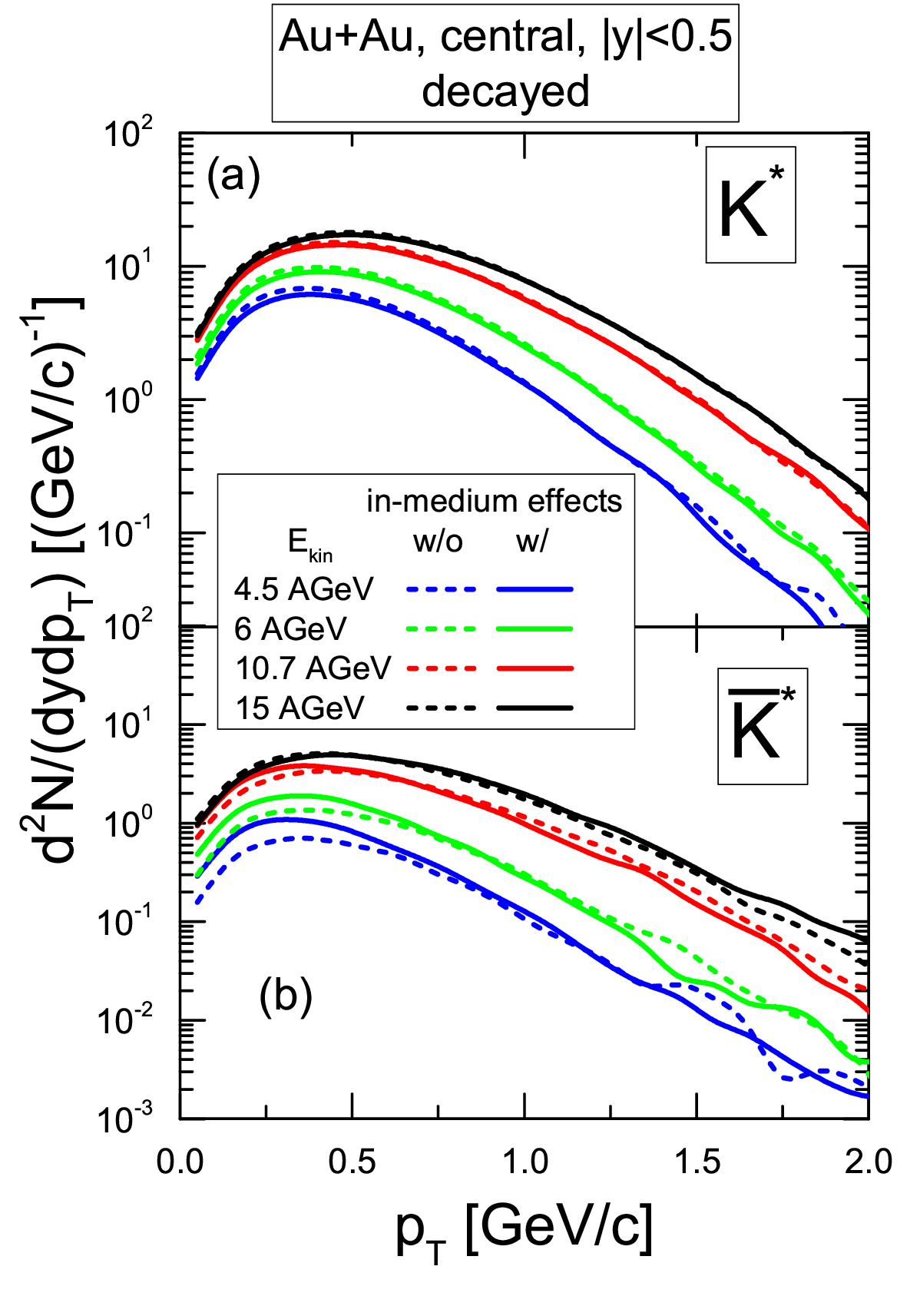}}
  \raggedright 
  {\includegraphics[width=7.0cm]{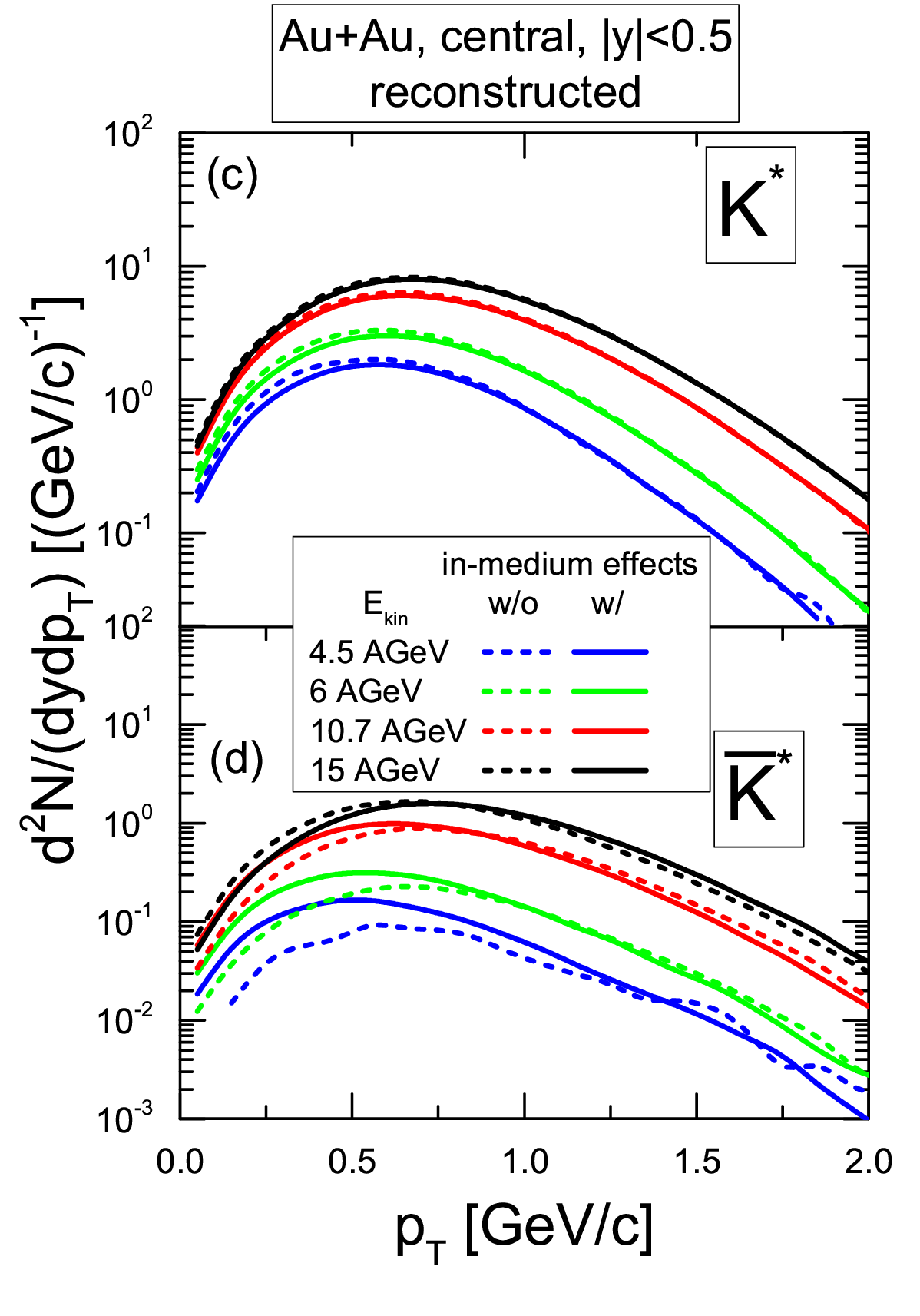}}
  \caption{PHSD predictions for the decayed (left panel)
  and  'reconstructed' (right panel) $p_T$-spectra 
of $K^*=K^{*+}+K^{*0}$ (upper panel (a),(c)) 
and $\bar K^*=K^{*-}+\bar K^{*0}$ (lower panel (b),(d)) 
for central Au+Au collisions at midrapidity $(|y|<0.5)$ at bombarding energies
of 4.5, 6, 8, 10.7 and 15 A GeV. The dashed lines show the PHSD results without
including the in-medium effects for $K^{*}$ and $\bar K^{*}$
while the solid lines stand for the case with in medium effects. }
\label{fig:ptFAIR}
 \phantom{a}\vspace*{1mm}
  \centering
 \raggedleft
 {\includegraphics[width=7.0cm]{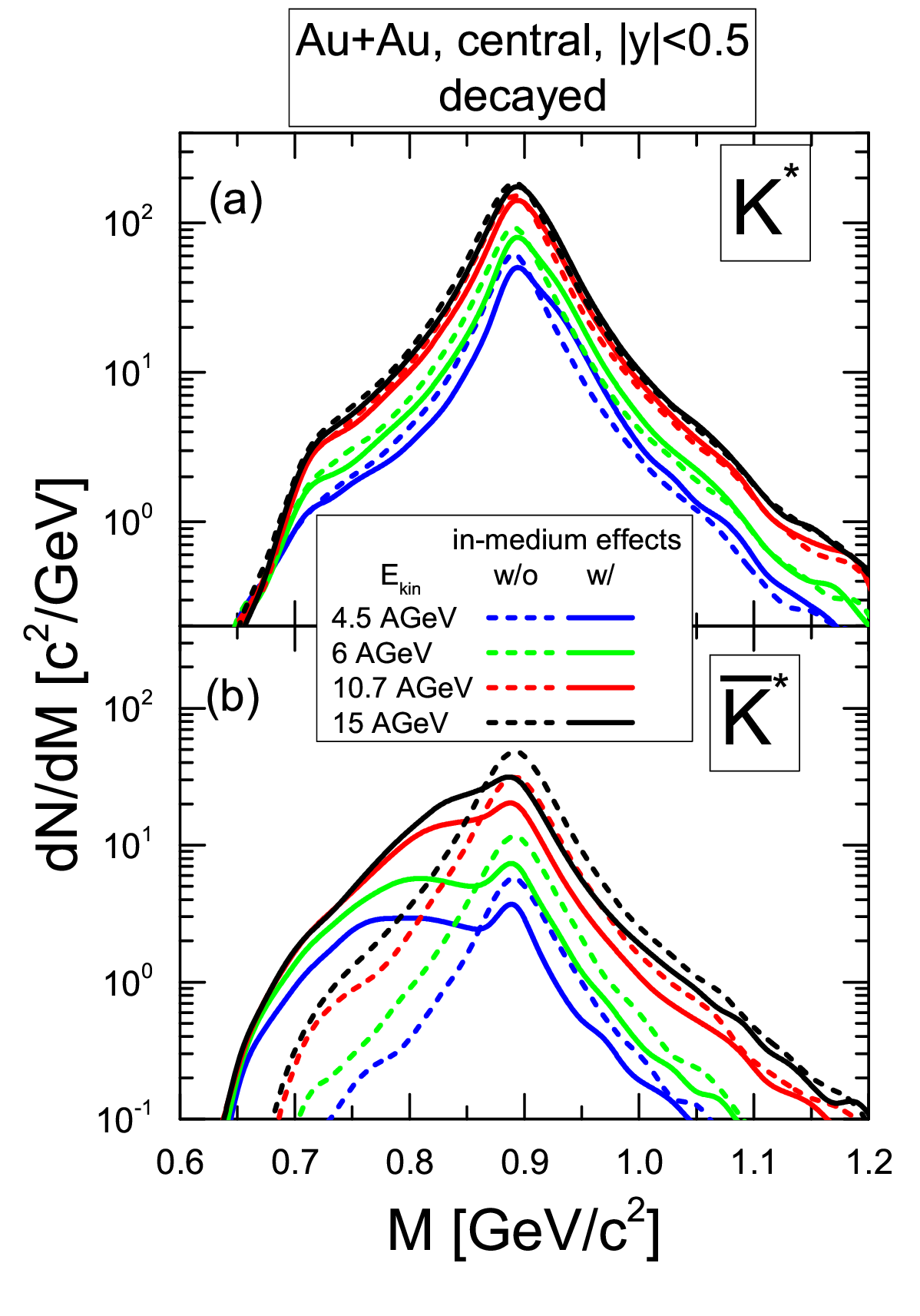}}
    \raggedright
 {\includegraphics[width=7.0cm]{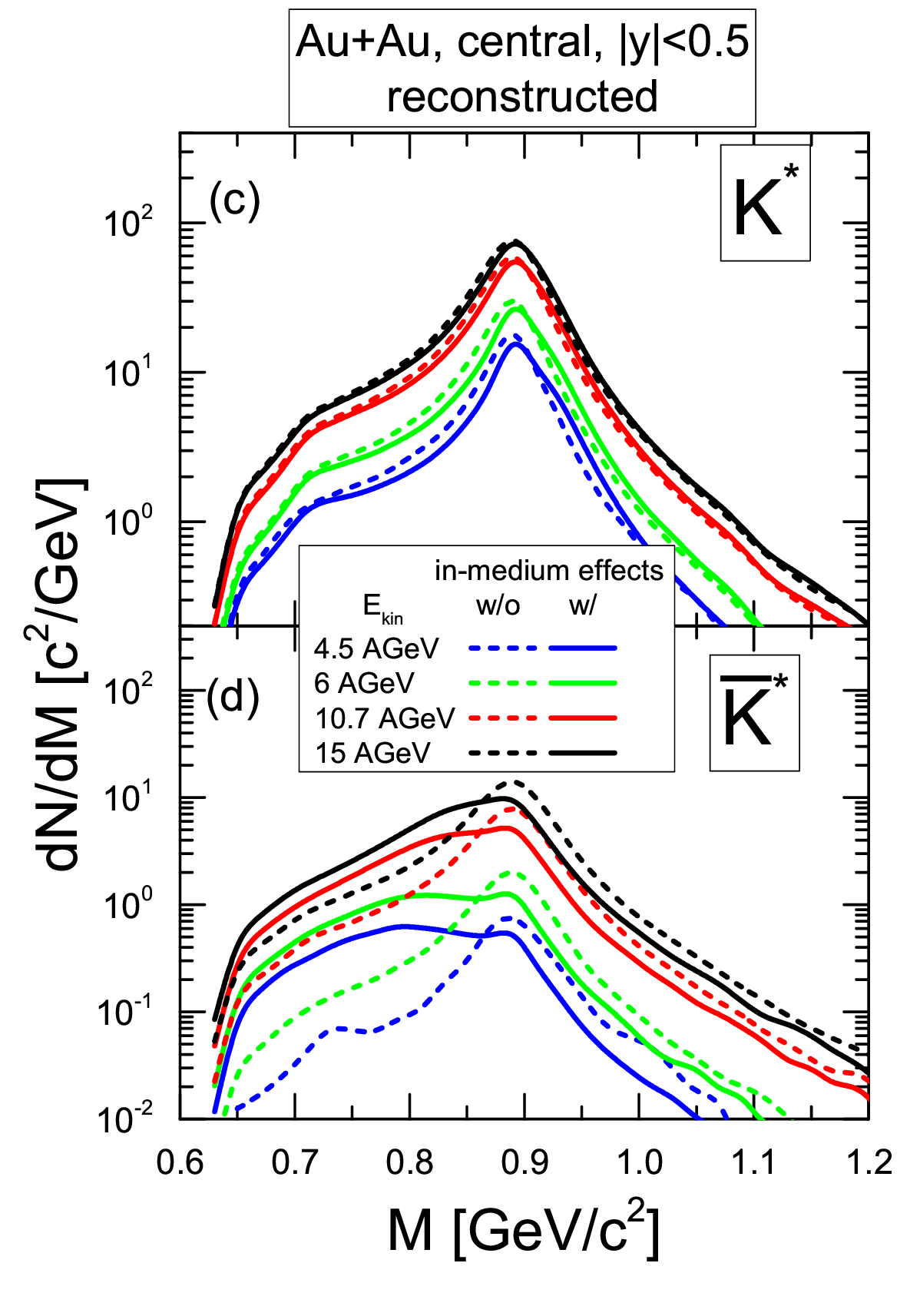}}    
  \caption{PHSD predictions for the decayed (left panel) and 'reconstructed'(right panel) mass distribution of $K^*=K^{*+}+K^{*0}$ (upper panel (a),(c)) 
and $\bar K^*=K^{*-}+\bar K^{*0}$ (lower panel (b),(d))  
for central Au+Au collisions at midrapidity $(|y|<0.5)$ at bombarding energies
of 4.5, 6, 8, 10.7 and 15 A GeV. The dashed lines show the PHSD results without
including the in-medium effects for $K^{*}$ and $\bar K^{*}$
while the solid lines stand for the case with in medium effects. }
\label{fig:MFAIR}
\end{figure*}

Now we step to the results with the in-medium modifications of strange vector mesons.
In Fig. \ref{fig:yFAIR} we show the PHSD predictions for the rapidity distributions  ('reconstructed' $y$-spectra) and in Figs. \ref{fig:ptFAIR}, \ref{fig:MFAIR} 
-- for the  $p_T$-spectra as well as the mass distribution (decayed and 'reconstructed'
spectra) of $K^*=K^{*+}+K^{*0}$ (upper panel (a),(c)) 
and $\bar K^*=K^{*-}+\bar K^{*0}$ (lower panel (b),(d)) 
for central Au+Au collisions at bombarding energies
of 4.5, 6, 8, 10.7 and 15 A GeV. 
The $p_T$- spectra and mass distributions are calculated at midrapidity $(|y|<0.5)$. 
The dashed lines show the PHSD results without including the in-medium effects 
for $K^{*}$ and $\bar K^{*}$ while the solid lines stand for the case with in medium effects.
We note that all calculations are performed with including the CSR effects. 

\begin{figure}[h!]
  \centerline{\includegraphics[width=10.0cm]{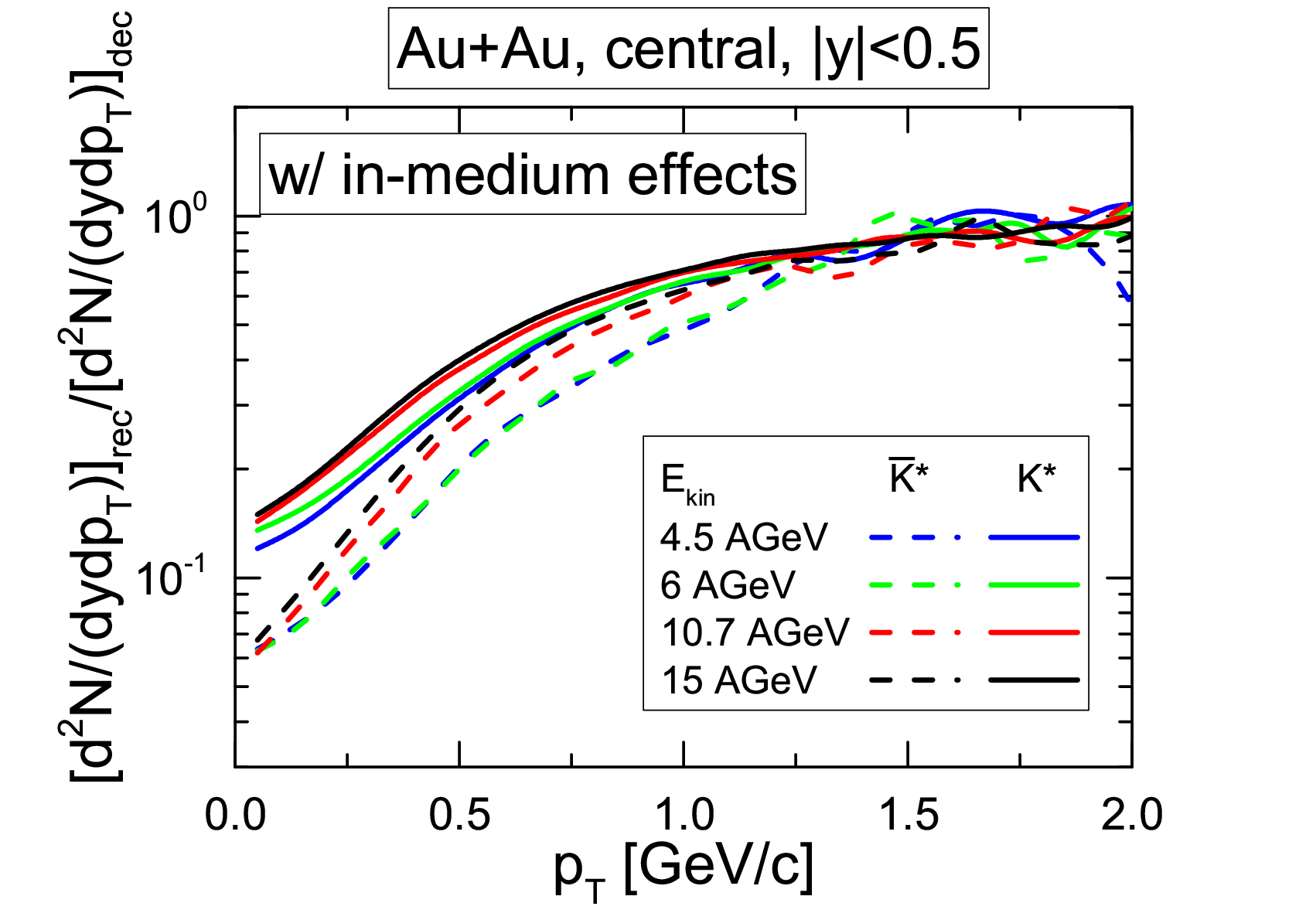}}
  \caption{The ratio of 'reconstructed' to 'decayed' $p_T$ spectra 
of $K^*=K^{*+}+K^{*0}$ (solid lines) and $\bar K^*=K^{*-}+\bar K^{*0}$ (dashed lines)  
for  central Au+Au collisions  at bombarding energies
of 4.5, 6, 8, 10.7 and 15 A GeV for the case with in-medium effects. }
\label{fig:Rat}
\end{figure}

As seen from Figs. \ref{fig:ptFAIR}, \ref{fig:yFAIR}, \ref{fig:MFAIR}, the in-medium
 effects on $p_T$- and $y$-spectra and the mass distribution increases with decreasing energy.
This is due to the  longer reaction time and slower expansion of the fireball at low energies such that:\\
i) the strange vector mesons are still produced in a baryon rich environment 
(cf. Fig. \ref{fig:dndrho}) which leads to a
pronounced modification of their spectral functions (cf. Fig. \ref{fig:specfun}).
The in-medium effects are stronger for the $\bar K^{*}$ than for the $K^{*}$ 
for all energies as expected from Fig. \ref{fig:specfun} which shows the substantial
modification of the $\bar K^{*}$ spectral function comparing in-medium and free cases
and only modest modifications of $K^{*}$ spectral functions. The pole of the
in-medium $\bar K^{*}$ spectral function is shifted to the low mass range due to the
attractive interaction of $\bar K^{*}$ with the baryonic medium while
the pole of the $K^{*}$ spectral function moves slightly to higher masses
due to the repulsive interaction.\\
ii) the decay of $K^*, \bar K^*$'s occurs in the hadronic medium such that the final 
mesons - (anti-)kaons and pions - rescatter with hadrons or are absorbed which leads 
to the distortion of the 'reconstructed' $K^*$ spectra due to the final-state iteration 
of the decay products - as discussed in Section V.A -
especially at low $p_T$. Again, the final $K^-,\bar K^0$ are interacting stronger with
baryons than $K^+,\bar K^0$.
The influence of the final-state hadronic interaction on the modification of the spectra
is illustrated in Fig. \ref{fig:Rat} which shows the ratio of 'reconstructed' to 'decayed' 
$p_T$ spectra of $K^*=K^{*+}+K^{*0}$ (solid lines) and $\bar K^*=K^{*-}+\bar K^{*0}$ (dashed lines)  
for  central Au+Au collisions  at bombarding energies
of 4.5, 6, 8, 10.7 and 15 A GeV for the case with in-medium effects as  presented in
Fig. \ref{fig:ptFAIR}. One can clearly see that rescattering of the final pions and kaons strongly affects the low $p_T$ part of the $K^*, \bar K^*$ spectra and it is larger for $\bar K^*$.

\begin{figure}[h!]
  \centerline{\includegraphics[width=8.0cm]{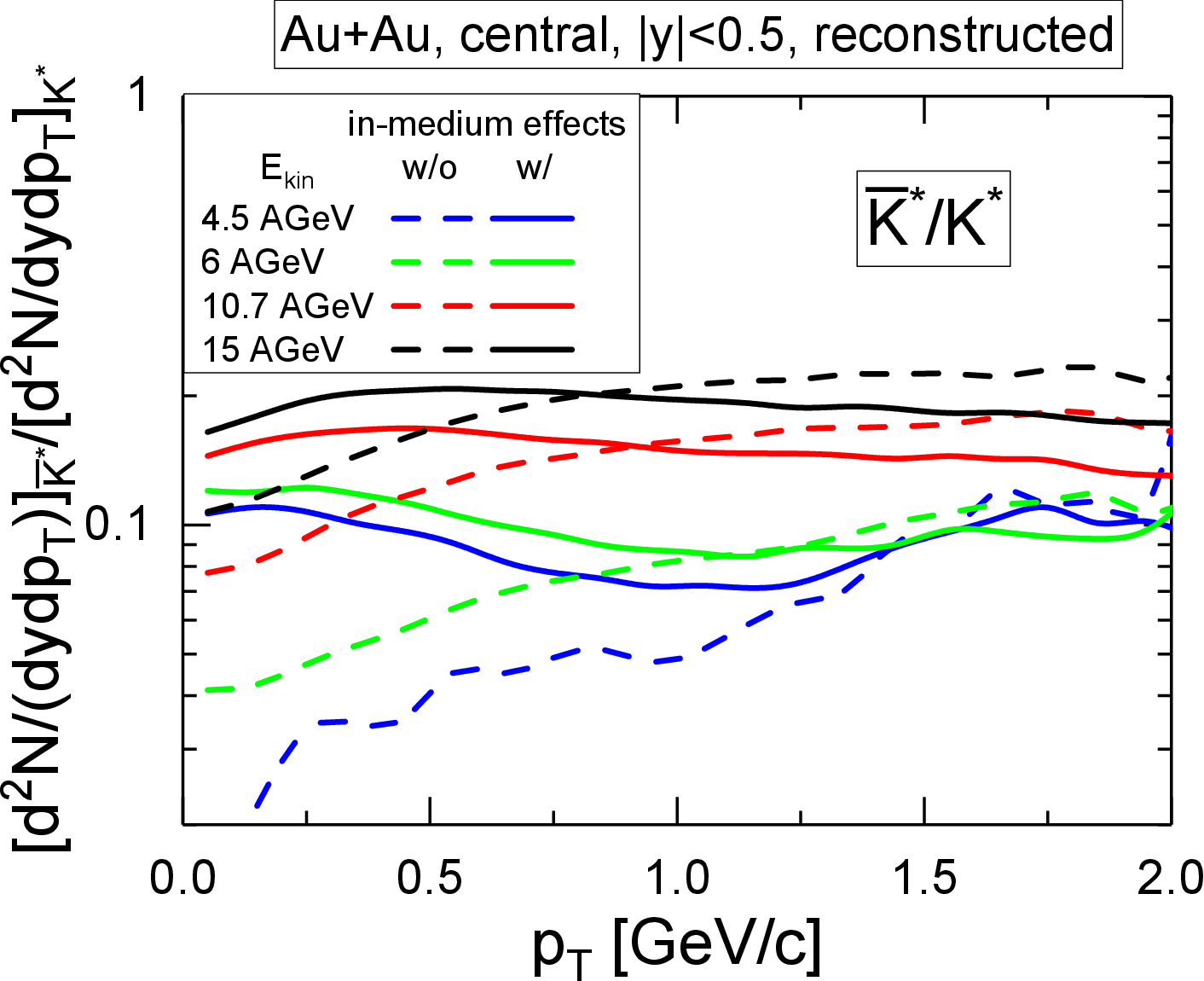}}
  \caption{PHSD predictions for the ratios of the 'reconstructed' $p_T$ spectra 
of $\bar K^{*}$ over $K^*$ for central Au+Au collisions at midrapidity $(|y|<0.5)$ 
at bombarding energies
of 4.5, 6, 8, 10.7 and 15 A GeV. The dashed lines show the PHSD results without
including the in-medium effects for $K^{*}$ and $\bar K^{*}$
while the solid lines stand for the case with in medium effects. }
\label{fig:RatptFAIR}
\end{figure}

As follows from Figs. \ref{fig:ptFAIR} a sizeable in-medium modification
of the $p_T$- spectra of $\bar K^{*}$'s is expected with decreasing bombarding
energies: the $p_T$- distribution is shifted to the low $p_T$ region,
such that the shift is about 0.1-0.15 GeV at 4.5 AGeV.  
Contrary to the $\bar K^{*}$ mesons, the $p_T$- spectra of $K^{*}$'s are only slightly 
shifted to the high $p_T$- region. Such shifts can be observed experimentally, 
e.g. by comparing the $K^{*}$ with $\bar K^{*}$ spectra. Such  Fig. 
\ref{fig:RatptFAIR} shows the ratios of the 'reconstructed' $p_T$ spectra 
of $K^{*}$ over $\bar K^*$ for central Au+Au collisions at midrapidity $(|y|<0.5)$ 
at bombarding energies
of 4.5, 6, 8, 10.7 and 15 A GeV. The dashed lines show the PHSD results without
including the in-medium effects for $K^{*}$ and $\bar K^{*}$
while the solid lines stand for the case with in medium effects.
One can see that there ratio $K^{*}/\bar K^{*}$ is approximately flat over $p_T$ 
for the in-medium cases while it decreases at low $p_T$ strongly, especially
at low energy.
\begin{figure}[h!]
  \centerline{\includegraphics[width=8.0cm]{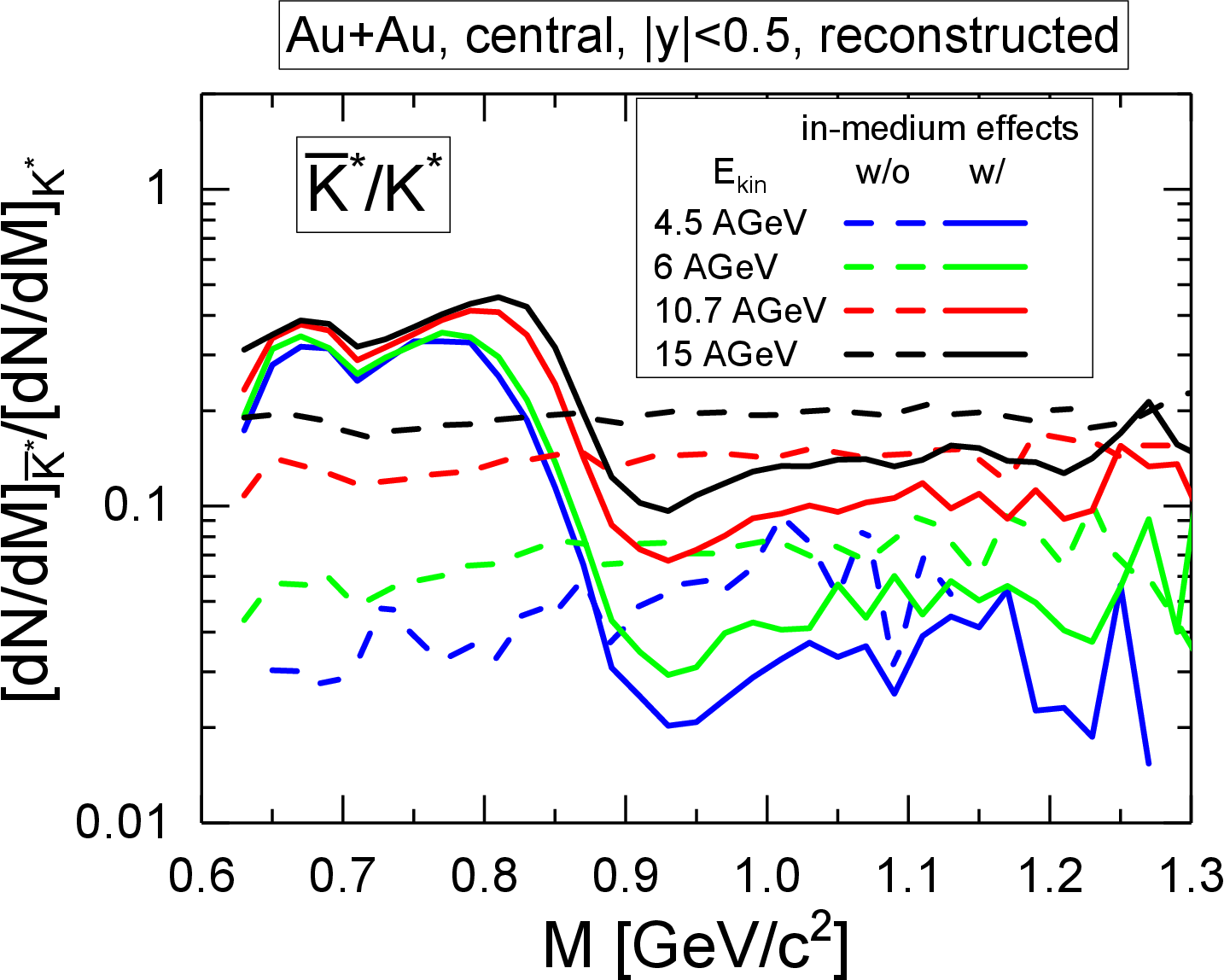}}
  \caption{PHSD predictions for the ratios of the 'reconstructed' mass distribution
of $\bar K^{*}$ over $K^*$ for central Au+Au collisions at midrapidity $(|y|<0.5)$ 
at bombarding energies
of 4.5, 6, 8, 10.7 and 15 A GeV. The dashed lines show the PHSD results without
including the in-medium effects for $K^{*}$ and $\bar K^{*}$
while the solid lines stand for the case with in medium effects. }
\label{fig:RatMFAIR}
\end{figure}

In-medium effects are even more pronounced  when looking at the mass distribution
in Fig. \ref{fig:MFAIR}. The shape of the $\bar K^{*}$ mass spectra (even integrated over
all $p_T$ as shown in Fig. \ref{fig:MFAIR}) are strongly modified -
it is getting flat at low $M$ with decreasing bombarding energy. 
This is mainly due to the in-medium modification of the $\bar K^{*}$ spectral functions 
and final-state interaction of the decay products. Here 
the modification of the 'reconstructed' $K^{*}$ mass distribution is less visible than for $\bar K^{*}$'s.
The medium 'distortion' of the $K^*/\bar K^*$ reconstructed mass spectra are
stronger at low $p_T$, thus, one can study experimentally the mass distribution 
of $K^*/\bar K^*$'s at different $p_T$ bins. Fig. \ref{fig:RatMFAIR} shows
the PHSD predictions for the ratios of the 'reconstructed' mass distribution
of $K^{*}$ over $\bar K^*$ for central Au+Au collisions at midrapidity $(|y|<0.5)$ 
at bombarding energies of 4.5, 6, 8, 10.7 and 15 A GeV. Here again the dashed lines 
show the PHSD results without including the in-medium effects for $K^{*}$ 
and $\bar K^{*}$ while the solid lines stand for the case with in medium effects.
One can see the strong enhancement of the ratio for $M < 1$ GeV$/c^2$ for the
in-medium scenarios while at larger $p_T$ the differences are small.

Finally, we conclude that the future facilities -- FAIR in Darmstadt as well as NICA 
and the fixed target BM\@N experiment at the Nuclotron in Dubna as well as the BES 
program at RHIC -- are well 
located in energy to study the in-medium effects related to high baryon density,
in particular the in-medium modification of the $K^*/\bar K^*$ spectral function.
We note, however, that it is a rather challenging experimental task: in spite
that the background is smaller at FAIR/NICA energies than at LHC or RHIC due 
to the lower pion and kaon abundances, the 'true' signal is also smaller,
and the final-state interaction of the decay products is still large. 
We will discuss in the next section how to perform the experimental analysis of the
mass spectra  in order to obtain the in-medium signal.

\section{Experimental procedure to extract mass and width of a resonance }

\label{sec:experim}
\subsection{Invariant mass distributions of 'decayed' and 'reconstructed' $K^{*0}$ and $\bar{K}^{*0}$ }

In this section we present in detail the experimental method to obtain 
information about the properties of resonances in the medium. 
We recall that experimental mass and width measurements of 
the $K^{*0}$ + $\bar{K}^{*0}$ at LHC and RHIC energies show a mass shift 
in the low momentum region for heavy-ion collisions 
\cite{Abelev:2014uua, Aggarwal:2010mt, Adams:2004ep}.
This result has been obtained by fitting the measured invariant mass spectra
with a fit function (or ansatz) constructed by a product of the free spectral function
with a constant width and a simplified Boltzmann factor weighted with some
prefactor aiming to account for the phase-space distribution of 
initial pions and kaons from which the $K^*$'s have been formed. 
However, one has to point out that the fitting function used first by the 
STAR Collaboration and later by the ALICE Collaboration has no further reference for validation. In spite of that we will use here the same procedure
for fitting our theoretical calculations with respect to invariant mass spectra
as in experiment, i.e. we will use the same ansatz. 
By that we show how sensitive the results of the fit (mass and width) depend 
on constraints of the fit parameters. 
This shows how difficult it is to extract any meaningful masses and widths  
from  final-state invariant mass spectra which are affected by additional 
signal loss due to resonance decay, final particle re-scattering and regeneration
of resonances at a later stage of the hadronic phase. We show also that in spite 
of such modification of the initial information between the decay point of the resonance and its reconstruction by the decay products (pions and kaons), 
the fit value for the $K^*$ mass is in agreement with the vacuum value when 
using the experimental ansatz for the fit.

We base our study on the PHSD results without and with an in-medium modification
and treat it in the same way as performed for the real experimental data.
Using the PHSD as a 'theoretical laboratory' we discuss also the consequences of
applying different experimental conditions, kinematic selections and 
statistical errors from the underlying combinatorial background. 
We will present four different fit scenarios
and add the experimental procedure to it step by step.

In order to shed light on the physical origin of this experimental observation, 
we investigate the mass and width of the $K^{*0}$ and $\bar{K}^{*0}$ mass distribution
from the PHSD approach calculated first at the decay point ('decayed') 
and apply the reconstruction procedure to the final-state particles - pions and kaons -  ('reconstructed') 
as discussed in the previous sections. In the next step we account for the statistical error in the experimental data. 
Furthermore, we compare the PHSD results with measure the $K^{*0}$ and $\bar{K}^{*0}$ 
from data form the ALICE experiment. We note, that for the PHSD results we use the 0-5\% most central Pb-Pb collisions 
at  $\sqrt{{s}_{NN}}=2.76$~TeV.

\subsubsection{Fitting ansatz for the mass distribution}

The mass and width information of the resonances is experimentally obtained from
the measured differential invariant mass spectra  $dN/dM$. 
One has to keep in mind that invariant mass spectra of resonances contain the information 
on the spectral function as well as on the occupation probability. 
Since both parts of the information cannot be separated 
experimentally from the measured spectra, one has to apply some assumptions. 
In the present study we follow the procedure used by the STAR and ALICE experiments
to fit the measured invariant mass spectra.

The simplest approximation is to assume that the resonances are produced 
in the equilibrated medium with a spectral function of Breit-Wigner form, 
i.e. the resonance mass spectrum is proportional to 
the Breit-Wigner spectral function $A$, as defined in section 3, Eq. (\ref{eq:sfrbw}),
weighted by the $K^*$'s occupation probability at given temperature $T$.

Following the experimental procedure used by the ALICE and STAR Collaborations
\cite{Abelev:2014uua,Adams:2004ep},
if one wants to differentiate the information further and account for the 
$p_T$ dependence, e.g. to consider the mass spectra in some fixed $p_T$ interval (experimental bin), the fit function $F$ is defined as:
\begin{eqnarray}
\label{fitf}
\left.{dN\over dM} \right|_{p_T} &=&
	F(M | C, M_V^{*}, \Gamma_V^*, p_{T}, T)  \nonumber \\
	&=& A(M | C, M_V^{*}, \Gamma_V^*)\cdot  f(M | p_{T}, T).	
\end{eqnarray}	
The notation $F(X|X_1,X_2,...,X_N)$ means that one fits the $X$-distribution by varying
the parameters $X_1,X_2,...,X_N$.

It should be understood as follows:
since the $F$ function is fitted for each $p_T$ bin, one can neglect the $p_T$ dependence
inside the bin and consider an "average" $p_T$ value as a parameter
which is varying only for different bins. Thus, inside each $p_T$ bin Eq. (\ref{fitf})
represents the so called mass distribution. The goal of the fit is to identify the shape
of this distribution in terms of chosen/assumed formula for the occupation probability 
and spectral function by varying free parameters charactering the shape
for each $p_T$ bin. Later on one can plot these parameters versus $p_T$ to identify the
dependences. The goal of this study as well as the experimental analysis by the ALICE
Collaboration, is to identify the in-medium mass of $K^*$'s (which is the $M_V^{*}$ 
parameter in the fit) at each $p_T$ bin.

Let's now come to the definition of ingredients in Eq. (\ref{fitf}) used for the fit.
Function $f(M | p_{T}, T)$ is the occupation probability at given temperature $T$. 
As mentioned above, the actual form is adopted in line with the ALICE and STAR analysis 
\cite{Abelev:2014uua,Adams:2004ep}:
\begin{align}
 \label{fitBF}
	f(M | p_{T}, T) &= \frac{M}{\sqrt{M^2 + p_{T}^2}}\exp{\left( - \frac{\sqrt{M^2 + p_{T}^2}}{T}\right)}.
\end{align}
The exponential term here accounts for the exponential phase-space distribution (spectrum) of the resonance  and follows from the Boltzmann distribution when neglecting 
the 'thermal' motion of resonances (see e.g. \cite{Shuryak2003}), 
i.e. replacing $p$ by $p_T$,
which should be an acceptable approximation for the mid-rapidity and is commonly used by experimental collaborations \cite{Abelev:2014uua,Adams:2004ep}.
Since in the thermal medium the $K^*$ resonances are produced dominantly via $K+\pi$ scattering, an extra factor (first term in Eq. (\ref{fitBF})) accounts for the phase-space population of parent pions and kaons and 
has been introduced in the experimental analysis in Refs. \cite{Abelev:2014uua,Adams:2004ep}.

In Eq.(\ref{fitf})  the $A(M | C, M_V^{*}, \Gamma_V^*)$ is a relativistic mass-dependent 
Breit-Wigner function for the strange vector resonance $V = K^{*0}$ or $\bar{K}^{*0}$, 
given by
\begin{align}
\label{fitfA}
	A(M | C, M_V^{*}, \Gamma_V^*)= C \frac{M^2 \Gamma_V^*}{(M^2 - M_V^{*2})^2 + (M\Gamma_V^*)^2}. 
\end{align}
For all fit cases considered below, the overall scale ($C$) and the mass peak ($M_V^*$) are free fit parameters. 
The quantity $\Gamma_V^*$  represents the total mass dependent width of the $V$- resonance in the medium which
relates to the imaginary part of the self energy in line with Eq.(\ref{eq:mass-width}).
However, for the experimental fit one adopts an approximation that the total mass dependent width in the spectral function 
can be expressed as a sum of the decay and collisional widths:
\begin{align}
\label{fitfG}
	\Gamma_V^*(M) = \Gamma_{V,dec}(M) + \Gamma_{coll}, 	 
\end{align}
where the mass dependent decay width of a strange vector resonance $V$ is defined by Eq. (\ref{eq:vmdw}) 
by assuming that the kaon spectral function can be replaced by the $\delta$-function, i.e. by
ignoring the in-medium modification of the kaon properties. Thus,
\begin{align}
 \label{fitGdec}
 \Gamma_{V,\, \textrm{dec}} (M) = \Gamma_{V}^{0} \left( \frac{M_{V}}{M} \right)^{2} 
  \left( \frac{q(M)}{q(M_V)} \right)^3. 
 \end{align}
Here, the momentum of the $V$-resonance with mass $M$ is 
$q(M) = \sqrt{\lambda(M, M_K, M_{\pi})}/ 2 M$.
The vacuum width and mass of the $K^{*0}$ and $\bar{K}^{*0}$ used for 
our fit is chosen to be the same as in the theoretical calculations: 
$\Gamma_V^0=42$~MeV and $M_V=892$~MeV;
$M_K$ is the kaon mass ($493.7$~MeV), $M_\pi$ is the pion mass ($139.6$~MeV).
In Eq. (\ref{fitf}) $\Gamma_{coll}$ stands for the collisional width which 
accumulates the effects of the in-medium modification to the total width 
and is one of the fit parameters.

Now  how  we select the temperature:
a temperature of $T= 160$ MeV, which is close to the chemical freeze-out,
is used for the resonances at decay point and $T=100$ MeV is used to describe the reconstructed resonance distribution close to the kinetic freeze-out. 
Before fixing $T$ we perform a fit treating $T$ as a free parameter. 
The temperature dependence in the fit is rather small and we achieve a good $\chi^2$ 
within 20 MeV around the selected fixed value for $T$.

For our fits, the width of the Breit-Wigner function is handled in three different ways: \\
1) In the first case -- the mass dependent width case, the vacuum decay width depends on mass, the collisional width $\Gamma_{coll}$ is the free fit parameter.  \\
2) For the second case the fits are performed assuming the total width of the Briet-Wigner spectral function ($\Gamma_V^*$) is a free fit parameter. \\
3) Finally we consider a fixed width case, where the total width is fixed to the constant value 50~MeV. \\

\subsubsection{Definition of four different fit procedures}

We define four different fitting procedures to describe the theoretical spectra 
which are treated as experimental data by applying/adding the experimental 
conditions. The four fitting options are listed below where the colors 
in parentheses indicate the corresponding curves in the figures.

{\bf Fit I.) }   {\bf Mass dependent width}  (black):\\
 A fit using the relativistic Breit-Wigner spectral function (\ref{fitfA}) 
 with mass dependent width defined by Eqs. (\ref{fitfG}), (\ref{fitGdec})
with $\Gamma_{coll}$ as a fit parameter.
The Boltzmann factor (\ref{fitBF}) with  $T=160$ MeV (for the fit to decayed spectra) 
or $T=100$ MeV (for the fit to reconstructed spectra)
is employed using the statistical error strictly based on the theoretical counts ($\sqrt{N}$ 
for each bin).

{\bf  Fit II.)}  {\bf Mass dependent width +5$\%$ error} (green):\\
The same as fit I but 5$\%$ error bars are added to the theoretical calculations.\\
This fit is used to  mimic the fitting conditions used for data obtained at ALICE. For that
we need to take into account the statistical errors present in the experimental data.  
Since the resonance mass peak sits on top of a large combinatorial background, 
the statistical errors of each bin in the resonance mass region
 are roughly the same  ($\approx 5\%$ of the peak bin).  

{\bf Fit III.)} {\bf Simple width + 5$\%$ error} (blue): \\
 A fit using the relativistic Breit-Wigner spectral function 
where the total width is a free parameter of the fit (no mass dependence).
The Boltzmann factor is defined with T=160 MeV ('decayed') or T=100 MeV ('reconstructed'). 
Additional 5$\%$ error bars are added for the PHSD calculations.
We note that this "simple width" case is mostly used in the experimental fitting procedures.

{\bf Fit IV.)}   {\bf Fixed vacuum width + 5$\%$ error} (red): \\
A fit using the relativistic Breit-Wigner spectral function where the total width 
is fixed to the PHSD vacuum width of $\Gamma_0=42$ MeV$/c^{2}$.
Also additional error bars of 5$\%$ are added to the theoretical calculations. 
We note that the experimental fit with a fixed width set to the vacuum value 
is used to search for any deviation from the vacuum widths. 
However, the $\chi^{2}$ value of the fit turned out to be reasonable within the statistical error. 
For the RHIC and LHC data the width was fixed to the vacuum width of ${50}~{MeV/c^{2}}$ to 
constrain the fit. This was done to stabilize the background (BG) fit to extract the resonance signal above the BG.\\
All fits have been performed within the invariant mass rage of 0.7~$GeV/c^{2}$ and 1.1~$GeV/c^{2}$.

\subsection{$ K^{*0}$ and $\bar K^{*0}$ mass and width for different fit options}

\begin{figure}[h!]
\includegraphics[width=8.9cm]{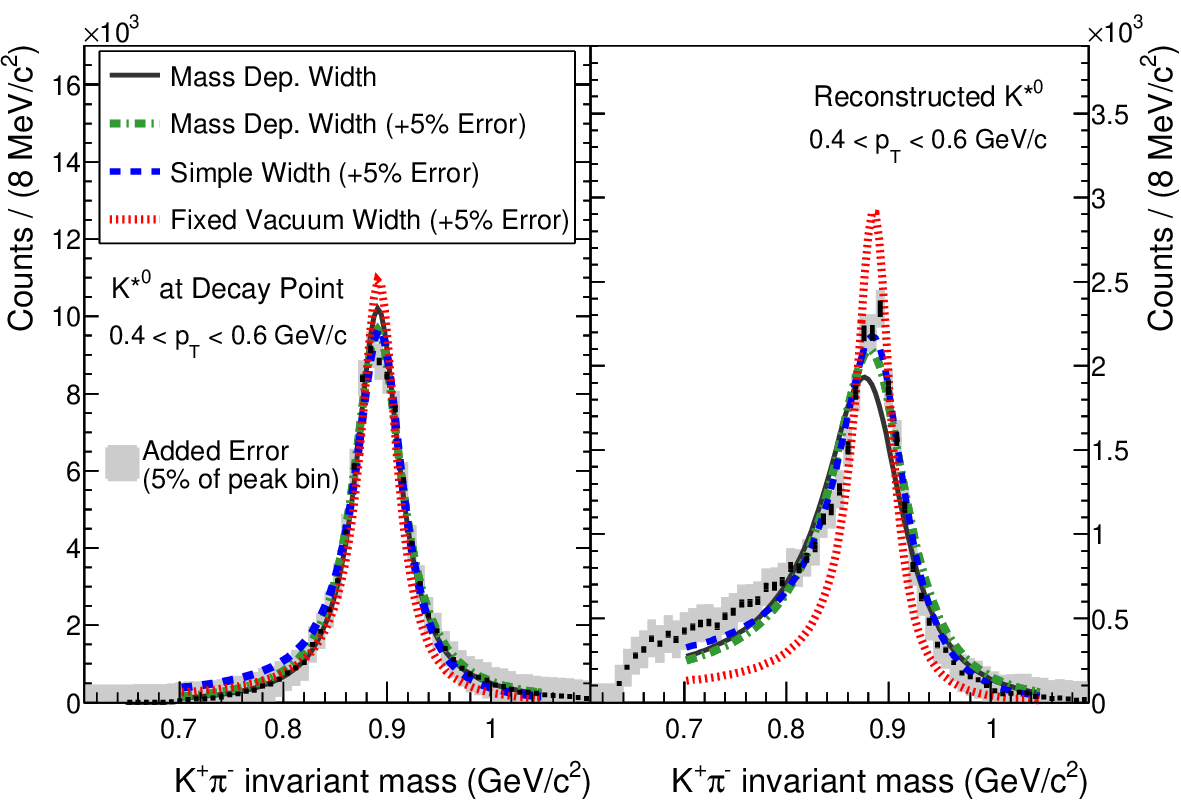}
  \caption{ Invariant mass distributions of the $K^{*0}$ from the PHSD calculations (with no in-medium modification)
  in the low momentum region (p$_{\rm T}$ = 0.4 - 0.6 GeV/c) at the decay point (left) 
  and for the 'reconstructed' case (right).}
  \label{fig:inv_kstar_decay_lowpt}
\end{figure}

\begin{figure}[h!]

\includegraphics[width=8.9cm]{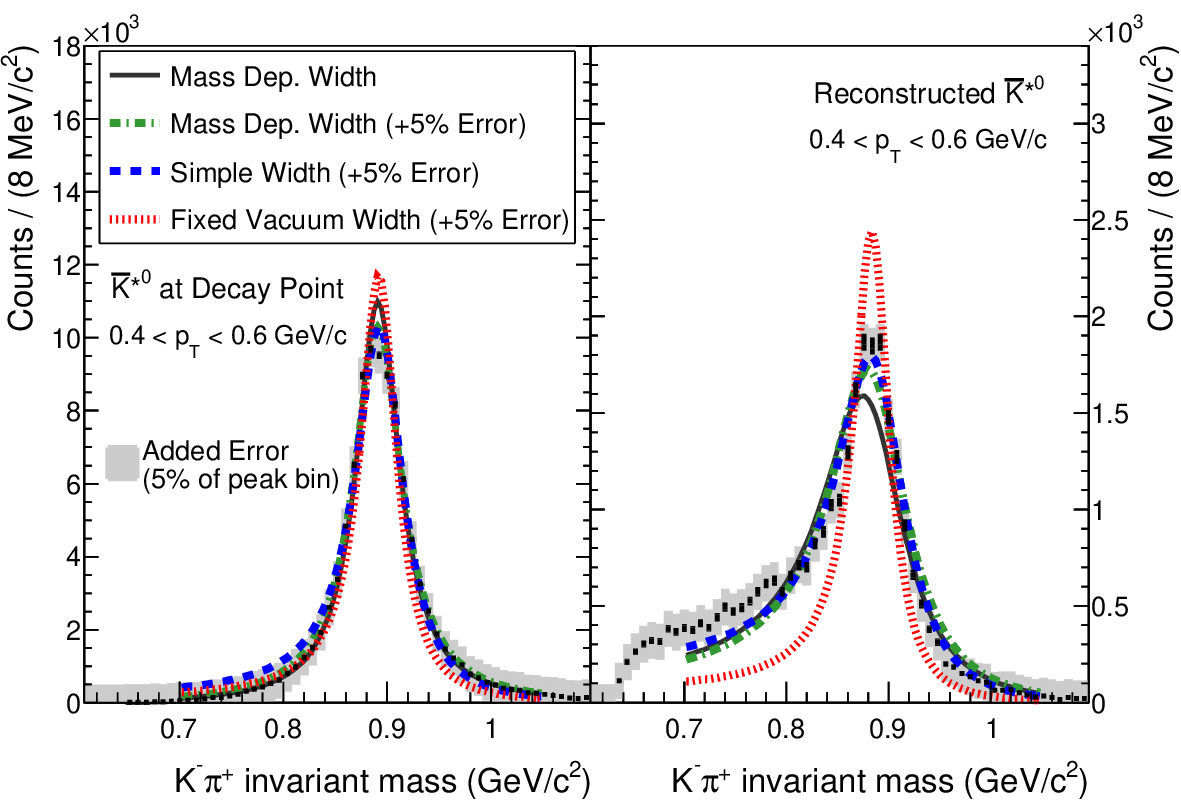}
  \caption{ Invariant mass distributions of the $\bar K^{*0}$ from the PHSD calculations (with no in-medium modification)
  in the low momentum region (p$_{\rm T}$ = 0.4 - 0.6 GeV/c) at the decay point (left) 
  and for the 'reconstructed' case (right).}
  \label{fig:inv_kstarbar_decay_lowpt}
\end{figure}

This PHSD calculations are done without the in-medium modification from CSR. Figure~\ref{fig:inv_kstar_decay_lowpt} shows the invariant mass distributions of the $K^{*0}$ in the low momentum region (p$_{\rm T}$ = 0.4-0.6 GeV/c) at the decay point (left) and the reconstructed (right). The reconstructed $K^{*0}$ resonances show a broadening of the invariant mass distribution to lower masses. This will result in a width broadening of the signal and a possible mass shift as an effect from the hadronic phase interactions. The same is visible for the antiparticle $\bar K^{*0}$ shown in figure~\ref{fig:inv_kstarbar_decay_lowpt}.

We investigate the initial masses and widths distribution of the $K^{*0}$ (and $\bar{K}^{*0}$)  
 due to the hadronic medium. This will answer the question as to how much of the initial vacuum spectral function is modified by the hadronic medium due to the hadronic interactions. In addition we investigate the effects 
of the experimental resonance signal extraction. And the statistical error 
on the experimental data due to the combinatorial background underneath the invariant mass signal
is studied.

\begin{figure}[h!]
\includegraphics[width=8.8cm]{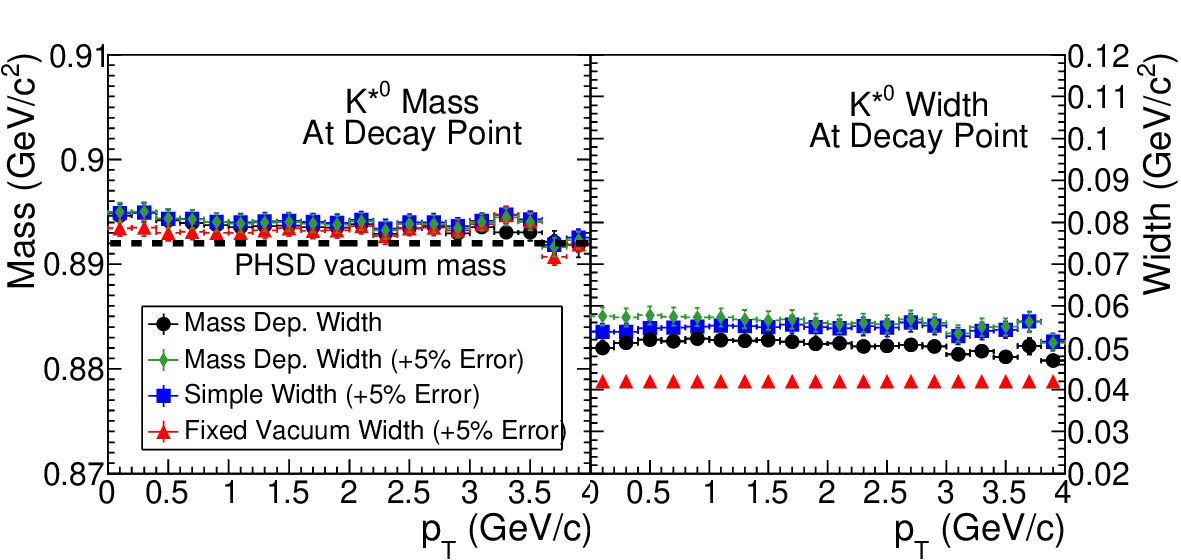}
  \caption{Mass and width for the different fits of the $ K^{*0}$ (same for $\bar K^{*0}$) at the decay point. The PHSD vacuum mass is 0.892 GeV/c$^{2}$ and vacuum width is 42~GeV/c$^{2}$.  }
  \label{fig:mass_width_kstar_decay}
\end{figure}

\begin{figure}[h!]
\includegraphics[width=8.8cm]{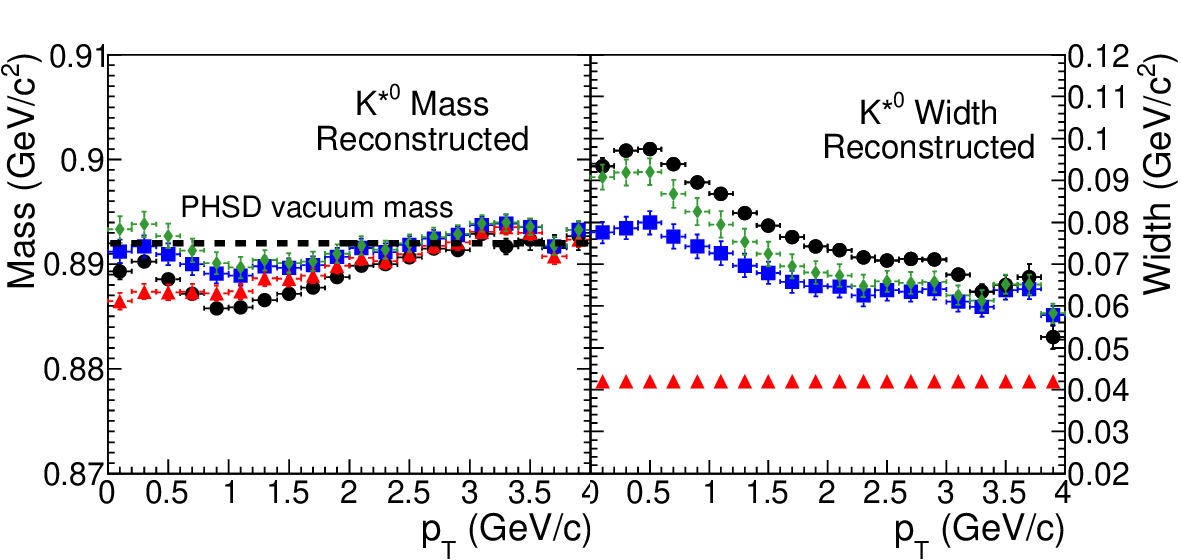}
\caption{Mass and width for the different fits of the $ K^{*0}$  (same for $\bar K^{*0}$)
for the reconstructed resonance case. The PHSD vacuum mass is 0.892 GeV/c$^{2}$.}
  \label{fig:mass_width_kstar_reco}
\end{figure}

 Figure~\ref{fig:mass_width_kstar_decay} shows the $K^{*0}$ (same for $\bar{K}^{*0}$) mass (left) and width (right) versus transverse momentum at the decay point without in-medium modification. The PHSD vacuum values for the mass  is 0.892 GeV/c$^{2}$ and for the width is 42~GeV/c$^{2}$.  A clear mass shift to higher masses is visible for the $K^{*0}$ and the width it is between $\Gamma_{K^{*0}}$ = 50-60 MeV/c$^{2}$.
 
 The reconstructed $K^{*}$ resonances from the final-state particles show a shift towards 
broader widths and smaller masses as shown in figure~\ref{fig:mass_width_kstar_reco}. 
The $K^{*0}$ (same for $\bar{K}^{*0}$) masses are shifted by 5-10 MeV/c$^{2}$ to lower masses in the low momentum region around 
$p_T$~=~1~GeV$/c$. The widths are increased in the low momentum regions by 20-40 MeV/c$^{2}$. 

This mass shift and width broadenings are effects from the hadronic phase interactions which changes the spectral function shape. This is mainly due to the regeneration of the resonance at later times and lower temperatures, which populate preferentially the low mass region.


 \begin{figure}[h!]
\includegraphics[width=8cm]{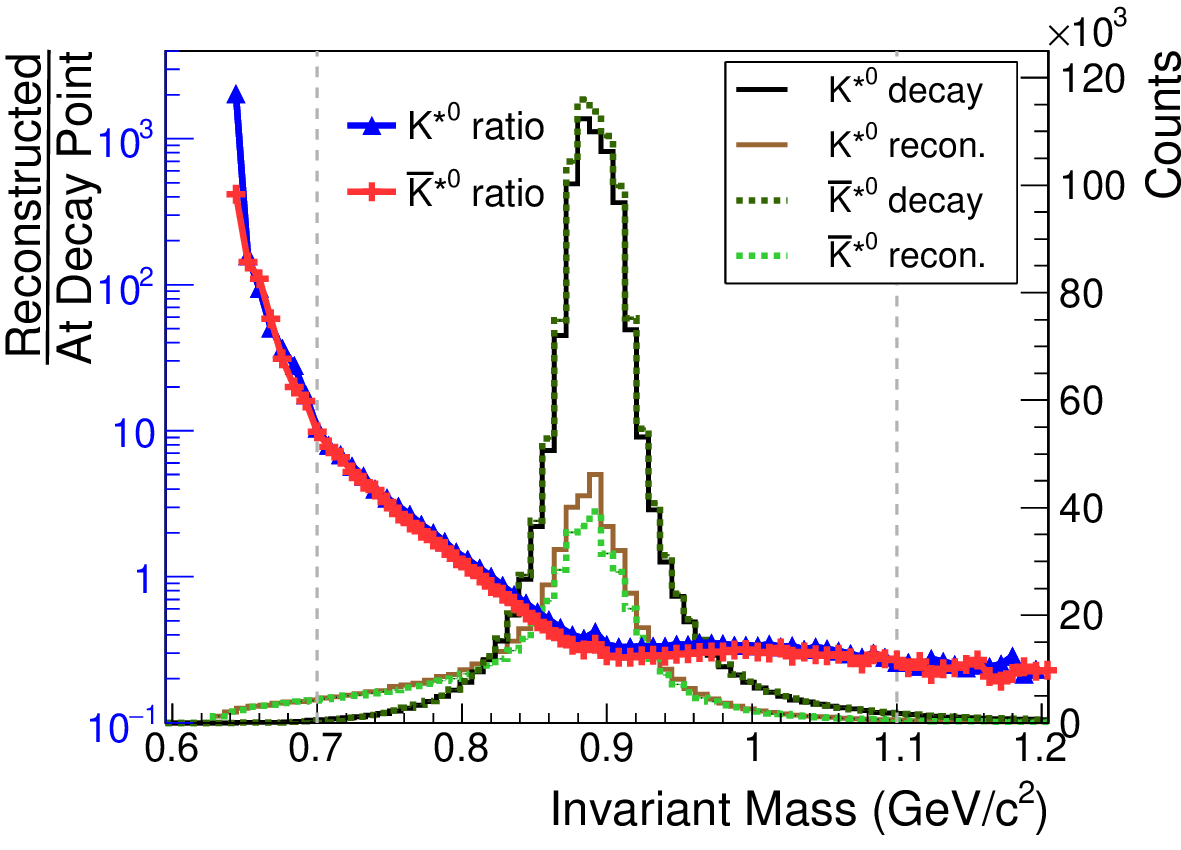}
  \caption{Ratios (left axis) of the reconstructed mass distributions 
  of the $K^{*0}$ (blue) and $\bar K^{*0}$ (pink) 
  resonances to the mass distribution at the decay point as a function of the invariant mass. 
  The corresponding mass distributions are also presented (right axis) :
  $K^{*0}$ reconstructed - black line, decayed - brown line; 
  $\bar K^{*0}$ reconstructed - olive line, decayed - green line.
  The results correspond to the PHSD calculations including no in-medium effects for 
  the $K^{*0}$ and $\bar K^{*0}$. }
  \label{fig:ratio}
\end{figure}

Figure \ref{fig:ratio} shows the ratios (left axis) of the reconstructed mass distributions of
the $K^{*0}$ and $\bar K^{*0}$ resonances to the mass distribution of the resonances
at the decay point as a function of the invariant mass. 
The corresponding mass distributions are also presented (right axis) :
  $K^{*0}$ reconstructed - black line, decayed - brown line; 
  $\bar K^{*0}$ reconstructed - olive line, decayed - green line.
It is clearly noticeable that the resonance mass distributions are 
shifted to lower masses throughout their interactions in the medium. 
The signal loss for the reconstructable resonances are mainly cause by further 
interactions of the decay particles in the hadronic phase.

\subsection{Comparison to experimental data}\label{sec:experimental-data-no-medium}

\begin{figure}[h!]
\includegraphics[width=7cm]{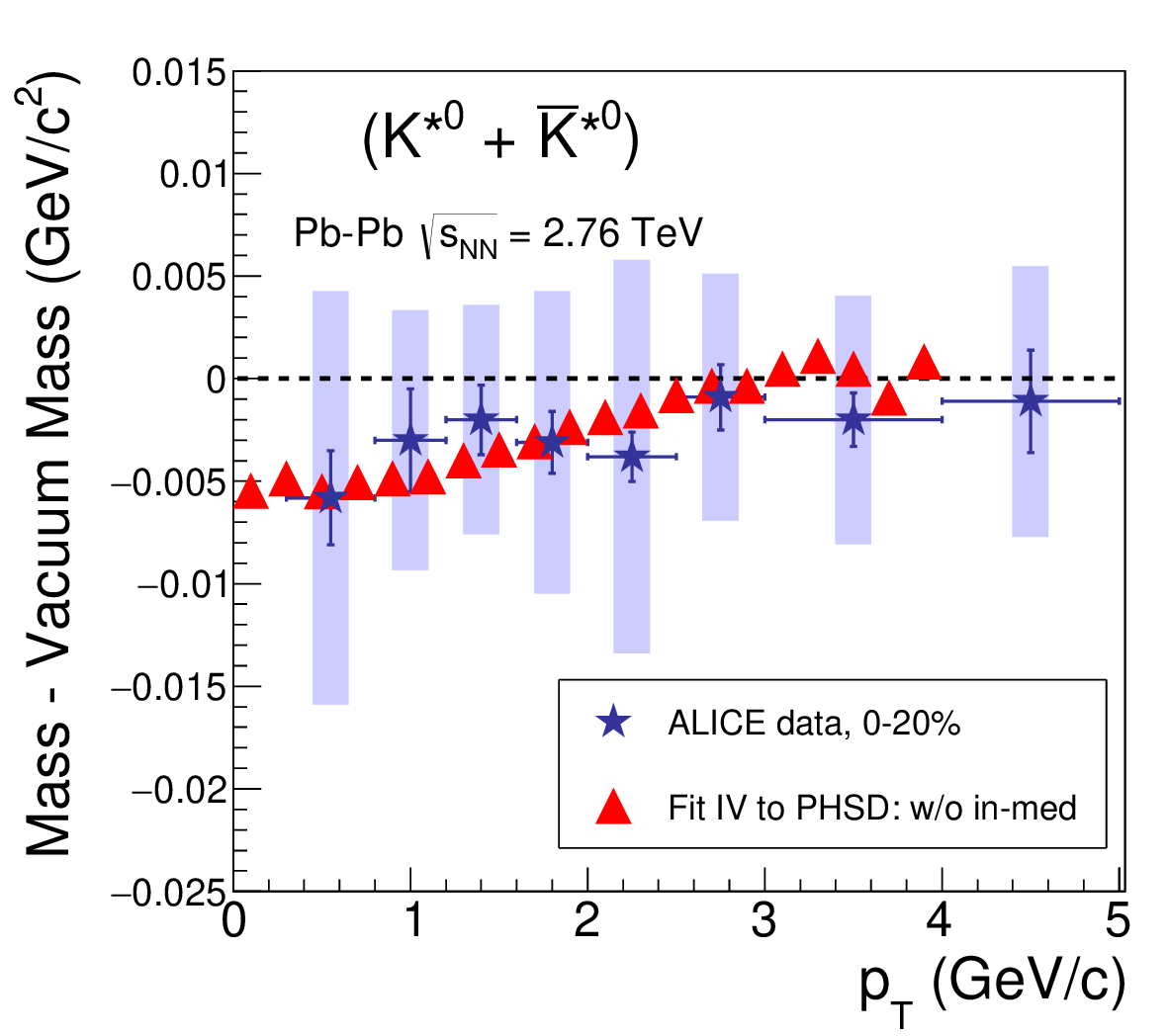}
  \caption{Deviations from the vacuum mass for the ALICE data \cite{Abelev:2014uua} and PHSD fitted mass spectrum of the reconstructed $K^{*0}$ + $\bar K^{*0}$ as a function of $p_T$. 
 The PHSD invariant mass distribution was fitted with {\bf Fit IV} (explained in the text) 
  including  statistical experimental error.}
  \label{fig:alice}
\end{figure}

 The comparison of the PHSD results (figure~\ref{fig:mass_width_kstar_reco} red line) 
 for the deviation from the vacuum mass of the fitted
mass spectrum of reconstructed $K^{*0}$ + $\bar K^{*0}$ mesons  
to the ALICE data \cite{Abelev:2014uua} is shown in Figure~\ref{fig:alice} 
as a function of the $p_T$. The ALICE data correspond to 
Pb+Pb, 0-20\% most-central collisions at $\sqrt{{s}_{NN}}=2.76$~TeV, while the PHSD data are from 0-5\% Pb-Pb collisions. Since the vacuum mass value for the $K^{*0}$ adopted in the PHSD
is 892 MeV/c$^{2}$ and the PDF value is 895.81 MeV/c$^{2}$, we present the mass as a deviation from its vacuum value. The vacuum width for the fixed width (Fit IV) in the fit is 50 MeV/c$^{2}$ for the ALICE data and 42 MeV/c$^{2}$ in PHSD. The theoretical calculations describe the data very well within the statistical and systematical errors. At RHIC energies a similar but statistically more significant mass shift is visible \cite{Aggarwal:2010mt,Adams:2004ep}. According to the PHSD calculations the mass shift is caused by several effects. One is the $K^*$ which survives that hadronic phase and the second is the fixed vacuum width fit. Remember the PHSD calculations here don't include the in-medium modification 
of the $K*/\bar K^*$ properties due to the coupling to baryonic medium discussed in Section III.

\subsection{Estimation for mass shift from an in-medium modification}\label{sec:in-medium}

As follows from Fig. \ref{fig:alice} the experimentally measured mass shift of strange vector mesons is reproduced by the PHSD
and attributed to the interactions in the hadronic phase as well as to the resonance reconstruction 
procedure itself. 
In order to understand the sensitivity of our experimental procedure for the investigation of the medium effects, 
we perform a 'model study' where we adopt an 'extreme' in-medium scenario for $K^*/\bar K^*$
by considering the in-medium effects only due to the coupling to the baryonic medium and discarding the coupling
to anti-baryons. As discussed previously such a scenario 
can not be realistic at the LHC energies  due to the large anti-baryon production rate at midrapidity which leads to a very low net-baryon density and small baryon chemical potential.
This is contrary to FAIR/NICA energies where the matter is baryon dominated.
However, this model study will help us get an idea for the sensitivity to measure the $K^{*0}$ in-medium modification 
at LHC energies and answer the question  if  mass-shift signatures from in-medium effects would survive 
the interactions in the  hadronic phase.
Thus, we consider for our 'model study' the in-medium modifications of $K^*/\bar K^*$ via coupling to the baryon density based on our in-medium scenario from Section III. 
Similar to low energies this will provide us a visible modification on the $\bar K^{*0}$ and very small change of $K^{*0}$ since the baryon density itself is nonzero at the
LHC energies contrary to the net-baryon density.
With this calculation we intent to simulate an upper limit for the possible in medium signature.

\begin{figure}[h!]
\includegraphics[width=8.9cm]{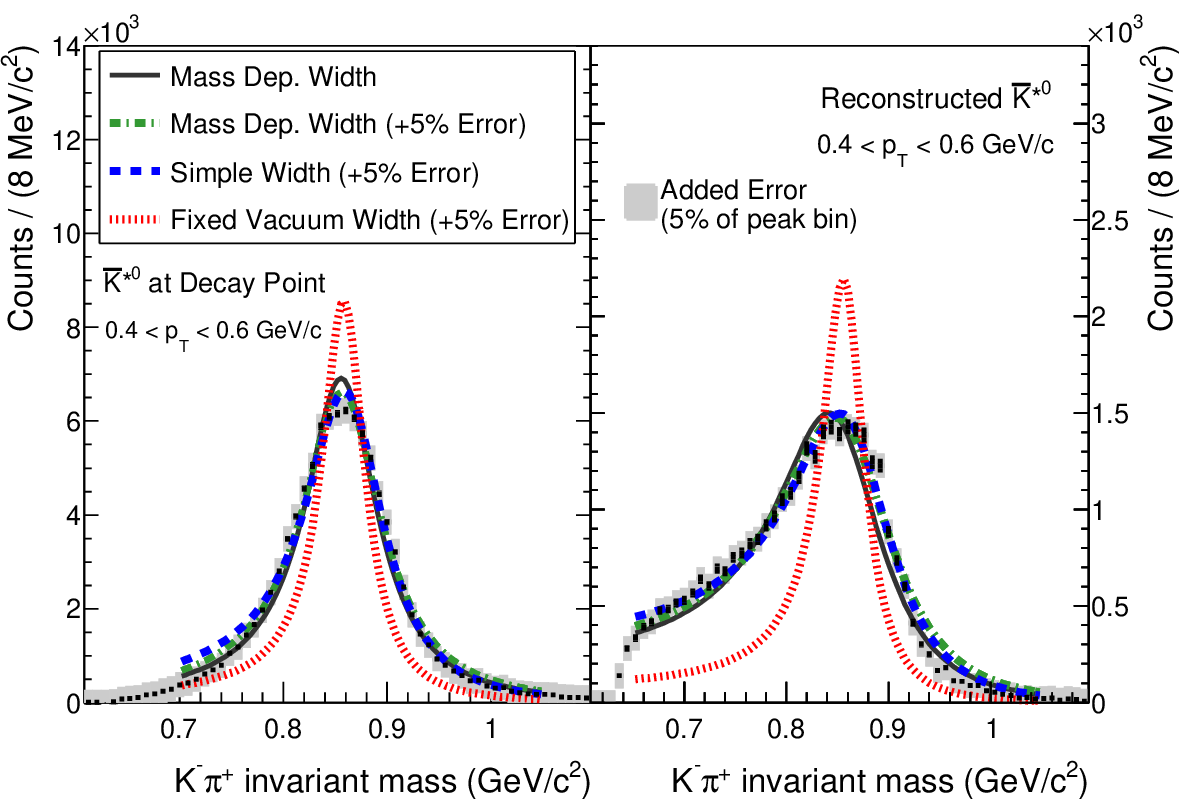}
 \caption{Invariant mass distributions of the $\bar K^{*0}$ from the PHSD calculations (with in-medium effects)
  in the low momentum region ($p_{T}$ = 0.4 - 0.6 GeV/c) at the decay point (left) 
  and for the 'reconstructed' case (right).}
  \label{fig:inv_kstarbar_decay_lowpt_medium}
\end{figure}

\begin{figure}[h!]
\includegraphics[width=8.8cm]{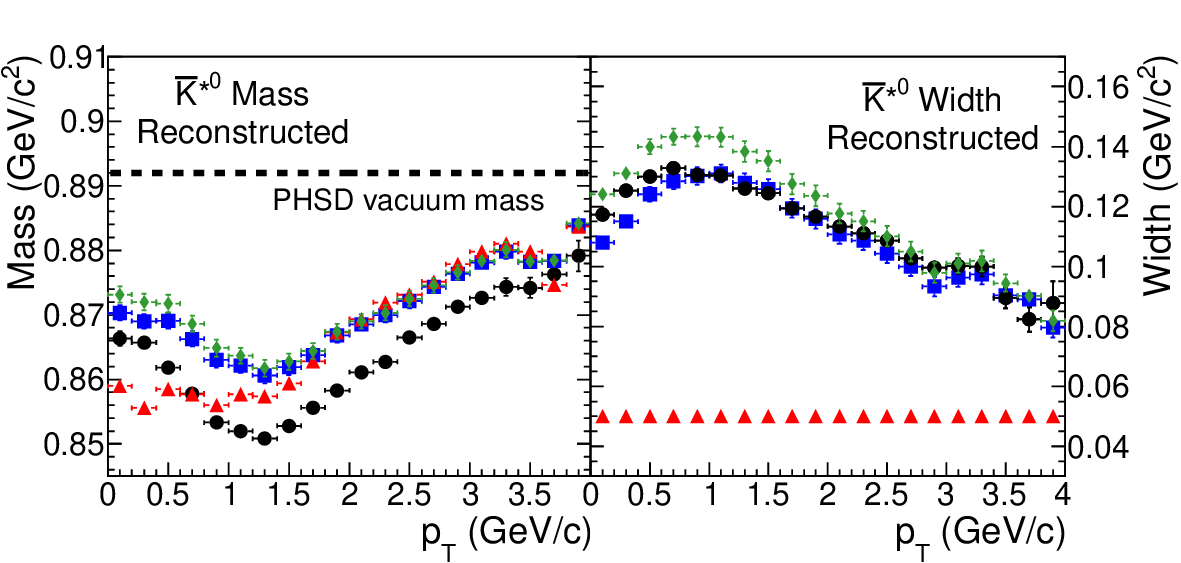}
\caption{Mass and width for the different fits of the $\bar K^{*0}$
for the reconstructed resonance case. The PHSD vacuum mass is 0.892 GeV/c$^{2}$.}
  \label{fig:mass_width_kstar_reco_medium}
\end{figure}

Figure~\ref{fig:inv_kstarbar_decay_lowpt_medium} shows the invariant mass distributions of the $\bar K^{*0}$ in the low momentum region (p$_{\rm T}$ = 0.4-0.6 GeV/c) at the decay point (left) and the reconstructed (right). The $\bar K^{*0}$ at decay point already shows a wider distribution  (compare to figure~\ref{fig:inv_kstarbar_decay_lowpt}) due to the implementation of the 
in-medium effects. The reconstructed $\bar K^{*}$ resonances (right) from the final-state particles show a shift towards broader widths and smaller masses. Figure~\ref{fig:mass_width_kstar_reco_medium} shows the corresponding $\bar{K}^{*0}$ mass shift of 20-30 MeV/c$^{2}$ to lower masses (left) and the width is increased to $\Gamma_{\bar K^{*0}}$ = 80-100 MeV/c$^{2}$ in the low momentum region of p$_{\rm T}$~=~0-2~GeV/c.

\begin{figure}[h!]
\includegraphics[width=7.0cm]{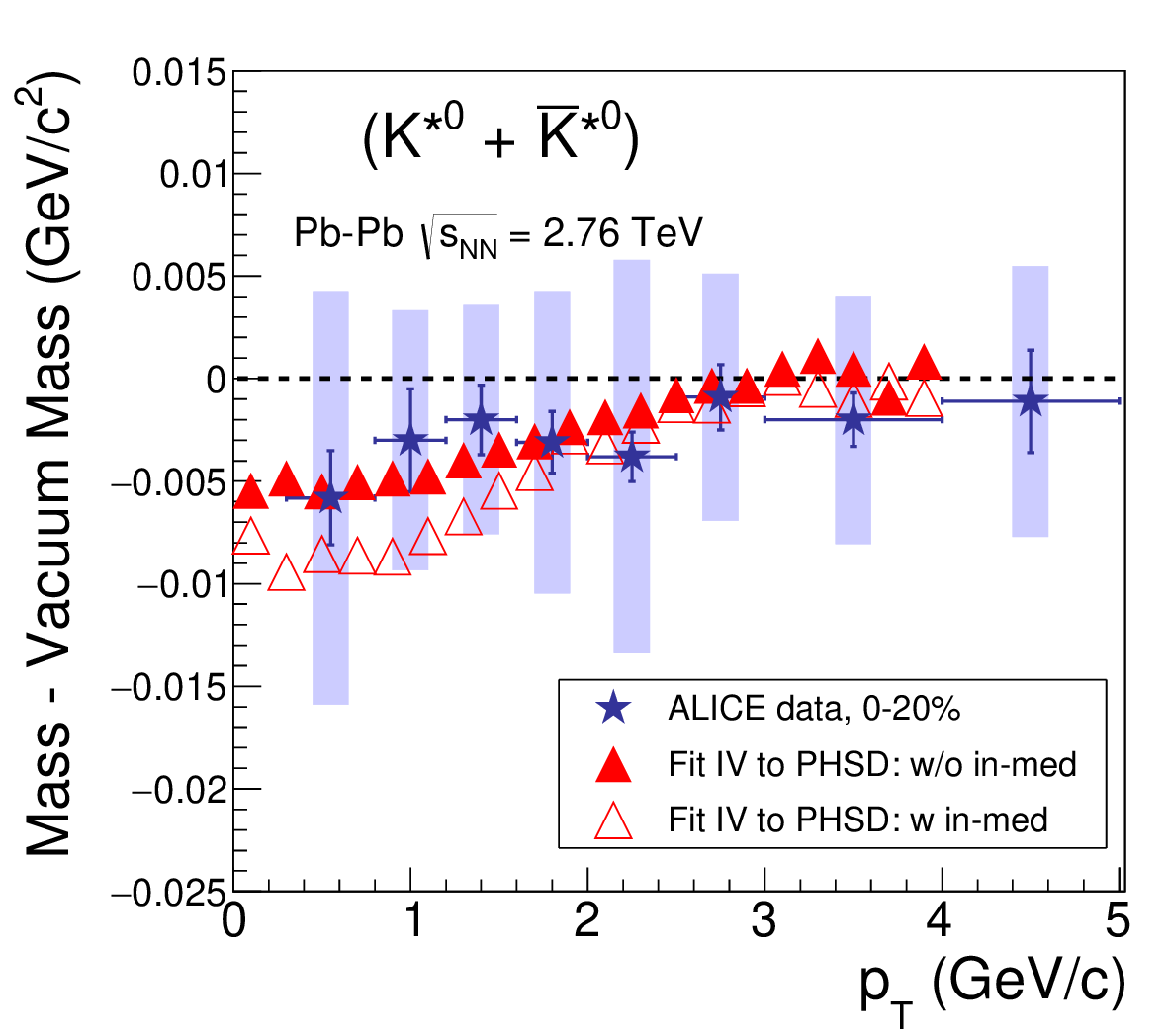}
  \caption{Deviations from the vacuum mass for the ALICE data \cite{Abelev:2014uua} and PHSD fitted mass spectrum of the reconstructed $K^{*0}$ + $\bar K^{*0}$ as a function of $p_T$. The red solid triangles show the PHSD calculation without the in medium modification and the red open triangels show the case for the in-medium modification.
 The PHSD invariant mass distribution was fitted with {\bf Fit IV} (explained in the text)  including statistical experimental error.}
  \label{fig:alice_med}
\end{figure}

The final comparisons to the experimental data are presented in figure~\ref{fig:alice_med} which shows the in-medium effects (open red triangles), which are very close to the calculation without the in-medium effects (solid red triangles). 
This shows that the presented way to 'measure' the invariant mass 
of the $\bar K^{*0}$ + $K^{*0}$ resonances  are not very sensitive to the in-medium 
scenarios. Even our 'extreme' scenario for the in-medium modification shows only a 5 MeV/c$^{2}$ mass shift in the low momentum region from p$_{\rm T}$~=~0-1.5~GeV/c. The data are consistent with both scenarios due to their large error bars.

  {
One of the important lessons from our study is related to the fact that the extraction 
of in-medium properties (masses and widths) of resonances is very sensitive 
to the fit procedure and especially to the ansatz for the mass distribution.
Here not only the form of the spectral function and the modelling of the in-medium masses and
widths is important, but also the modelling of the occupation probability which
describes the Boltzmann shape of the spectra. The fact that a large fraction of the
resonances is produced during the latest (hadronic) stage of the heavy-ion collisions
by annihilation of pions and kaons - requires to account for their phase-space distribution,
which is beyond the scope of simple thermal fits and requires a modelling of the
time evolution of the system which is, by the way, realized by default in transport
approaches.}
 Thus, the most straightforward way to extract the information on the in-medium modification would be a direct comparison of the experimental data with 
transport simulations.

\section{Summary}\label{sec:summary}

We have studied the strange vector-meson dynamics (for ${K}^{*}$ and $\bar{K}^{*}$) 
in p+p and heavy-ion collisions at relativistic energies based on the in-medium 
effects and off-shell propagation from 'G-Matrix' calculations within the framework 
of the PHSD transport approach. We have used the self-energies obtained in our previous study
\cite{Ilner:2013ksa} and implemented them in the form of mass shifts and widths into 
relativistic Breit-Wigner functions. On the basis of the widths and spectral functions we have
implemented the resulting cross-sections into the hadronization and $K \pi$ annihilation 
channels into PHSD  for all isospin channels of the ${K}^{*}$ and $\bar{K}^{*}$ mesons
\cite{Ilner:2016xqr}.
Since both the QGP and the hadronic phase are fully covered within the PHSD transport approach 
we have followed up the origin of the ${K}^{*}/\bar K^*$'s - created during the collision - 
as well as their properties. We have calculated differential spectra 
(as well as particle ratios) and compared the results to experimental data 
obtained by the ALICE Collaboration at the LHC. Furthermore, we have also obtained 
results on $K^*/\bar K^*$ in-medium effects from PHSD for lower energies that create systems 
of higher net-baryon density and will be studied  at the future FAIR and NICA.

Our findings are:

\begin{itemize}
\item At LHC energies - similar to the RHIC energies - the main production channel 
of the ${K}^{*}/\bar K^*$ mesons is the resonant annihilation  of $\pi + K(\bar K)$ pairs 
in the final hadronic phase.

\item Only a small fraction of ${K}^{*}/\bar K^*$ mesons, which is created during the 
hadronization of the QGP, contributes to the final spectra.

\item At high energies, e.g. RHIC energies of $\sqrt{{s}_{NN}}=200$~GeV or LHC energies 
of $\sqrt{{s}_{NN}}=2.76$~TeV, most ${K}^{*}/\bar K^*s$ are produced at rather low  baryon densities.
Consequently,  in-medium effects on the $K^*/\bar K^*$ spectral functions do not play 
a sensible role.

\item On the other hand, the ${K}^{*}/\bar K^*$'s created at lower bombarding energies can probe  
net-baryon densities of up to $\rho/{\rho}_{0} \approx 1.5$. This is due to the fact that 
many ${K}^{*}/\bar K^*$'s come from the annihilation of (anti-)kaons and pions in the hadronic phase. 
Although ${K}^{*}/\bar K^*$'s are created at all stages in the collision, a few of them stem 
from high baryon-density regions.

\item The medium at high energies expands very fast and only low baryon density 
regions can be reached because the system is dominated by the more abundant mesons 
rather than baryons. This is the case for RHIC and even more for LHC energies.

\item The PHSD results match the data for strange vector mesons in p+p collisions at LHC energies very well.

\item The transverse momentum spectra in Pb+Pb collisions at LHC energies are 
reproduced very well in the lower transverse momentum region while the PHSD 
results show a softer spectrum in the high $p_T$ region. 
Other observables like particle ratios and average momenta are in a reasonable 
agreement with experimental data.

\item Nevertheless, it is rather difficult to extract the $K^*/\bar K^*$ in-medium
properties  from the present experimental spectra at the LHC. 
By comparing the PHSD results for the 'true' $K^*/\bar K^*$ spectra
(calculated at the decay point of $K^*/\bar K^*$'s) with the 'reconstructed' spectra (obtained 
by matching the final pions and (anti-)kaons coming from the $K^*/\bar K^*$ decay), we have demonstrated how
the distortion of the spectra occurs due to i) the detector acceptance, the cuts in 
the invariant mass spectrum in the determination of the background spectra and 
due to ii) the rescattering and absorption of the (anti-)kaons and pions in the hadronic medium.

\item
In spite of the difficulties discussed above we have presented experimental 
procedures to obtain the information on the medium effects 
by performing fits of the  $K^*/\bar K^*$ 'reconstructed' mass spectra 
and extracting the mass shift and in-medium widths.
 We also demonstrated the sensitivity of  results to the actual ansatz used 
for the fit function. In order to pin down the robust conclusions on the in-medium effects 
from the measured mass distributions, a comparison with dynamical models is needed
to account for the dynamical origin of resonance formation during the time evolution 
of the expanding systems.
This will help to interpret the experimental measurement on the mass shift.

\item 
We have presented PHSD predictions  for strange vector meson
production at lower energies - from a few AGeV to few tens of AGeV,
achievable by the future FAIR, NICA and the BES program at RHIC.
Here the expected in-medium effects are large, especially for $\bar K^*$'s, 
due to the longer reaction time and higher baryon densities at the production 
of $K^*/\bar K^*$. However, similar to the high energy regime of LHC and RHIC, 
the 'reconstructed' spectra are distorted due to the final-state interaction 
of kaons and pions.

\item 
We find that the low momentum mass shift of  $K^{*0}$+$\bar{K^{*0}}$ at LHC and RHIC 
energies is in agreement with the 
theoretical calculations based on the PHSD transport approach. 
The mass shift is dominated by the later hadronic interactions and the exact reconstruction method used, such that even in the extreme model of large coupling to a baryonic medium the overall mass shift is within experimental errors.

\end{itemize}

Thus, our present analysis for the LHC energies together with our earlier study 
in Ref. \cite{Ilner:2013ksa} for RHIC energies and our new predictions for 
the FAIR/NICA/BES-RHIC energies show that the $K^*/\bar K^*$ resonances are 
much better suited to probe the final hadronic phase rather than the QGP at freeze-out. 
Moreover, the distortion of the initial shape of the $K^*$ spectra
at the decay point compared to the final spectra (reconstructed from $\pi + K$ 
directly) indicate a strong final-state interaction (rescattering and absorption) during 
the hadronic stage of HIC.
We note that the influence of hadronic interactions on the final observables 
has been also discussed in the recent study in Ref. \cite{Steinheimer:2017vju}.
Finally, the hadronic phase plays a dominant role for the $K^*/\bar K^*$
resonance dynamics in the heavy-ion collisions
and has to be accounted for the interpretation of the experimental results.

\vspace*{3mm}
\section*{Acknowledgements}\label{sec:ack}

The authors acknowledge inspiring discussions with J. Aichelin, W.~Cassing, A. Knospe, P. Moreau, A. Palmese, T. Song, L. Tolos and V. Voronyuk.
A.I. acknowledges support by HIC for FAIR and HGS-HIRe for FAIR. 
This work was supported by BMBF, HIC for FAIR and by U.S. Department of Energy Office of Science under contract number DE-SC0013391.
The computational resources have been provided by the LOEWE-CSC at the Goethe University Frankfurt.
The authors acknowledge the Texas Advanced Computing Center (TACC) at the University of Texas at Austin for providing computing resources that have contributed to the research results reported within this paper. 
URL: http://www.tacc.utexas.edu.

\end{document}